\pdfoutput=1
\documentclass{aastex63}

\newcommand{\beq}{\begin{equation}}
\newcommand{\beqa}{\begin{eqnarray}}
\newcommand{\eeq}{\end{equation}}
\newcommand{\eeqa}{\end{eqnarray}}

\newcommand{\simgt}{\lower.5ex\hbox{$\; \buildrel > \over \sim \;$}}
\newcommand{\simlt}{\lower.5ex\hbox{$\; \buildrel < \over \sim \;$}}

\newcommand{\bd}[1]{\mbox{\boldmath $#1$}}

\newcommand{\ms}[1]{\textcolor{black}{#1}}

\submitjournal{ApJ}

\shorttitle{Pairwise velocity between dark matter halos}
\shortauthors{Shirasaki et al.}
\graphicspath{{./}{figures/}}
\begin{document}

\title{A semi-analytic model of pairwise velocity distribution between dark matter halos}

\correspondingauthor{Masato Shirasaki}
\email{masato.shirasaki@nao.ac.jp}

%\author[0000-0002-0786-7307]{Greg J. Schwarz}
%\affiliation{American Astronomical Society \\
%1667 K Street NW, Suite 800 \\

\author{Masato Shirasaki}
\affiliation{National Astronomical Observatory of Japan, Mitaka, Tokyo 181-8588, Japan}
%\affiliation{Jet Propulsion Laboratory, California Institute of Technology, 4800 Oak Grove Drive, Pasadena, CA, USA}
%\affiliation{Overseas Research Fellowships by the Japan Society for the Promotion of Science (JSPS)}
\affiliation{The Institute of Statistical Mathematics,
Tachikawa, Tokyo 190-8562, Japan}

%\collaboration{1}{(AAS Journals Data Scientists collaboration)}

\author{Eric M.~Huff}
\affiliation{Jet Propulsion Laboratory, California Institute of Technology, 4800 Oak Grove Drive, Pasadena, CA, USA}

\author{Katarina Markovic}
\affiliation{Jet Propulsion Laboratory, California Institute of Technology, 4800 Oak Grove Drive, Pasadena, CA, USA}

\author{Jason D.~Rhodes}
\affiliation{Jet Propulsion Laboratory, California Institute of Technology, 4800 Oak Grove Drive, Pasadena, CA, USA}
\affiliation{Kavli Institute for the Physics and Mathematics of the Universe (WPI), The University of Tokyo Institutes for Advanced Study (UTIAS), The University of Tokyo, Chiba 277-8583, Japan }



\begin{abstract}
We study the probability distribution function (PDF) 
of relative velocity between two different dark matter halos 
(i.e.~pairwise velocity) with high-resolution cosmological 
$N$-body simulations.
We revisit a non-Gaussian framework 
to predict pairwise velocity statistics developed in Tinker (2007).
%and 
%calibrate a statistical relationship between the pairwise velocity and 
%an environmental mass density around halos in the framework.
We investigate the pairwise velocity PDFs over a wide range of halo masses 
of $10^{12.5} \simlt M\, [h^{-1}M_{\odot}] \simlt 10^{15}$ and redshifts of $0<z<1$.
At a given set of masses, redshift and the separation length between two halos, our model requires three parameters to set the pairwise velocity PDF, whereas previous non-Gaussian models in the literature assume four or more free parameters.
At the length scales of $5<r\, [h^{-1}\, \mathrm{Mpc}] <40$,
our model predicts the mean and 
dispersion of the pairwise velocity for dark matter halos with their masses of $10^{12.5} \simlt M\, [h^{-1}M_{\odot}] \simlt 10^{13.5}$ at $0.3 < z < 1$ with a 5\%-level precision.%, while the model precision reaches a 20\% level (mostly a 10\% level) for other masses and redshifts explored in the simulations.
We demonstrate that our model of the pairwise velocity PDF provides an accurate mapping of the two-point clustering of massive-galaxy-sized halos 
at the scales of $O(10)\, h^{-1}\mathrm{Mpc}$
between redshift and real space
for a given real-space correlation function.
For a mass-limited halo sample with their masses greater than $10^{13.5}\, h^{-1}M_{\odot}$ at $z=0.55$, 
our model can explain the monopole and quadropole moments 
of the redshift-space two-point correlations with 
a precision better than 5\% at the scales of $5-40$ and 
$10-30\, h^{-1}\mathrm{Mpc}$, respectively.
Our model of the pairwise velocity PDF will give 
a detailed explanation of statistics of massive galaxies
at the intermediate scales in redshift surveys. 
%including the non-linear redshift-space distortion effect in two-point correlation functions
%and the measurements of the kinematic Sunyaev-Zel'dovich effect.
\end{abstract}



\section{Introduction} \label{sec:intro}

% standard cosmology
Accelerating expansion of the late-time Universe is
a long-standing mystery in modern astronomy \citep[e.g.][for a review about observational probes]{2013PhR...530...87W}. 
%\textbf{JASON: I would remove 'recent' for a paper from 2013}
There are two leading physical models 
to solve the most surprising cosmological discovery in many decades. One is the dark energy model 
which assumes an exotic form of energy in the Universe, and 
the other requires a modification of General Relativity at long-length and weak-force regimes.
To distinguish the two models in an observational way, one needs
detailed measurements of cosmic mass density and velocity
over a large volume in the Universe.
A modification of gravity can induce the scale dependence on the gravitational growth and flow of cosmic mass density even at linear scales, while 
the large-scale density growth and flow can be uniquely determined by the expansion rate of the Universe alone in the presence of a smooth uniform dark energy under General Relativity \citep[but see, e.g.][for more detailed discussions]{2008PhRvD..78f3503J}.

% galaxy redshift survey and challenge in the current cosmological framework
Among observational probes, redshift surveys of distant galaxies are one of the most promising approaches to investigating density and velocity fields.
In the standard theory of formation of large-scale structures,
galaxies are thought to be a tracer of underlying cosmic mass density.
Their clustering properties contain rich cosmological information in principle.
The main challenges to use galaxies for cosmological studies are that galaxies are a biased tracer of density fields and a galaxy bias may depend on various factors. 
Throughout this paper, we consider a halo-based model for the galaxy bias.
Dark matter halos are gravitationally-bounded objects formed in cosmic matter distributions and a building block of large-scale structures \citep[see,][for a review of the halo model]{2002PhR...372....1C}.
In the halo model, one commonly assumes that dark matter halos host some galaxies and the number of galaxies in single halos depends on the halo mass alone.
These assumptions enable us to explain the observed clustering of a variety of galaxies in a precise way 
\citep[e.g.][]{2005ApJ...630....1Z, 2006MNRAS.365..842C, 2011ApJ...736...59Z}.
Nevertheless, the common halo-galaxy connection may cause serious systematic errors in galaxy-based cosmological analyses if the galaxy bias depends on other properties
\citep[e.g.][]{2007MNRAS.374.1303C, 2014MNRAS.443.3044Z}.

The two-point correlation function $\xi_{\mathrm{gg}}$ 
is a common observable in galaxy redshift surveys to measure the clustering of galaxies.
In a universe with statistical isotropy, the two-point correlation depends on the separation length between galaxies alone. However, actual surveys rely on the observation of redshifts to infer the line-of-sight distances to individual galaxies. A redshift $z$ in the spectrum of each galaxy is caused by cosmic expansion as well as the peculiar velocity of the galaxy itself. Hence, the spatial coordinate inferred by the observation of redshifts is different from the true counterpart. This effect is known as redshift-space distortions and the observed spatial separation (i.e. in redshift space) for a given pair of galaxies is then expressed as
\beqa
s_{p} &=& r_{p}, \\
s_{\pi} &=& r_{\pi} + \frac{(1+z)\, v_{z}}{H(z)},
\eeqa
where 
$H(z)$ is the Hubble parameter at $z$,
$r_{p}$ and $r_{\pi}$ represents the perpendicular and parallel components with respect to the line-of-sight direction, $s_{p}$ and $s_{\pi}$ are the counterparts in redshift space,
and $v_{z}$ is the relative velocity between two galaxies (i.e. the pairwise velocity) along the line of sight.
In redshift space, the line of sight toward each galaxy is a special direction and the observed two point correlation function depends on both of $s_{p}$ and $s_{\pi}$.
The relation of the two-point correlation between real and redshift space is given by \citep[e.g.][]{1980lssu.book.....P, 2004PhRvD..70h3007S},
\beqa
1+\xi^{S}_{\mathrm{gg}}(s_p, s_{\pi})
 = \int_{-\infty}^{\infty} \frac{H(z)\, \mathrm{d}r_{\pi}}{(1+z)} \, 
 {\cal P}_{\mathrm{g}}\left(v_z=\frac{H(z)(s_{\pi}-r_{\pi})}{(1+z)} \, \Bigl|\, s_p, r_{\pi}\right)\, \left[1+\xi_{\mathrm{gg}}(\sqrt{s^2_p+r^2_{\pi}})\right],
 \label{eq:2pcf_real_to_redshift}
\eeqa
where ${\cal P}_{\mathrm{g}}$ is the probability distribution function (PDF) of the line-of-sight pairwise velocity of galaxies, and $\xi^{S}_{\mathrm{gg}}$ is the two-point correlation function in redshift space.
Therefore, it is essential to develop an accurate theoretical model of the pairwise velocity statistics as well as the real-space correlation function for cosmological analyses with redshift surveys.

% small-scale measurements of pairwise velocity
% - assembly bias, modified gravity theory
% New observables will be available soon, e.g. kSZ and phase-space density 
Measurements of small-scale streaming motions of galaxies
may bring meaningful information to improve our understanding 
of the halo-galaxy relationship.
\citet{2018MNRAS.479.1579X} have shown that the dispersion of the pairwise velocity between two dark matter halos depends on properties other than the halo mass, while a similar effect has been found in a semi-analytic galaxy formation model \citep{2019MNRAS.486..582P}.
Apart from the halo-galaxy relationship, 
numerical simulations have shown that a modification of gravity can change the streaming motion between two massive-galaxy-sized halos with a separation length of $\sim1-5\, \mathrm{Mpc}$ \citep{2014PhRvL.112v1102H} as well as the infall velocity of galaxies to massive clusters
\citep{2012PhRvL.109e1301L, 2014MNRAS.445.1885Z}.
To infer the streaming motion of galaxies from observables in galaxy redshift surveys, we require a detailed modeling 
of the pairwise velocity statistics \citep[e.g.][also see Eq.~\ref{eq:2pcf_real_to_redshift}]{2004PhRvD..70h3007S}.
Furthermore, measurements of the secondary anisotropy of cosmic microwave background caused by the bulk motion of tracers of large-scale structures, referred to as the kinematic Sunyaev-Zel'dovich effect, have a great potential as a direct probe of the streaming motion of dark matter halos with upcoming experiments \citep[e.g.][]{2017JCAP...01..057S, 2018MNRAS.475.3764S, 2018arXiv181013423S}.
Therefore, it is timely and important to develop an accurate model of the pairwise velocity statistics of dark matter halos.

% A summary of previous velocity PDFs
% The aim of this paper
In this paper, we develop a semi-analytic model of the pairwise velocity distribution of dark matter halos with different masses, redshifts, and separations.
We pay a special attention to massive-galaxy-sized dark matter halos in this paper. 
This is because the precise measurements of small-scale two-point clustering are already available \citep[e.g.][]{2014MNRAS.444..476R, 2015MNRAS.446..578G}
and the statistical detections of the kinematic Sunyaev-Zel'dovich effect have been reported \citep[e.g.][]{2012PhRvL.109d1101H, 2017JCAP...03..008D, 2018MNRAS.475.3764S}.
Using the latest cosmological $N$-body simulations, we calibrate a physically-intuitive and efficient model of the pairwise velocity developed by \citet{2007MNRAS.374..477T} over a range of halo masses
$10^{12.5}<M\, [h^{-1}\, M_{\odot}] < 10^{15}$ at $0<z<1$.
We then validate that our model can reproduce the two-point correlation function in redshift space including the non-linear distortion effects due to the peculiar velocity of galaxies.
We also study information contents of the pairwise velocity statistics in galaxy clustering analyses.

The paper is organized as follows. 
In Section~\ref{sec:model},
we present an overview of the Tinker model and introduce our revised model.
In Section~\ref{sec:data}, we describe the %data of 
$N$-body simulations, mock galaxy catalogs, and clustering statistics used in this paper. 
We summarize our calibration process of the model parameters based on three non-zero moments in the pairwise
velocity in Section~\ref{sec:calibration}.
The results are presented in Section~\ref{sec:results}
and we mention the limitations in our model in Section~\ref{sec:limitations}.
Finally, the conclusions and discussions are provided in Section~\ref{sec:conclusion}.

\section{Pairwise velocity distribution of dark matter halos} \label{sec:model}

In this section, we briefly introduce a theoretical framework to predict the pairwise velocity statistics of dark matter halos proposed in \citet{2007MNRAS.374..477T}. We then present our new model with some modifications.
Table~\ref{tab:table_model_parameters} summarizes the model parameters in the framework and those are dependent on halo masses, redshifts, and separation lengths.

\subsection{Setup}

Consider a pair of two halos with their masses of $M_1$ and $M_2$ at a given redshift $z$. The pairwise velocity for the halo pair is then defined as the relative velocity between the two halos, 
\beqa
\bd{v}_{12}(\bd{r}) \equiv \bd{v}_{1}(\bd{r}_{1}) - \bd{v}_{2}(\bd{r}_{2}),
\eeqa
where $\bd{v}_{i}$ and $\bd{r}_{i}$ represent the velocity and position of the $i$-th halo $(i=1,2)$, 
and $\bd{r} \equiv \bd{r}_{1}-\bd{r}_{2}$.
Throughout this paper, 
we define the position of a given halo in the comoving coordinate, while the velocity is defined 
in the physical coordinate.
Assuming a spherically symmetric phase-space distribution of halos, we simplify $\bd{v}_{12}(\bd{r}) = \bd{v}_{12}(r)$ where $r = |\bd{r}|$.
In this paper, we seek a physically-motivated and efficient model of the probability distribution function (PDF) of the pairwise velocity. In particular, we study the dependence of the pairwise velocity PDF on halo masses ($M_1$ and $M_2$), redshifts $z$ and the separation length $r$.
Since the line-of-sight component of the pairwise velocity is relevant to statistical analyses in redshift surveys in practice (also see Section~\ref{subsec:clustering}), we focus on two different components in the three-dimensional velocity $\bd{v}$. One is the radial component of $v_r$ and another is the half of the tangential components $v_t$ at each radius $r$.
We define these two components as
\beqa
v_{r} &\equiv& \bd{v}_{12} \cdot \bd{r}/r, \\
v_{t} &\equiv& v_{x} \cos \theta \cos \phi + v_{y} \cos \theta \sin \phi - v_{z} \sin \theta,
\eeqa
where we set a Cartesian coordinate system of 
$\bd{r}/r = (\sin \theta \cos \phi, \sin \theta \sin \phi, \cos \theta)$, the angle $\theta$ is defined by the opening angle between the line-of-sight direction and the vector $\bd{r}$,
$v_{x}$ represents the $x$-axis component of $\bd{v}$ in the Cartesian coordinate and so on.
Note that $v_{z}$ corresponds to the line-of-sight velocity 
in this notation and it holds $v_{z} = v_{r} \cos \theta - v_{t} \sin \theta$.

\subsection{The Tinker model}

\citet{2007MNRAS.374..477T} proposed an analytic model of 
the joint PDF of ${\cal P}(v_r, v_t \, | \, r, M_1, M_2, z)$ for a pair of dark matter halos by assuming the following conditions:

\begin{enumerate}
\item[(i)] There exists a latent variable to make 
the joint PDF non-Gaussian, but the PDF at a given latent variable can be approximated as Gaussian. 
\item[(ii)] At a given latent variable, $v_{r}$ and $v_t$ are assumed to be independent.
\item[(iii)] The latent variable for modeling of ${\cal P}(v_r, v_t \, | \, r, M_1, M_2, z)$ is set to the local environmental mass overdensity $\delta$ around a pair of halos.
\end{enumerate}

These conditions allow us to express the joint PDF as the following functional form:
\beqa
{\cal P}(v_r, v_t \, | \, r, M_1, M_2, z)
= \int \mathrm{d}\delta \, {\cal N}\left(v_t\, |\, \mu_{t} \left(\delta\right), \Sigma_{t}\left(\delta\right)\right)\,
{\cal N}\left(v_r\, | \, \mu_{r}\left(\delta\right), \Sigma_{r}\left(\delta\right) \right)\, {\cal F}(\delta\, | \, r, M_1, M_2, z), \label{eq:T07_model_joint_PDF}
\eeqa
where ${\cal N}(u \, | \, \mu, \Sigma)$ represents
a Gaussian PDF of a one-dimensional random field $u$
with the mean of $\mu$ and the variance of $\Sigma^2$
and ${\cal F}(\delta\, | \, r, M_1, M_2, z)$ is the conditional PDF of the mass over-density field $\delta$ when one finds the halo pair with their masses of $M_1$ and $M_2$ at the redshift $z$ within the radius of $r$.
In Eq.~(\ref{eq:T07_model_joint_PDF}), \citet{2007MNRAS.374..477T} sets the mean $v_t$ at a given $\delta$ to be zero ($\mu_t = 0$), while the variables of $\Sigma_{t}$, $\mu_r$, and $\Sigma_{r}$ can depend on
the halo masses, redshift and radius as well as the environmental density $\delta$. 
In the following, we summarize key ingredients in Eq.~(\ref{eq:T07_model_joint_PDF}).

\subsubsection{Conditional PDF of mass overdensity}

The unconditional PDF of smoothed mass density at a smoothing scale of $r$ can be described by a log-normal distribution \citep[e.g.][]{1991MNRAS.248....1C, 1994ApJ...420...44K, 2001ApJ...561...22K}, while \citet{2007MNRAS.374..477T} found the conditional PDF ${\cal F}$
%finding a halo pairs within the radius of $r$ 
in numerical simulations is well fitted by
\beqa
{\cal F}(\delta\, | \, r, M_1, M_2, z) = 
{\cal A}\, \exp\left[-\frac{\tilde{\rho}_{0}(r,M_1,M_2,z)}{1+\delta}\right]\, P_{\mathrm{ln}}(\delta\, | \, r), \label{eq:cond_PDF}
\eeqa
where ${\cal A}$ is a normalization constant so that $\int \mathrm{d}\delta\, {\cal F} = 1$, $\tilde{\rho}_{0}$ is a density cutoff scale to be calibrated with $N$-body simulations, and $P_{\mathrm{ln}}$ is the log-normal distribution given by
\beqa
P_{\mathrm{ln}}(\delta\, | \, r) = 
\frac{1}{\sqrt{2\pi} \sigma_{\mathrm{ln}}}
\exp\left[-\frac{\{\ln(1+\delta)+\sigma^2_{\mathrm{ln}}/2\}^2}{2\sigma^2_{\mathrm{ln}}}\right] \frac{1}{1+\delta}. \label{eq:Lognormal_PDF}
\eeqa
In Eq.~(\ref{eq:Lognormal_PDF}), 
the variance $\sigma^2_{\mathrm{ln}}$ is set to be $\sigma^2_{\mathrm{ln}}(r,z) = \ln(1+\sigma^2_{\mathrm{NL}}(r,z))$ where $\sigma^2_{\mathrm{NL}}(r,z)$ is the non-linear mass variance smoothed by a top-hat filter at the scale of $r$ at the redshift $z$. 
To be specific, the top-hat mass variance is given by
\beqa
\sigma^2_{\mathrm{NL}}(r,z) &=& \int_{0}^{\infty}\, 
\frac{4\pi k^2\mathrm{d}k}{(2\pi)^3} \, W^2_{\mathrm{TH}}(kr)\, P_{\mathrm{NL}}(k,z), \\
W_{\mathrm{TH}}(x) &=&\frac{3\, (\sin x -x \cos x)}{x^3},
\eeqa
where $P_{\mathrm{NL}}(k,z)$ is the non-linear matter power spectrum at $z$.
\citet{2007MNRAS.374..477T} assumed that the density cutoff scale $\tilde{\rho}_{0}$ takes the form
\beqa
\tilde{\rho}_{0} (r,M_1,M_2,z) = \tilde{\rho}_{1}
\left(b_{\mathrm{L}}(M_1, z)+b_{\mathrm{L}}(M_2, z)\right)
+ \left(\frac{r}{r_0}\right)^{\alpha}, \label{eq:density_cut_off}
\eeqa
where $b_{\mathrm{L}}(M, z)$ is the linear halo bias at the redshift $z$, and three parameters of $\tilde{\rho}_{1}$,
$r_0$, and $\alpha$ have been calibrated with a set of $N$-body simulations. \citet{2007MNRAS.374..477T} found the simulation results can be explained by the form of Eq.~(\ref{eq:cond_PDF}) when $\tilde{\rho}_{1}=1.41$, $\alpha=-2.2$, and $r_{0} = 9.4 \times {\rm MAX}(R_{{\rm 200b},1}, R_{{\rm 200b}, 2})$ 
where $R_{{\rm 200b}, i}$ represents a spherical over-density radius of the halo of $M_{i}$. 
We here define the halo mass by $M\equiv M_{\mathrm{200b}}=(4\pi/3) \, 200\bar{\rho}_{\mathrm{m0}}\, R^3_{\mathrm{200b}}$,
where $\bar{\rho}_{\mathrm{m0}}$ is the mean cosmic mass density today and the halo radius $R_{\mathrm{200b}}$ is defined in the comoving coordinate.

Throughout this paper, we adopt the linear halo bias model in \citet{2010ApJ...724..878T}, while we compute the non-linear matter power spectrum by using the fitting formula calibrated by a set of $N$-body simulations \citep{2012ApJ...761..152T}. Note that we use the linear matter power spectrum {\it without} the baryon acoustic oscillations \citep{1998ApJ...496..605E} when computing $\sigma^2_{\mathrm{NL}}$ to avoid any oscillations in the predicted velocity moments at large scales of $\simgt10 \, h^{-1}\, \mathrm{Mpc}$. 

\subsubsection{Mean and variance at a given environmental density}

For a given halo pair at the separation of $r$ and the environmental density $\delta$, \citet{2007MNRAS.374..477T} developed a model of the mean infall velocity $\mu_r(\delta, r)$ by combining linear theory and the spherical collapse model. 
The linear theory predicts the relation of the velocity and overdensity fields as
\beqa
\mu_{\mathrm{lin}}(\delta, r, z) = -\frac{H(z)}{1+z} \, r \, f(z) \, \frac{\delta}{3},  \label{eq:mean_v_lin}
\eeqa
where $H(z)$ is the Hubble parameter at $z$, 
$f(z) = \mathrm{d}\ln\, D/\mathrm{d}\ln (1+z)^{-1}$ where $D(z)$ is the linear growth factor at $z$\footnote{We normalize $D(z)=1$ at $z=0$ throughout this paper.}.
At non-linear scales, the spherical collapse model can provide a reasonable approximation of the mean infall velocity. In 
the Einstein-de Sitter universe, one can derive the relation of the velocity and density perturbations as
\beqa
\mu_{\mathrm{sc}}(\delta, r, z) = \frac{H(z)}{1+z}\, r \, f(z)\, {\cal G}(\delta), \label{eq:mean_v_sc}
\eeqa
where ${\cal G}(\delta)$ is expressed in a parametric form as\footnote{Equation (17) in \citet{2007MNRAS.374..477T} misses the factor of $1/2$ to compute $1+{\cal G}$.}
\beqa
\delta = \frac{9}{2}\frac{(\gamma - \sin \gamma)^2}{(1-\cos \gamma)^3} - 1, \,\,\,\,\,\,\,
{\cal G} = \frac{3}{2} \frac{\sin \gamma (\gamma-\sin \gamma)}{(1-\cos \gamma)^2} -1.
\eeqa
\citet{2007MNRAS.374..477T} then proposed a model by combining Eqs.~(\ref{eq:mean_v_lin}) and (\ref{eq:mean_v_sc}): 
\beqa
\mu_{r}(\delta, r, M_1, M_2, z) 
= \left\{
\begin{array}{ll}
w(r) \mu_{\mathrm {sc}}(\delta, r, z) \, \exp\left[-\left(\frac{4.5}{r(1+\delta)}\right)^2\right]
+ 
\left[1-w(r)\right] \mu_{\mathrm{lin}}(\delta, r, z) & 
\, (r>R_{\mathrm{cut}}) \\
\mu_{\mathrm {sc}}(\delta_{\mathrm{cut}}, r, z) \, \exp\left[-\left(\frac{4.5}{r(1+\delta_{\mathrm{cut}})}\right)^2\right] & (r\le \, R_{\mathrm{cut}}) \\
\end{array} \right., \label{eq:mean_v_T07}
\eeqa
where $R_{\mathrm{cut}} = {\rm MAX}(R_{{\rm 200b},1}, R_{{\rm 200b}, 2})$,
$1+\delta_{\mathrm{cut}} = \exp(-\sigma^{2}_{\mathrm{ln}}/2)$, and the weight function $w(r)$ and the exponential cutoff have
been calibrated against the numerical simulations. 
\citet{2007MNRAS.374..477T} found that the following weight function shows a reasonable fit to the simulation results,
\beqa
w(r) 
= \left\{
\begin{array}{ll}
1 & \,\,\,\, (r\, [h^{-1}\mathrm{Mpc}]\le 4) \\
1.86-0.62 \ln r & \,\,\,\, (4 < r \, [h^{-1}\mathrm{Mpc}]\le 20) \\
0 & \,\,\,\, (20 < r\, [h^{-1}\mathrm{Mpc}])
\end{array} \right. .
\eeqa

For the velocity dispersions $\Sigma_{t,r}$, \citet{2007MNRAS.374..477T} introduced the following 
parametric form of 
\beqa
\Sigma_{t,r} = 200 \, [\mathrm{km}/\mathrm{s}] \, \left(\frac{\Omega_{\mathrm{m}}(z)}{0.3}\right)^{0.6}\, \left(\frac{D(z)\, \sigma_8}{0.8}\right)\,
\left(\frac{1+\delta}{\tilde{\rho}_{t,r}}\right)^{\beta}, \label{eq:Sigma_v_T07}
\eeqa
where $\Omega_{\mathrm{m}}(z) = \Omega_{\mathrm{m0}}(1+z)^3 / [\Omega_{\mathrm{m0}}(1+z)^3+(1-\Omega_{\mathrm{m0}})]$ ($\Omega_{\rm m0}$ is the mass-density parameter today),
$\sigma_8$ is the mass variance for the linear overdensity field at $z=0$ when smoothed by the top-hat filter 
at $8\, h^{-1}\mathrm{Mpc}$, 
and three parameters $\tilde{\rho}_{t,r}$ and $\beta$
have been calibrated with the simulation results as a function of $r$, $M_{1}$ and $M_{2}$. 
Note that the scaling with $\Omega_{\mathrm{m}}(z)$ and $\sigma_8$ in Eq.~(\ref{eq:Sigma_v_T07}) is motivated 
by the linear theory (recall $f\simeq \Omega^{0.6}_{\rm m}(z)$). For $M_{1}\ge M_{2}$, 
the fitting formulas are summarized as
\beqa
\beta(r) &=& \left(\frac{r}{35\, h^{-1}\mathrm{Mpc}}\right)^{0.1}, \label{eq:Sigma_power_law}\\
\tilde{\rho}_{t}(r, M_1, M_2) &=& 
\left(\frac{r}{7.2 R^{1/2}_{\mathrm{200b, 1}}}\right)^{-2.5}
+
\left(\frac{r}{12.6 R^{1/2}_{\mathrm{200b, 0}}}\right)^{-0.8}
+0.48, \label{eq:Sigma_rho_t_T07} \\
\tilde{\rho}_{r}(r, M_1, M_2) &=& 
\left(\frac{r}{5.0 R^{1/2}_{\mathrm{200b, 1}}}\right)^{-4.0}
+
\left(\frac{r}{11.5 R^{1/2}_{\mathrm{200b, 0}}}\right)^{-1.3}
+0.50, \label{eq:Sigma_rho_r_T07}
\eeqa
where $R_{\mathrm{200b},0} = R_{\mathrm{200b},1}+R_{\mathrm{200b},2}$ in comoving $h^{-1}\, \mathrm{Mpc}$.

\subsection{New model}

The model by \citet{2007MNRAS.374..477T} is physically-intuitive and efficient to compute the pairwise velocity PDF for dark matter halos, but we find that it does not provide a reasonable fit to the latest high-resolution simulation results as shown in Section~\ref{sec:results}.
There may be several reasons why the model can not reproduce the simulation results today. 
A major concern about the model of \citet{2007MNRAS.374..477T} 
is that its parameter calibration relies on the results of $N$-body simulations in a $\Lambda$CDM cosmology with the spectral index $n_s=1$ and a larger amplitude of the initial density power spectrum at $k=0.05\, \mathrm{Mpc}^{-1}$
than the inferred value from Planck \citep{2016A&A...594A..13P}. This can affect the kinematics of dark matter halos even at large scales, because the linear velocity in Fourier space scales with $\delta / k$ where $k$ is the wave number. In addition, the simulations in \citet{2007MNRAS.374..477T} assume $\Omega_{\rm m0}=0.1$
and $\sigma_8 = 0.95$ at $z=0$, which may result in sizable differences in the non-linear evolution 
of cosmic mass density.
Furthermore, the simulations consist of 
$360^3$ particles in a volume of $253^3\, [h^{-1}\mathrm{Mpc}]^3$ and the mass resolution may be less sufficient to study the halo-galaxy connection in a modern manner.
In fact, recent observations of massive galaxies in the Sloan Sky Digital Sky Survey III (SDSS III) have shown that the kinematics of galaxies closely relate to the phase-space density in the inner regions of their host dark matter halos \citep{2014MNRAS.444..476R, 2015MNRAS.446..578G},
while the halo velocity in \citet{2007MNRAS.374..477T} 
is defined by the center-of-mass velocity.
Detailed simulations show that
halo cores are not at rest relative to the halo bulk \citep{2013ApJ...762..109B}.
High-resolution and large-volume cosmological simulations would be needed to re-calibrate the model of \citet{2007MNRAS.374..477T} and this is the scope of this paper. 
%\textbf{JASON:  Do you mean this is out of scope (instead of our scope)?}

Our new model follows the basic concept in \citet{2007MNRAS.374..477T}, but we introduce minor revisions so that the final model can reproduce the latest simulation results over a wide range of halo masses, redshifts, and separation lengths.
For the conditional PDF of finding a halo pair given a mass density,
we adopt the exponential cutoff as in Eq.~(\ref{eq:cond_PDF}) to effectively include the environmental dependence of halo formation and parametrize the density cutoff scale as in Eq.~(\ref{eq:density_cut_off}), but we allow a more complicated mass and redshift dependence:
\beqa
\tilde{\rho}_{0} (r,M_1,M_2,z) = {\cal A}_{\rho}(M_1, M_2, z)
\left(b_{\mathrm{L}}(M_1, z)+b_{\mathrm{L}}(M_2, z)\right)
+ {\cal B}_{\rho}(M_1, M_2, z)\, \left(\frac{r}{{\rm MAX}(R_{{\rm 200b},1}, R_{{\rm 200b}, 2})}\right)^{{\cal C}_{\rho}(M_1, M_2, z)}, \label{eq:density_cut_off_ours}
\eeqa

Also, we modify the functional form of mean radial velocity at a given overdensity, $\mu_{r}(\delta)$, as
\beqa
\mu_r(\delta, r, z) = -\frac{H(z)}{1+z} \, r \, \frac{f(z)}{3}\, \delta_{c} \left[(1+\delta)^{1/\delta_c}-1\right], \label{eq:mean_v_our}
\eeqa
where $\delta_{c}=1.686$. 
Eq.~(\ref{eq:mean_v_our}) is the approximate solution of the infall velocity for the spherical collapse model
when the inital condition of the radial shells is set by the Zeldovich approximation \citep{1996ApJS..103....1B, 2006ApJ...645..783S, 2008MNRAS.389.1249L}.
Therefore, it naturally reduces to the linear-theory prediction (Eq.~[\ref{eq:mean_v_lin}]) at $\delta \rightarrow 0$ and we do not introduce any weight functions to stitch the solution between non-linear and linear regimes.

For the velocity dispersions, we keep the functional form of Eq~(\ref{eq:Sigma_v_T07}), but 
we generalize the dependence of 
$\tilde{\rho}_{t,r}$ 
on halo masses and radius by using a double power-law form of
\beqa
\tilde{\rho}_{t,r}(r, M_1, M_2, z) = 
{\cal C}^{(0)}_{t,r}(M_1, M_2, z) \, 
r^{p_{t,r}(M_1, M_2, z)}
+ {\cal C}^{(1)}_{t,r}(M_1, M_2, z) \, 
r^{q_{t,r}(M_1, M_2, z)} + 
{\cal C}^{(2)}_{t,r}(M_1, M_2, z), \label{eq:Sigma_rho_t_r_ours}
\eeqa
where we introduce five functions of ${\cal C}^{(i)} \, (i=0-2)$, $p$ and $q$ for each velocity dispersion.
Note that the power-law index of Eq.~(\ref{eq:Sigma_v_T07}) is fixed to Eq.~(\ref{eq:Sigma_power_law}) 
in the new model as well.

The detailed forms of ${\cal A}_{\rho}$, ${\cal B}_{\rho}$, ${\cal C}_{\rho}$, 
${\cal C}^{(i)}_{t,r}\, (i=0-2)$, $p_{t,r}$, and $q_{t,r}$
are found in Appendix~\ref{apdx:model_params}.
We also provide the details of our calibration process 
to find the forms of various functions in Section~\ref{sec:calibration}.

\begin{table*}[t!]
\renewcommand{\thetable}{\arabic{table}}
\centering
\caption{A short summary of parameters in the model of pairwise velocity PDF of dark matter halos.} \label{tab:table_model_parameters}
\begin{tabular}{c|c|c|l}
\tablewidth{0pt}
\hline
\hline
Model parameters 
&
\citet{2007MNRAS.374..477T} 
&
This paper
& 
Reference
\\
\hline
\decimals
$\tilde\rho_{0}$ 
& 
Eq.~(\ref{eq:density_cut_off}) 
&
Eq.~(\ref{eq:density_cut_off_ours})
&
Density cutoff scale on the halo formation as in Eq.~(\ref{eq:cond_PDF})
\\
\hline
\decimals
$\mu_{r}$ 
& 
Eq.~(\ref{eq:mean_v_T07}) 
&
Eq.~(\ref{eq:mean_v_our})
&
Mean radial velocity at a given environmental density
\\
\hline
\decimals
$\tilde\rho_{t,r}$ 
& 
Eqs.~(\ref{eq:Sigma_rho_t_T07}) \& (\ref{eq:Sigma_rho_r_T07}) 
&
Eq.~(\ref{eq:Sigma_rho_t_r_ours})
&
Scale density on the dispersion-density relation as in Eq.~(\ref{eq:Sigma_v_T07})
\\
\hline
\end{tabular}
\end{table*}

%\begin{figure*}[t!]
%\gridline{\fig{sig2_comp.pdf}{1.\textwidth}{}}
%\caption{
%The effect of baryon acoustic oscillations on the mass variance.
%\label{fig:mass_variance}}
%\end{figure*}

\section{Data} \label{sec:data}

\subsection{$N$-body simulations and halo catalogs}

To study the pairwise velocity statistics of dark matter halos, we use a set of publicly available halo catalogs provided by the $\nu^2$GC collaboration\footnote{The data are available at \url{https://hpc.imit.chiba-u.jp/~nngc/}.}.
\citet{2015PASJ...67...61I} performed a series of high-resolution cosmological (dark-matter-only) $N$-body simulations with various combinations of mass resolutions and volumes on the basis of the $\Lambda$CDM cosmology consistent with observational results obtained by the Planck satellite. Among them, we use the halo catalogs based on the largest-volume run called $\nu^2$GC-L run, which consists of $8192^3$ dark matter particles in a box of $1.12\, h^{-1}\mathrm{Gpc}$. The corresponding mass resolution is $2.2\times10^{8}\, h^{-1}M_{\odot}$, allowing us to study the core velocity of dark matter halos in a robust way.
The simulations were performed by a massive parallel TreePM code of GreeM$^3$ \citep{2009PASJ...61.1319I, 2012arXiv1211.4406I} on the K computer at the RIKEN Advanced Institute for Computational Science, and Aterui super-computer at Center for Computational Astrophysics (CfCA) of National Astronomical Observatory of Japan.
The authors generated the initial conditions by a publicly available code, 2LPTic\footnote{\url{http://cosmo.nyu.edu/roman/2LPT/}}, using second-order Lagrangian perturbation theory \citep[e.g.][]{2006MNRAS.373..369C}, as well as 
the online version of CAMB\footnote{\url{http://lambda.gsfc.nasa.gov/toolbox/tbcambform.cfm}} \citep{Lewis:1999bs} to set 
the linear power spectrum at the initial redshift of $z=127$.
In the simulations, the following cosmological parameters were adopted: $\Omega_{\mathrm{m0}}= 0.31$,
$\Omega_{\mathrm{b0}}= 0.048$,
$\Omega_{\Lambda}=1-\Omega_{\mathrm{m0}} = 0.69$,
$h= 0.68$,
$n_s= 0.96$, and $\sigma_8= 0.83$.
These are consistent with Planck \citep{2016A&A...594A..13P}.

In this paper, we use the halo catalogs produced with the ROCKSTAR halo finder \citep{2013ApJ...762..109B} at four different redshifts of $z=0, 0.30, 0.55$ and $1.01$. 
We focus on parent halos identified by the ROCKSTAR algorithm and exclude any subhalos in the following analyses. The halo position is defined by the center-of-mass location of a subset of member particles in the inner halo density, while the velocity is computed by the average particle velocity
within the innermost 10\% of the virial radius.
We keep the halos with $M>10^{12.5}\, h^{-1}M_{\odot}$ as a very conservative choice to study the halo properties (i.e. the smallest halos in the analysis consist of $\sim14000$ dark matter particles).
To study the mass dependence, we divide the halos into six subgroups by their masses: 
$M\, [h^{-1}M_{\odot}] = 10^{12.5-13}, 10^{13-13.5}, 10^{13.5-14}, 10^{14-14.5}$ and $10^{14.5-15}$.
Table~\ref{tab:table_halos} summarizes the number of dark matter halos in each subgroup of interest.
We use these subgroups to calibrate the model parameters as in Section~\ref{sec:calibration}.

\subsection{Mock galaxy catalogs}\label{subsec:mock}

To test our model of the pairwise velocity distribution 
of dark matter halos, we produce a set of 
mock galaxy catalogs. 
For the simplest model, we consider a mass-limited sample with the halo mass above $M_{\mathrm{th}}$ at different redshifts.
For the mass-limited sample, we consider two different mass thresholds of $M_{\mathrm{th}} = 10^{12.5}$ and $10^{13.5}\, h^{-1}M_{\odot}$, which are typical halo masses of massive early-type galaxies at $z<1$ \citep[e.g.][]{2007ApJ...667..760Z, 2009ApJ...707..554Z, 2009ApJ...698..143R, 2012ApJ...744..159L, 2017ApJ...839..121T}.
In the mass-limited sample, we do not include satellite galaxies in their host halos and assume that there exist single galaxies at the center of their hosts.
These mass-limited samples enable us to examine our interpolation scheme over the halo masses in the model of the pairwise velocity distribution. 

\begin{table*}[t!]
\renewcommand{\thetable}{\arabic{table}}
\centering
\caption{The number of dark matter halos analyzed in this paper. Note that the halo mass is defined by the mass of a spherical overdensity, with 200-times the mean density of the universe.} \label{tab:table_halos}
\begin{tabular}{c|c|c|c|c}
\tablewidth{0pt}
\hline
\hline
Halo mass
&
$z=0$
&
$z=0.30$
& 
$z=0.55$
&
$z=1.01$
\\
\hline
\decimals
$10^{12.5} \le M\, [h^{-1}\, M_{\odot}]<10^{13}$ 
& 1,514,560
& 1,406,363
& 1,289,149
& 1,014,952
\\
\hline
\decimals
$10^{13} \le M\, [h^{-1}\, M_{\odot}]<10^{13.5}$ 
& 513,892
& 444,888
& 377,278
& 244,833
\\
\hline
\decimals
$10^{13.5} \le M\, [h^{-1}\, M_{\odot}]<10^{14}$ 
& 155,681
& 118,600
& 87,235
& 40,634
\\
\hline
\decimals
$10^{14} \le M\, [h^{-1}\, M_{\odot}]<10^{14.5}$ 
& 36,977
& 22,295
& 12,805
& 3,339
\\
\hline
\decimals
$10^{14.5} \le M\, [h^{-1}\, M_{\odot}]<10^{15}$ 
& 5,136
& 2,038
& 767
& 68
\\
\hline
\end{tabular}
\end{table*}

For a more realistic catalog, we employ the halo occupation distribution (HOD) method that allows us to populate hypothetical galaxies into halos in the simulations. 
The HOD, denoted by $\langle N_{\mathrm{gal}}\rangle_M$, gives the mean number of galaxies in host halos with mass $M$.
As a representative example, we consider the spectroscopic sample of massive galaxies in the SDSS-III Baryon Oscillation Spectroscopic Survey (BOSS). There are two targets of galaxies in the BOSS, but we focus on the sample referred to as CMASS. The CMASS sample is designed to be a roughly volume-limited sample of massive, luminous galaxies \citep{2011MNRAS.418.1055M} and has a large galaxy bias of $b\sim2$, showing that most galaxies reside in the dark matter halos of $M\sim10^{13}\, h^{-1}M_{\odot}$ \citep{2011ApJ...728..126W}.

For the HOD of the CMASS sample, we adopt the model in \citet{2014MNRAS.444..476R} with the form of
\beqa
\langle N_{\mathrm{gal}}\rangle_M &=& 
\langle N_{\mathrm{cen}}\rangle_M + 
\langle N_{\mathrm{sat}}\rangle_M, \label{eq:HOD} \\
\langle N_{\mathrm{cen}}\rangle_M &=&
\frac{1}{2}\left[1+\mathrm{erf}\left(\frac{\log_{10} M - \log_{10} M_{\mathrm{min}}}{\sigma_{\log_{10} M}}\right)\right], \label{eq:HOD_cen}\\
\langle N_{\mathrm{sat}}\rangle_M &=&
\langle N_{\mathrm{cen}}\rangle_M 
\left(\frac{M - M_{\mathrm{cut}}}{M_1}\right)^{\alpha_M} {\cal H}(M-M_{\mathrm{cut}}),\label{eq:HOD_sat}
\eeqa
where ${\cal H}(x)$ is the Heaviside step function, 
$\langle N_{\mathrm{cen}}\rangle_M $ and 
$\langle N_{\mathrm{sat}}\rangle_M $
represent the HODs for the central and the satellite galaxies, respectively. 
We adopt the best-fit parameters in \citet{2014MNRAS.444..476R}:
$\log_{10}M_{\mathrm{min}}=13.031$, 
$\sigma_{\log_{10} M} = 0.38$,
$\log_{10}M_{\mathrm{cut}}=13.27$,
$\log_{10}M_1 = 14.08$,
and $\alpha_M = 0.76$.
Using the HOD in Eqs.~(\ref{eq:HOD})-(\ref{eq:HOD_sat}),
we populate the $\nu^2$GC-L halos with hypothetical CMASS galaxies at $z=0.55$ in the following manner.

\begin{enumerate}
    \item[(i)] We populate halos with central CMASS galaxies by randomly selecting halos according to the probability distribution, $\langle N_{\mathrm{cen}}\rangle_M$ (Eq.~[\ref{eq:HOD_cen}]). In this step, we assume that each central galaxy resides at the halo center \ms{and is at rest with respect to the host halo.}
    \item[(ii)] When halos have central galaxies, we then randomly populate the halos with satellite galaxies assuming a Poisson distribution with the mean of $\lambda_{M} =\left[(M - M_{\mathrm{cut}})/M_1\right]^{\alpha_M} 
    {\cal H}(M-M_{\mathrm{cut}})$. We assume that the radial distribution of satellites on average follows that of dark matter in each host halo. 
    We simply assume the analytical Navarro-Frenk-White (NFW) profile \citep{1996ApJ...462..563N}, where we use the concentration-mass-redshift relation in \citet{2015ApJ...799..108D}, 
    to compute the density profile for each host halo.
    We set the halo-centric radius of each satellite by drawing a random variable $q$ which follows $M_{\mathrm{NFW}}(<q)/M_{\mathrm{NFW}}(<R_{\mathrm{200b}})$. Here $M_{\mathrm{NFW}}(<r)$ represents the enclosed mass predicted by the NFW profile as a function of radius $r$.
    \item[(iii)] For each satellite galaxy, we assign a virial random motion by using a Gaussian random variable with the zero mean and the variance of $\sigma^2_{\mathrm{vir}} = (1+z) \, G M / (2 R_{\mathrm{200b}})$. Note that the halo radius is defined in the comoving coordinate in this paper. 
\end{enumerate}

\ms{After adopting the above procedures, we find the number density of our mock galaxies to be $4.37\times10^{-4}\, [(h\, \mathrm{Mpc}^{-1})^3]$.
This density is in good agreement with the value in \citet{2014MNRAS.444..476R}\footnote{The difference between two is a 6\% level.}.} 
Note that in principle we could
use the concentration of individual halos measured by ROCKSTAR to set the NFW distribution of satellites. Instead we simply adopt the model of \citet{2015ApJ...799..108D} to ignore a possible impact 
of scatter in the halo concentration. Our primary purpose is to validate if our model can be suitable to model the two-point correlation function in redshift space at a scale of $\sim10\, h^{-1}\mathrm{Mpc}$ where 
the dominant contribution to any clustering observable is expected to come from pairs of central galaxies \citep[e.g.][]{2016MNRAS.458.4015Z}.

\subsection{Clustering statistics}\label{subsec:clustering}

For a given catalog of mock galaxies in Section~\ref{subsec:mock}, we perform a two-point clustering analysis in redshift space to test if our model is useful for the most widely used statistics in redshift surveys.
The two-point correlation function of galaxies 
is formally defined by
\beqa
\langle n_{\mathrm{g}}(\bd{r}_1) n_{\mathrm{g}}(\bd{r}_2)\rangle =  \bar{n}^2_{\mathrm{g}}\left(1+\xi_{\mathrm{gg}}(\bd{r})\right),
\eeqa
where $\bd{r}=\bd{r}_1-\bd{r}_2$, $n_{\mathrm{g}}$ represents the number density field of galaxies of interest, $\bar{n}_{\mathrm{g}}$ is the mean density, and $\xi_{\mathrm{gg}}(\bd{r})$ is the two-point correlation function. 
Since distances to individual galaxies are affected by the redshift-space distortion, the galaxy two-point correlation must be anisotropic as in Eq.~(\ref{eq:2pcf_real_to_redshift}) in practice.
Eq.~(\ref{eq:2pcf_real_to_redshift}) also shows that the anisotropy in the observed galaxy clustering is set by the pairwise velocity PDF for a given cosmology. According to this fact, we shall validate our model of pairwise PDFs by studying the mapping of two-point correlation functions between real and redshift space.

In this paper, we measure the two-point correlation functions, $\xi^{S}_{\mathrm{gg}}$ and $\xi_{\mathrm{gg}}$ for a given mock catalog. For comparison, we then predict $\xi^{S}_{\mathrm{gg}}$ by using Eq.~(\ref{eq:2pcf_real_to_redshift}) with the true $\xi_{\mathrm{gg}}$ in the simulations and our model 
of ${\cal P}_{\mathrm{g}}(v_z)$.
We derive an analytic expression of ${\cal P}_{\mathrm{g}}(v_z)$ for the HOD-based model in Appendix~\ref{apdx:halo_streaming_model}.
\ms{In the simulations, 
we adopt the distant-observer approximation.
For the line-of-sight direction, 
we set an axis in the Cartesian coordinate system 
applied to the simulation box.}
We measure the two point correlation function by using the natural estimator of $\mathrm{DD}/\mathrm{RR}-1$ where $\mathrm{DD}$
and $\mathrm{RR}$ represent the number of pairs of galaxies and random points at a given separation, respectively\footnote{
The Landy-Szalay estimator \citep{1993ApJ...412...64L} is often adopted in the literature and has a different form from $\mathrm{DD}/\mathrm{RR}-1$.
However, the difference between two is only important for large scales of $\sim100 \,\mathrm{Mpc}$.
Since we are working on smaller scales ($\sim10\, \mathrm{Mpc}$), 
it should not be necessary for our purpose.}. In the periodic box without boundaries, we can compute the number of random points analytically. For the measurement of $\xi_{\mathrm{gg}}(r)$, 
we employ the logarithmic binning in the range of $r=0.01\, h^{-1} \mathrm{Mpc}$ to $100\, h^{-1}\mathrm{Mpc}$ with the number of bins being 40. For the redshift-space correlation, we measure $\xi^{S}_{\mathrm{gg}}(s_p, s_{\pi})$ in the linearly-spaced bins over $0 < s_{p, \pi} \, [h^{-1}\, \mathrm{Mpc}]< 50$ with the number of bins in each direction being 50.
\ms{We measure $\xi^{S}_{\mathrm{gg}}$ 
while changing three axes in the Cartesian coordinate system.
We then take the average over 3 realizations 
to have final results of $\xi^{S}_{\mathrm{gg}}$.}

In practice, it is more common to compress the information in the two-point correlation in redshift space by using the  Legendre expansion:
\beqa
\xi_{\ell}(s) \equiv \frac{2\ell+1}{2}
\int_{-1}^{1}\, \mathrm{d}\mu \, 
\xi^{S}_{\mathrm{gg}}(s_p, s_{\pi})\, {\cal L}_{\ell}(\mu),
\label{eq:clustering_moments}
\eeqa
where $s=(s^2_p+s^2_{\pi})^{1/2}$, 
$\mu = s_{\pi}/s$, ${\cal L}_{\ell}(\mu)$ is the Legendre polynomial of order $\ell$.
We measure the first three non-zero moments $(\ell=0,2,4)$
for a given mock catalog. We evaluate $\xi_{\ell}(s)$ by using the measurement of $\xi^{S}_{\mathrm{gg}}(s_p, s_{\pi})$
with 20 logarithmic bins in the range of $0.5 < s \, [h^{-1}\, \mathrm{Mpc}]< 50.0$ and 
40 linear bins of $\mu$ with the width of $\Delta \mu=0.05$.
For comparison purposes, we estimate the variance of $\xi_{\ell}(s)$ by dividing the data volume into 
$2^3$ sub-volumes and measuring $\xi_{\ell}(s)$ for galaxies in each sub-volume. We then compute the variance of $\xi_{\ell}(s)$ at a given $s$ as
\beqa
{\rm Var}[\xi_{\ell}(s)] = \frac{V_{\rm sub}}{V_{\rm full}}\frac{1}{N-1}\sum_{i=1}^{N}
\left[\xi_{\ell}(s;i)-\bar{\xi}_{\ell}(s)\right]^2, \label{eq:var_xis}
\eeqa
where $N=8$, $V_{\rm sub}$ is the sub-volume, $V_{\rm full}$ is the full data volume, $\xi_{\ell}(s;i)$ represents the clustering multipole for the $i$-th subsample, and $\bar{\xi}_{\ell}(s)$ is the average multipole over 8 sub-volumes. Note that $V_{\rm full} = 8 \, V_{\rm sub}$.
Eq.~(\ref{eq:var_xis}) provides a rough estimate of the sample variance of the measurements of $\xi_{\ell}(s)$ for the survey volume of $(1.12)^3 \simeq 1.40\, [h^{-1}\, \mathrm{Gpc}]^3$.

\section{Calibration of the model parameters} \label{sec:calibration}

In this section, we summarize how to calibrate the model parameters in Section~\ref{sec:model} with the pairwise velocity statistics in the simulations. In principle, 
we can determine the functions of $\tilde{\rho}_{0}(r, M_1, M_2, z)$, $\tilde{\rho}_{t}(r, M_1, M_2, r)$ and $\tilde{\rho}_{r}(r, M_1, M_2, r)$ by using the smoothed density distribution from $N$-body particles and the statistics of halo pairs. Unfortunately, the particle data in the $\nu^2$GC-L run are not saved because of the hard drive shortage.
%\textbf{JASON: I didn't understand the previous sentence.  Was the particle data not saved?  If so just say that.} 
Hence, we assume the specific forms of three functions $\tilde{\rho}_{0}(r, M_1, M_2, z)$, $\tilde{\rho}_{t}(r, M_1, M_2, r)$ and $\tilde{\rho}_{r}(r, M_1, M_2, r)$ as in Eqs~(\ref{eq:density_cut_off_ours})
and (\ref{eq:Sigma_rho_t_r_ours}), but we attempt to find the parameters in the functions so that the model can reproduce the first three non-zero moments of the pairwise velocity in the simulations.

Given the joint PDF in Eq.~(\ref{eq:T07_model_joint_PDF}), one can find the first three non-zero moments as
\beqa
\langle v_{r} \rangle(r, M_1, M_2, z) &\equiv& 
\int \mathrm{d}v_{t}\, \mathrm{d}v_{r}\, v_{r}\, 
{\cal P}(v_t, v_r\, |\, r, M_1, M_2, z) \nonumber \\
&=&\int \mathrm{d}\delta\, \mu_{r}(\delta, r, z)\, 
{\cal F}(\delta\, |\, r, M_1, M_2, z), \label{eq:mean_v_final}
\\
\sigma^2_{t}(r, M_1, M_2, z) &\equiv&
\int \mathrm{d}v_{t}\, \mathrm{d}v_{r}\, v^2_{t}\, 
{\cal P}(v_t, v_r\, |\, r, M_1, M_2, z) \nonumber \\
&=&\int \mathrm{d}\delta\, \Sigma^2_{t}(\delta, r, M_1, M_2, z)\, {\cal F}(\delta\, |\, r, M_1, M_2, z), \label{eq:sigma2_t_final}\\
\sigma^2_{r}(r, M_1, M_2, z) 
&\equiv&
\int \mathrm{d}v_{t}\, \mathrm{d}v_{r}\, v^2_{r}\, 
{\cal P}(v_t, v_r\, |\, r, M_1, M_2, z) -\left[\langle v_{r}\rangle(r, M_1, M_2, z)\right]^2 \nonumber \\
&=&\int \mathrm{d}\delta\, \Sigma^2_{r}(\delta, r, M_1, M_2, z)\, {\cal F}(\delta\, |\, r, M_1, M_2, z)
\nonumber \\
\qquad \qquad \qquad \qquad \qquad \qquad
&&
+ \int \mathrm{d}\delta\, \mu^2_{r}(\delta, r, z)
\, {\cal F}(\delta\, |\, r, M_1, M_2, z)
- \left[\langle v_{r}\rangle(r, M_1, M_2, z)\right]^2,
\label{eq:sigma2_r_final}
\eeqa
where $\tilde{\rho}_{0}(r, M_1, M_2, z)$ sets the functional form of ${\cal F}$, while $\tilde{\rho}_{t,r}(r, M_1, M_2, z)$
is involved in the functions of $\Sigma^2_{t,r}$.
Because our model assumes the functional form of $\mu_r$ is known as Eq.~(\ref{eq:mean_v_our}), Eqs~(\ref{eq:mean_v_final})-(\ref{eq:sigma2_r_final}) 
may provide sufficient information to determine the model parameters in $\tilde{\rho}_{0}$ and $\tilde{\rho}_{t,r}$.
In this paper, we first find the parameters in $\tilde{\rho}_{0}$ (i.e. ${\cal A}_{\rho}$, ${\cal B}_{\rho}$, and ${\cal C}_{\rho}$ in Eq.~[\ref{eq:density_cut_off_ours}])
by the least square fitting of the profile $\langle v_r \rangle(r, M_1, M_2, z)$ for a given set of masses and redshift ($M_1, M_2$, and $z$).
After finding the best-fit values of ${\cal A}_{\rho}$, ${\cal B}_{\rho}$, and ${\cal C}_{\rho}$ by fitting of $\langle v_r \rangle(r)$, we then find the best-fit parameters in Eq.~(\ref{eq:Sigma_rho_t_r_ours}) by comparing the profiles of $\sigma^2_{t}(r)$ and $\sigma^2_{r}(r)$ with the predictions as in Eqs~(\ref{eq:sigma2_t_final}) and (\ref{eq:sigma2_r_final}).

\begin{figure*}[t!]
\gridline{\fig{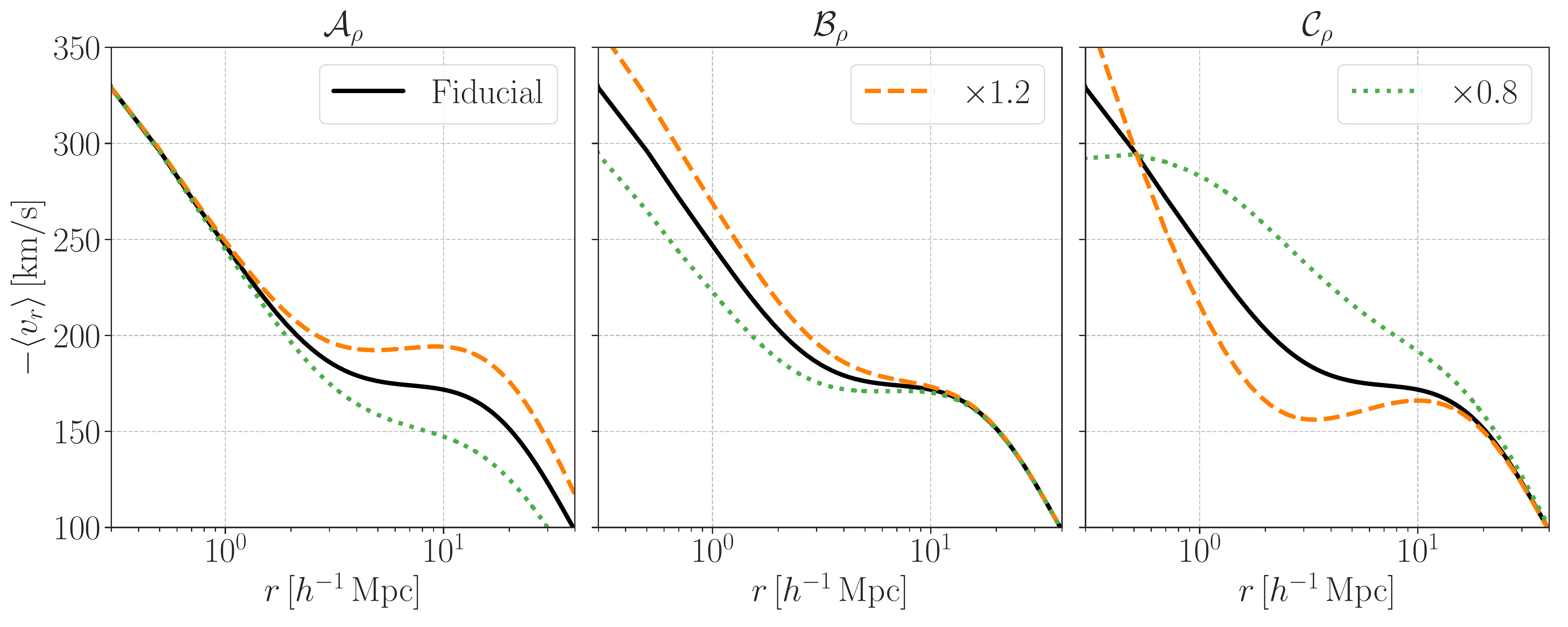}{0.8\textwidth}{}}
\caption{
Dependence of the mean radial velocity profile on the model parameters in Section~\ref{sec:model}. In each panel, 
the solid line shows the profile when we adopt the parameters proposed in \citet{2007MNRAS.374..477T}, while the dashed (dotted) line stand for the cases when varying a given parameter by a factor of 1.2 (0.8). 
The left, middle, right panels show the dependence on ${\cal A}_{\rho}$, ${\cal B}_{\rho}$, and ${\cal C}_{\rho}$, respectively. In this figure, we consider the halo pairs of $M_1=M_2=10^{13}\, h^{-1}M_{\odot}$ at $z=0$.
\label{fig:mean_r_model_depend}}
\end{figure*}

Figure~\ref{fig:mean_r_model_depend} shows the model prediction of the mean radial velocity profile as a function of model parameters ${\cal A}_{\rho}$, ${\cal B}_{\rho}$, and ${\cal C}_{\rho}$ in Eq.~(\ref{eq:density_cut_off_ours}).
In the figure, we consider a pair of halos with their masses of $M_1=M_2=10^{13}\, h^{-1}\, M_{\odot}$.
The figure represents a good flexibility of our model for fitting of $\langle v_r \rangle$.
For the velocity dispersions, 
we fix four parameters in the fitting process to avoid a complex degeneracy among parameters. 
To be specific, we fix ${\cal C}^{(2)}_{t} = 0.45$, $q_t=-0.9$, $p_{r}=-4.0$, and $q_{r}=-1.3$ in this paper.
Even if we reduce the number of degree of freedoms, 
the model prediction is found to be sufficiently flexible to fit the profiles of $\sigma^2_{t}$ and $\sigma^2_{r}(r)$ in the simulation. 
%(see Figure~\ref{fig:std_model_depend}).

\if0
\begin{figure*}[t!]
\gridline{\fig{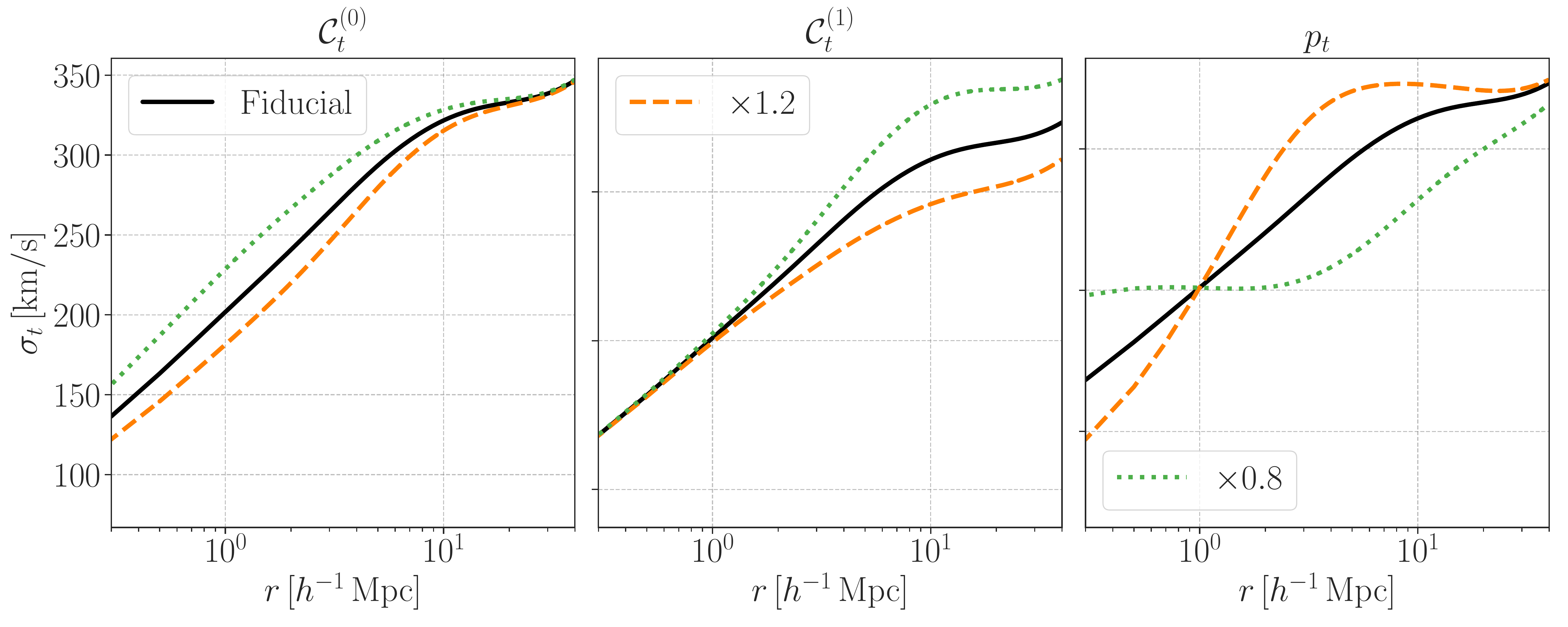}{0.8\textwidth}{}}
\gridline{\fig{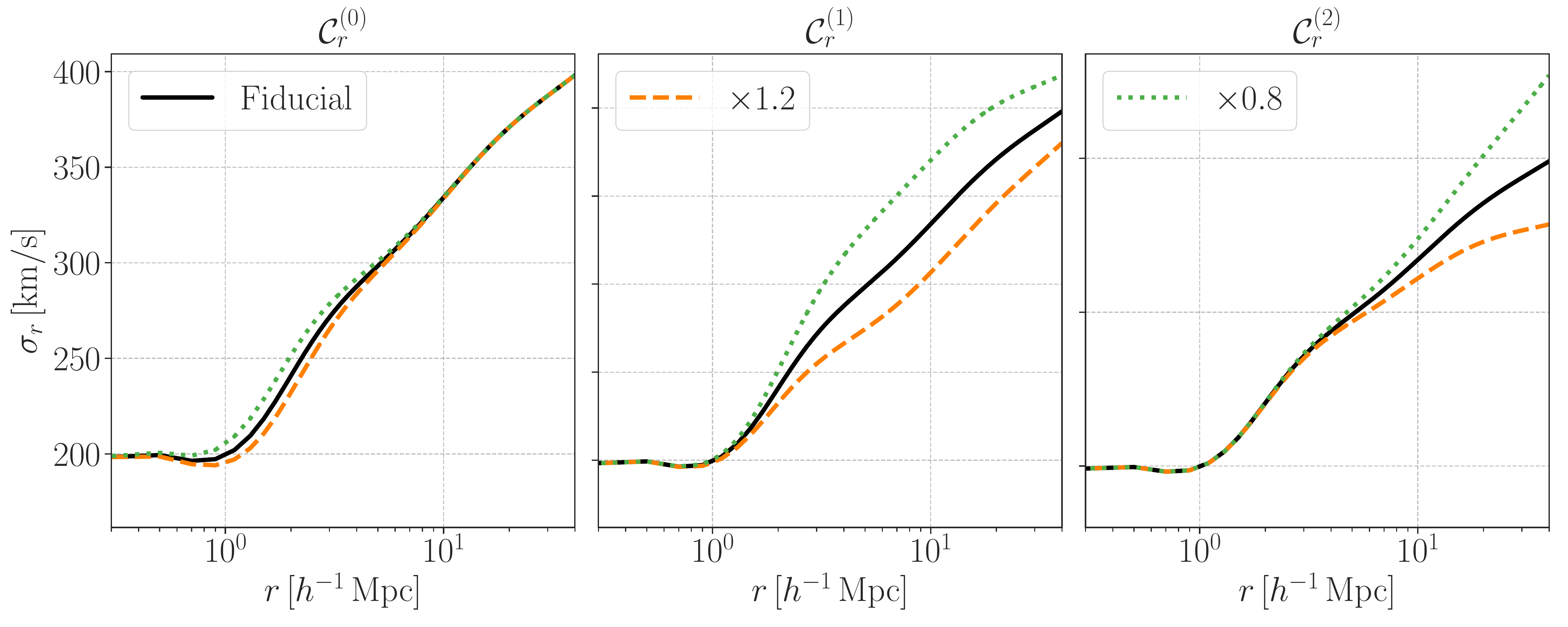}{0.8\textwidth}{}}
\caption{
Similar to Figure~\ref{fig:mean_r_model_depend}, but we show the dependence of the velocity dispersion profile on the model parameters in Section~\ref{sec:model}.
The top left, middle, right panel show the dependence of $\sigma_{t}(r)$ on ${\cal C}^{(0}_{t}$, 
${\cal C}^{(1)}_{t}$, and $p_{t}$, respectively.
In the bottom panels, the dependence of $\sigma_{r}(r)$
on ${\cal C}^{(0}_{r}$, 
${\cal C}^{(1)}_{r}$, and ${\cal C}^{(2)}_{r}$ are shown
from left to right.
\label{fig:std_model_depend}}
\end{figure*}
\fi

In the fitting processes, we measure the three moments of the pairwise velocity for the samples in Table~\ref{tab:table_halos}. For the measurements, 
we employ a linear-space binning in the range of $0<r\,[h^{-1}\,\mathrm{Mpc}]<40$ with 200 bins. 
For a given halo sample, 
we then find the best-fit values of 
$\cal A_{\rho}$, ${\cal B}_{\rho}$ and ${\cal C}_{\rho}$
in a given $M_1$, $M_2$, and $z$ bin
by minimizing the following $\chi^2$ statistic:
\beqa
\chi^2(\bd{p}_{\mathrm{mean}}\, |\, M_1, M_2, z) = \sum_{i} \frac{\left[\langle v_{r,\mathrm{sim}} \rangle(r_i)-\langle v_{r,\mathrm{mod}} \rangle(r_i\, |, \bd{p}_{\mathrm{mean}}) \right]^2}
{\sigma^2_{r,\mathrm{sim}}(r_i)/N_{\mathrm{pairs}}(r_{i})}, \label{eq:chi2_mean}
\eeqa
where $\langle v_{r, \mathrm{sim}} \rangle(r_i)$ is the mean radial velocity profile at the $i$-th radius in the simulation, 
$\langle v_{r,\mathrm{mod}} \rangle(r_i)$ is the counterpart of our model prediction, 
$\bd{p}_{\mathrm{mean}} = ({\cal A}_{\rho}, {\cal B}_{\rho}, {\cal C}_{\rho})$,
$\sigma_{r, \mathrm{sim}}(r_i)$ is the 
dispersion of $v_{r}$ at the $i$-th radius,
and $N_{\mathrm{pair}}(r_i)$ is the number of pairs in the $i$-th radius.
Once the best-fit $\bd{p}_{\mathrm{mean}}$ is found,
we then minimize other $\chi^2$ quantities to find the best-fit parameters in Eq.~(\ref{eq:Sigma_rho_t_r_ours}):
\beqa
\chi^2(\bd{p}_{\mathrm{\alpha}}\, |\, M_1, M_2, z) = \sum_{i} \frac{\left[\sigma^2_{\alpha, \mathrm{sim}}(r_i)-\sigma^2_{\alpha, \mathrm{mod}}(r_i\, |, \bd{p}_{\alpha}) \right]^2}
{2\sigma^4_{\alpha,\mathrm{sim}}(r_i)/N_{\mathrm{pairs}}(r_{i})}, \label{eq:chi2_var}
\eeqa
where $\alpha=t\, \mathrm{or}\,r$, 
$\sigma^2_{\alpha, \mathrm{sim}}(r_i)$ is the velocity dispersion profile at the $i$-th bin radius, 
$\sigma^2_{\alpha, \mathrm{mod}}(r_i)$ is the our model prediction,
$\bd{p}_{t} = ({\cal C}^{(0)}_{t}, {\cal C}^{(1)}_{t}, p_t)$
and 
$\bd{p}_{r} = ({\cal C}^{(0)}_{r}, {\cal C}^{(1)}_{r}, {\cal C}^{(2)}_{r})$.

\ms{In the calibration process, we find that the minimum $\chi^2$ per the number of degree of freedoms ranges from $1$ to $10^{4}$, while it depends on halo masses and redshifts.
Note that our fitting assumes zero covariances among different radii and Gaussian errors for velocity moments.
Because we work on non-linear scales of ${\cal O}(10)\, h^{-1}\mathrm{Mpc}$, these assumptions are expected to be invalid to have an appropriate $\chi^2$.
The goodness-of-fit based on $\chi^2$ will be meaningful when the error bars are precisely estimated.
Hence, the minimum $\chi^2$ value in our fitting should be taken as just a reference.}

Given the sets of $\bd{p}_{\mathrm{mean}}$, $\bd{p}_{t}$, $\bd{p}_{r}$ as a function of $M_1$, $M_2$, and $z$,
we then find an appropriate form to smoothly interpolate 
the data points after trial and error.
For an example, we assume that the form of $B_{\rho}(M_1, M_2, z)$ is given by $B_{0}(z)\, [(M_1+M_2)/10^{13}\, h^{-1}M_{\odot}]^{B_{1}(z)}$.
We then find the best-fit $B_{0}$ and $B_{1}$ for a given $z$
by a least-square fitting with the measured $B_{\rho}(M_1, M_2, z)$. The redshift dependence of $B_0$ and $B_1$ is then derived by a quadratic function fit.
The details of our functional forms for other parameters are provided in Appendix~\ref{apdx:model_params}\footnote{We also make our pipeline for the calibration process publicly available at \url{https://github.com/shirasakim/Fitting_velocity_moments_T07}.}.
It is obvious that our calibration process can be affected by details of the interpolation of $\bd{p}_{\mathrm{mean}}$, $\bd{p}_{t}$, and $\bd{p}_{r}$ (e.g. a choice of the functional form). This point is discussed in Section~\ref{subsec:interpolation_error}.

\begin{figure*}[t!]
\gridline{\fig{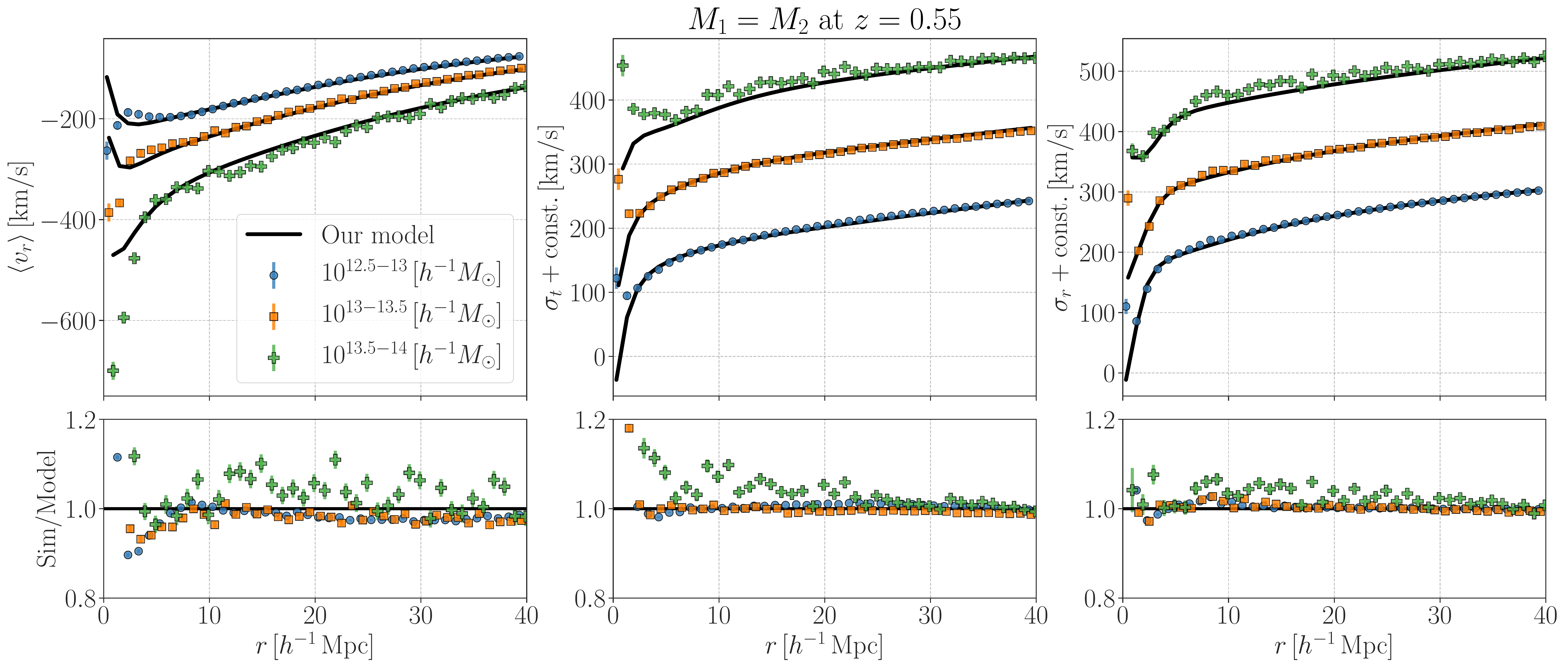}{1.\textwidth}{}}
\caption{
The mean and dispersions of pairwise velocities 
of dark matter halos at $z=0.55$. We here focus on the halo samples with $M=M_1=M_2=10^{12.5-13}, 10^{13-13.5}$, and $10^{13.5-14}\, h^{-1}M_{\odot}$.
In each panel, the blue circle, orange sqaure, and green plus symbols represent the simulation results for  $M=10^{12.5-13}, 10^{13-13.5}$, and $10^{13.5-14}\, h^{-1}M_{\odot}$, respectively.
The model predictions are shown in the solid lines.
The error bar in each panel shows the Gaussian error at a given radius.
The upper three panels present the profiles of mean radial velocity, the dispersions of the tangential and radial components, while the lower panels show the ratio between the simulation results and our model prediction.
Note that the Gaussian error in each panel is too small to plot for most cases. 
For visualization purpose, we shift the profiles of $\sigma_{t,r}$ by $-100$ and $+100\, \mathrm{km/s}$ for the sample of $M=10^{12.5-13}$
and $10^{13.5-14}\, h^{-1}M_{\odot}$, respectively.
\label{fig:mean_std_fine_mass_bins}}
\end{figure*}

\section{Results} \label{sec:results}

Here we present the comparison of the pairwise velocity statistics from our model with the simulation results.
We pay a special attention to the results at $z=0.55$, 
because it is relevant to the CMASS sample in the SDSS-III BOSS. 
We then discuss the information content in redshift-space clustering in terms of the pairwise velocity statistics.

\subsection{Velocity moments for galaxy- and group-sized halos}

We first show the results of the first three non-zero moments 
of the pairwise velocity for the halo samples with their masses of $10^{12.5-13}, 10^{13-13.5}$, and $10^{13.5-14}\, h^{-1} M_{\odot}$. Figure~\ref{fig:mean_std_fine_mass_bins} summarizes the comparisons of $\langle v_r \rangle$, $\sigma_t$, and $\sigma_{r}$ from our model with the simulation results. 
In the range of $5\simlt r \, [h^{-1}\, \mathrm{Mpc}]<40$,
our model can reproduce the mean and dispersion of the pairwise velocity of the simulated halos with $10^{12.5} 
< M \, [h^{-1}M_{\odot}] < 10^{13.5}$ within a $5\%$-level precision, while still providing a reasonable fit to the results for group-sized halos with $10^{13.5}<M\, [h^{-1}M_{\odot}]<10^{14}$.
According to the Gaussian error estimate, 
our measurement of velocity moments in simulations is precise with a level of $<5\%$ 
for $M<10^{14}\, h^{-1}M_{\odot}$ and $z<1$. 
The number of halo pairs including halos of $M>10^{14}\, h^{-1}M_{\odot}$ at $z=1.01$ becomes small, but our measurements reach a $15\%$-level precision even at halo pairs with their masses of $<10^{14.5}\, h^{-1}M_{\odot}$ at $z=1.01$.

The comparisons in Figure~\ref{fig:mean_std_fine_mass_bins} 
demonstrate that our model in Section~\ref{sec:model} is efficient and flexible enough at $z=0.55$ and a selection of mass bins to explain the radial profiles of the mean and dispersion of the pairwise velocity in the simulated halos
for all but the small scales of 
$r<5\, [h^{-1}\, \mathrm{Mpc}]$.
For other redshifts and halo masses, we summarize the comparisons in Appendix~\ref{apdx:performance_model}.
We confirm that our model can reproduce the velocity-moment profiles with a 5\%-level precision for $10^{12.5} < M \, [h^{-1}M_{\odot}] < 10^{13.5}$ at $0.3<z<1$, and the model precision reaches a 20\% level at the very worst in other ranges of halo masses and redshift.

\subsection{Mass-limited samples}

We then move onto the comparisons of various velocity statistics for the mass-limited halo samples.
Because our calibration process includes the interpolation of model parameters as a function of halo mass and redshift,
it is important to check if our model still works for samples with a wider range of halo masses.

\subsubsection{Pairwise-velocity distribution and its moments}

\begin{figure*}[t!]
\gridline{\fig{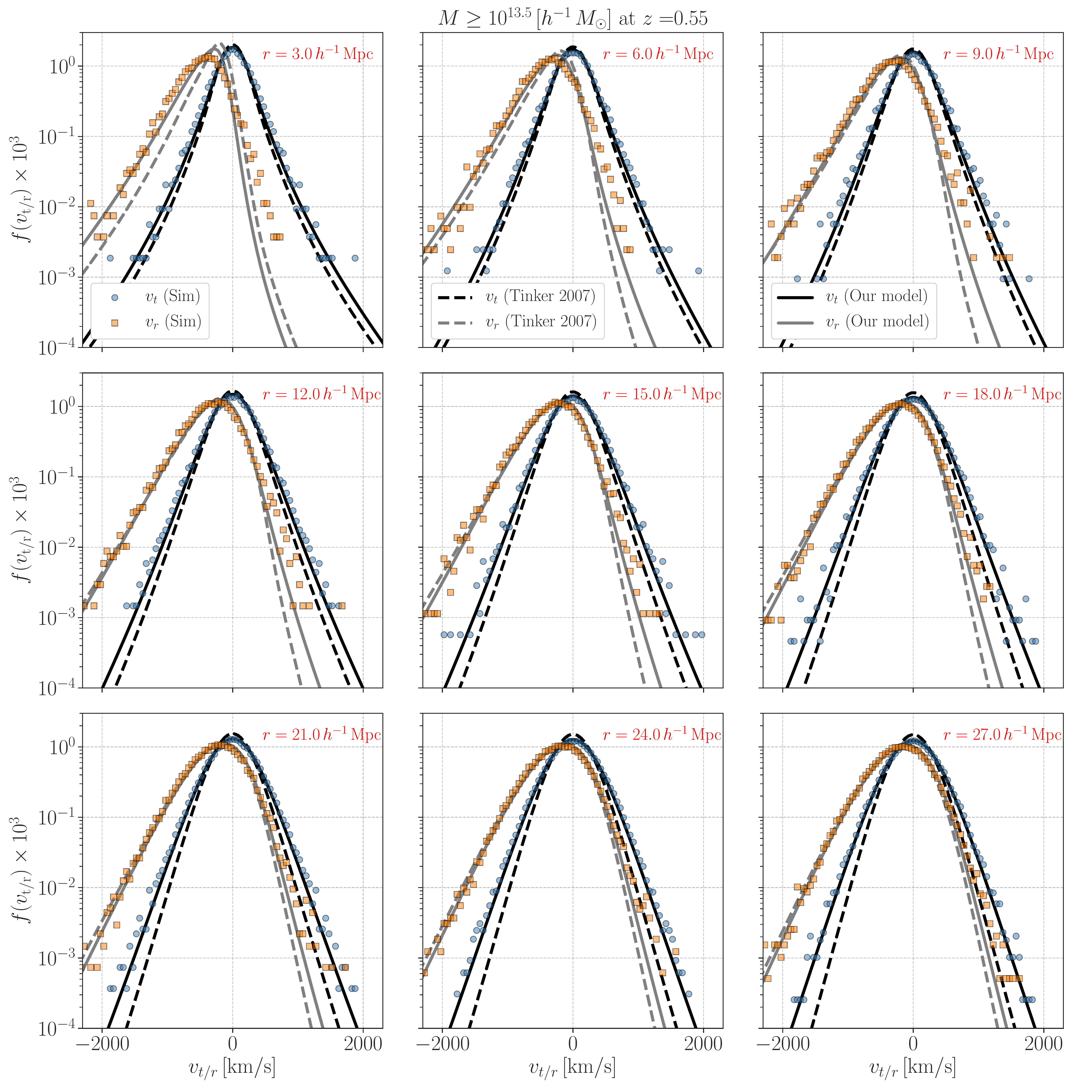}{1.\textwidth}{}}
\caption{
The probability distribution functions (PDFs) 
of the pairwise velocity of dark matter halos with their masses greater than $10^{13.5}\, h^{-1} M_{\odot}$ at 
the redshift of $0.55$.
The nine panels show the PDFs at different separation length, $r$ (labeled at the top right in each panel).
In each panel, the blue circle and orange square symbols
show the PDFs of $v_t$ and $v_r$ in the simulation, respectively.
Our model predicitons are shown in the black and gray solid lines in each panel, while the dashed lines represent the models in \citet{2007MNRAS.374..477T}.
Our model improves the precision for the velocity dispersion of $v_t$ as well as the mean velocity of $v_r$ compared to the model in \citet{2007MNRAS.374..477T}.
Note that we do not introduce any parameters to explain the skewness and kurtosis in our model, and any non-Gaussian features in the PDFs come from non-Gaussianity in the cosmic mass density (see Section~\ref{sec:model} for our model). 
\label{fig:v_pdf_logM_135}}
\end{figure*}

Figure~\ref{fig:v_pdf_logM_135} shows the comparisons of the velocity PDFs in the simulation with our model prediction.
In each panel, the gray and black solid lines represent
the model predictions for $v_r$ and $v_t$, respectively.
For a comparison, we show the predictions by the model in \citet{2007MNRAS.374..477T} by the gray 
and black dashed lines.
Although our model has been calibrated by the measurements of the three velocity moments 
as in Eqs.~(\ref{eq:mean_v_final})-(\ref{eq:sigma2_r_final}), 
the non-Gausssian tails in the PDFs can be explained by our model in a reasonable way.
This indicates that a large part of the non-Gaussianity in the velocity PDFs can be related with the non-Gaussianity in the cosmic mass density \citep[see][for further discussion]{2007MNRAS.374..477T}.

\begin{figure*}[t!]
\gridline{\fig{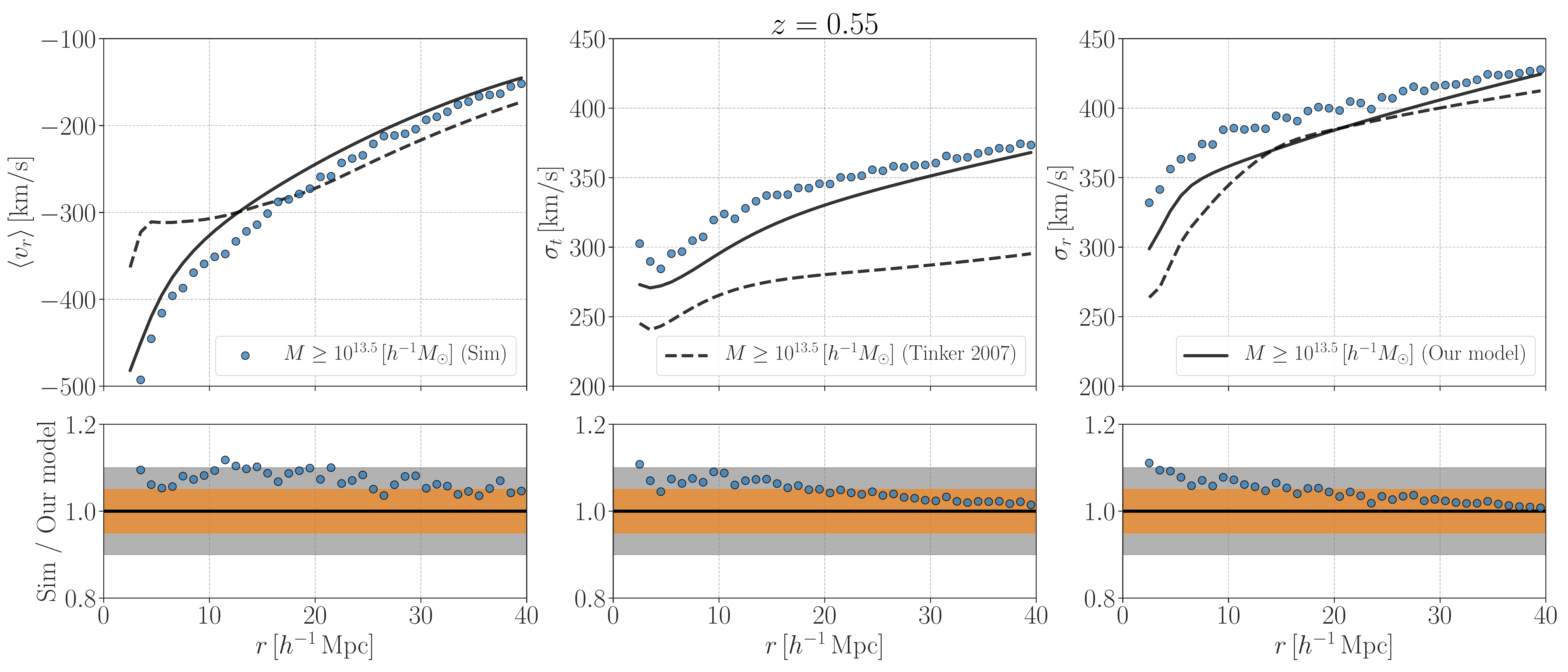}{0.9\textwidth}{}}
\caption{
Similar to Figure~\ref{fig:mean_std_fine_mass_bins}, but
for a mass-limited sample of halos. Here we plot the mean and dispersion of pairwise-velocity of dark matter halos with their masses greater than $10^{13.5}\, h^{-1} M_{\odot}$ at the redshift of $0.55$.
In the upper three panels, the blue circles show the simulation results, while the solid and dashed lines represent the predictions by our model and \citet{2007MNRAS.374..477T}, respectively.
In the lower panels, we show the ratio between the simulation result and our model. For a reference, the gray regions show
$\pm10\%$ levels, while the yellow one stands for 
$\pm5\%$ levels.
 \label{fig:v_moments_logM_135}}
\end{figure*}

Compared to the previous work, our model can provide a better fit to the velocity dispersion of $v_t$ as well as the long tails in PDFs for a wide range of $r$.
We find that the mean and velocity dispersion can be explained by our model within a 10\% level for the mass-limited sample with $M\ge10^{13.5}\, h^{-1}M_{\odot}$.
This is clearly shown in Figure~\ref{fig:v_moments_logM_135} and valid for the redshifts of $z=0.3$ and 1.01.
%Figure~\ref{fig:v_moments_diff_logM_z05} shows the comparisons of the mean and dispersion profiles for the lower halo masses.
For the mass-limited sample with $M\ge10^{12.5}\, h^{-1}M_{\odot}$, we find that our model is in good agreement with the simulation results and the precision reaches a 5\% level in the range of $5<r\, [h^{-1}\mathrm{Mpc}]<40$.

\subsubsection{Two-point clustering analyses in redshift space}

\begin{figure*}[t!]
\gridline{
\fig{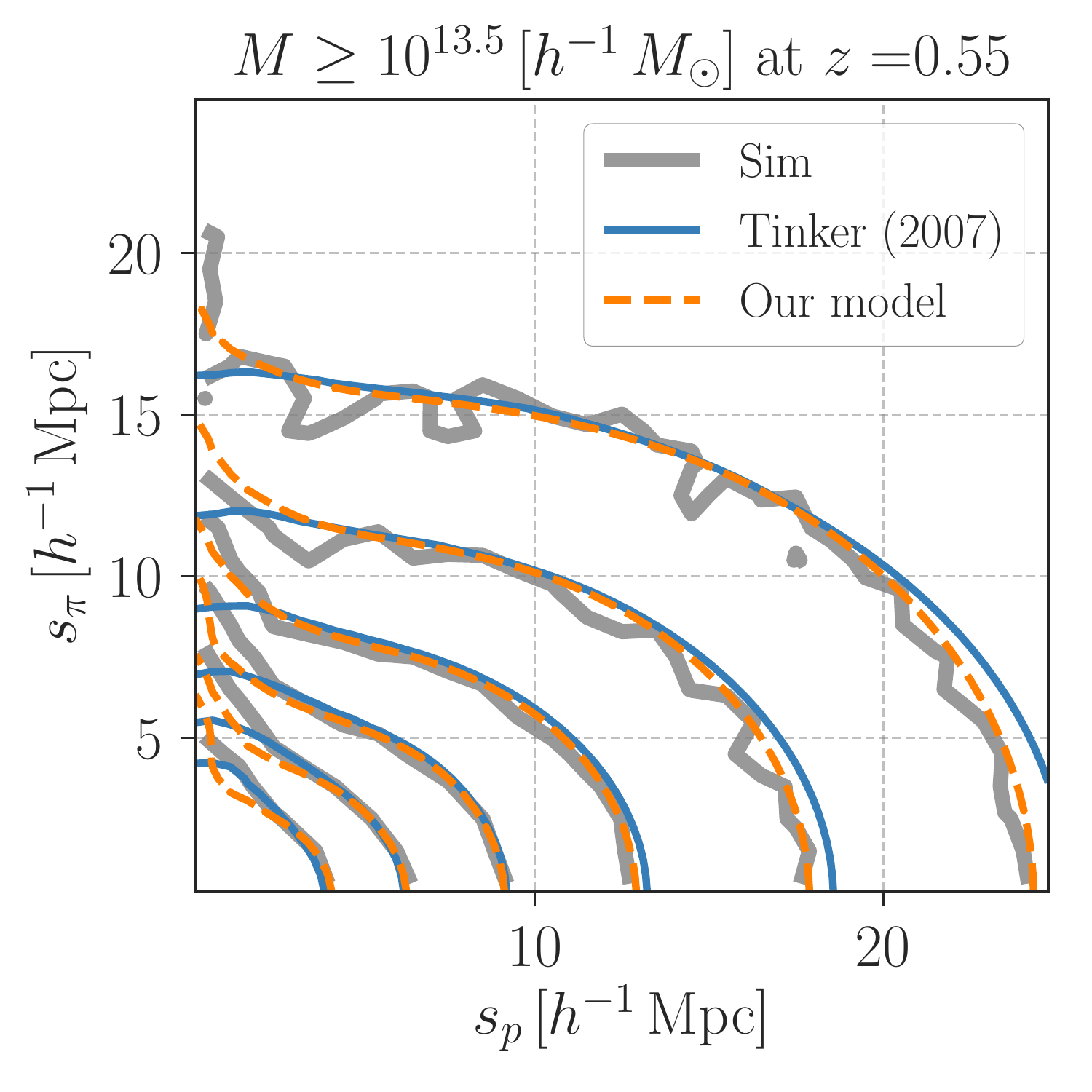}{0.4\textwidth}{}
\fig{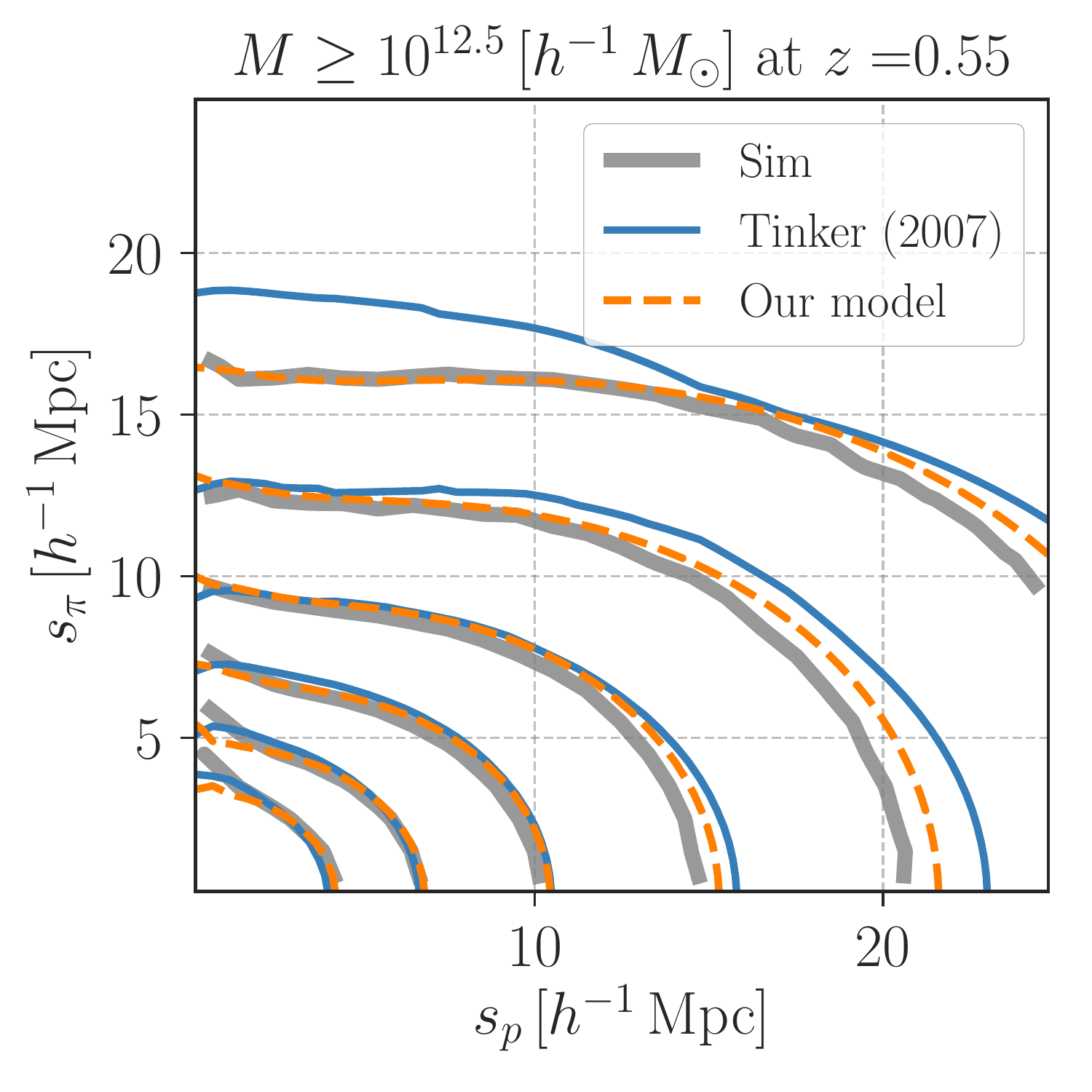}{0.4\textwidth}{}
}
\caption{
Two-dimensional redshift-space correlation functions of mass-limited halo samples at $z=0.55$.
The left panel shows the result for $M\ge10^{13.5}\, h^{-1}M_{\odot}$, while the right stands for $M\ge10^{12.5}\, h^{-1}M_{\odot}$.
In each panel, the gray lines show the contours of the redshift-space correlation of $\xi^{S}_{\mathrm{gg}}(s_p, s_{\pi})$, and the orange dashed lines are the model predictions by our model. For a comparison, the blue sold lines shows the model by \citet{2007MNRAS.374..477T}.
In each panel, the contours are separated by factors of 1.9 for clarity.
The outermost represents $\xi^{S}_{\mathrm{gg}}(s_p, s_{\pi})=1.9^{-1.5}$ and $1.9^{-3.5}$ in the left and right panels, respectively.
\label{fig:xi_2D}}
\end{figure*}

We next examine a more practical analysis of the two-point correlation function in redshift space, denoted by $\xi^{S}_{\mathrm{gg}}(s_p, s_{\pi})$,
for the mass-limited samples. Note that we suppose the real-space correlation function is known in this paper, while we study the mapping of the correlation function between real and redshift space by our model of the velocity PDFs (see Eq.~[\ref{eq:2pcf_real_to_redshift}]).
The details about modeling of the velocity PDF with a HOD are summarized in Appendix~\ref{apdx:halo_streaming_model}.

Figure~\ref{fig:xi_2D} shows the comparisons of $\xi^{S}_{\mathrm{gg}}$ between the simulation results and the model predictions. For the model prediction of $\xi^{S}_{\mathrm{gg}}$, we use the real-space correlation function measured in the simulations and interpolate the data points over separation lengths $r$.
We find that our model can provide a more reasonable fit to $\xi^{S}_{\mathrm{gg}}$ in the simulations than the models by \citet{2007MNRAS.374..477T}. In particular, our model improves the mapping at small $s_{\pi}$, because our model shows a better fit to the velocity dispersion of $v_{t}$ in the simulations and $v_{t}$ is relevant to the line-of-sight velocity at small $s_{\pi}$.

\begin{figure*}[t!]
\gridline{\fig{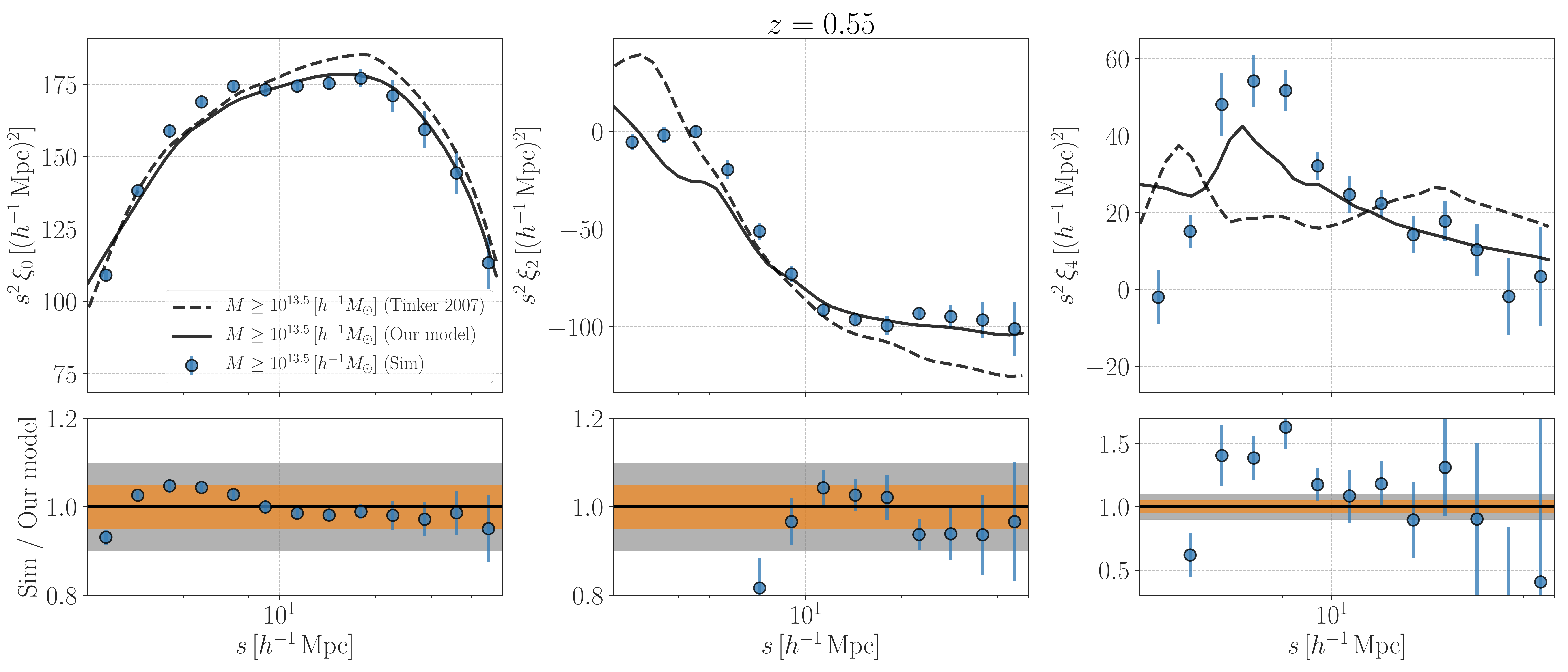}{0.9\textwidth}{}}
\caption{
Redshift-space clustering multipoles of the mass-limited halo sample with $M\ge10^{13.5}\, h^{-1}M_{\odot}$
at $z=0.55$.
The upper panels show the multipole moments 
of $\xi_{0}$, $\xi_{2}$ and $\xi_{4}$ from left to right.
In each upper panel, the blue points with error bars show the simulation results, while the solid and dashed lines represent the predictions by our model and \citet{2007MNRAS.374..477T},
respectively.
In the lower panels, we show the ratio of the simulation results and our model predictions. The yellow and gray filled regions show $\pm5\%$- and $\pm10\%$-level differences.
\label{fig:xi_multipoles}}
\end{figure*}

For a more quantitative view, we show comparisons of the clustering multipoles defined in Eq.~(\ref{eq:clustering_moments}).
Figure~\ref{fig:xi_multipoles} summarizes comparisons for the mass-limited sample with $M\ge10^{13.5}\, h^{-1}M_{\odot}$.
The figure clearly shows that our model can provide an accurate mapping of the two-point correlation function between real and redshift space at intermediate scales of 
$\sim10\, h^{-1}\mathrm{Mpc}$.
For the lowest-order moment, our model can provide an excellent fit to the simulation results within a 5\% level 
over $5<s\, [h^{-1}\,\mathrm{Mpc}]<40$.
Even for the higher-order moments, we find that our model can explain $\xi_{2}(s)$ and $\xi_{4}(s)$
in the range of $10-30\, h^{-1}\mathrm{Mpc}$ 
with a 10\%- and 50\%-level precision, respectively.
%Figure~\ref{fig:xi_multipoles_diff_M} shows 
For the mass-limited sample with $M\ge10^{12.5}\, h^{-1}M_{\odot}$, the agreement is found to be worse compared to the samples with $M\ge10^{13.5}\, h^{-1}M_{\odot}$.
Nevertheless, the lowest moment $\xi_0$ can still be reproduced by our model within a 5\% level precision over $3<s\, [h^{-1}\,\mathrm{Mpc}]<30$ even for the sample with 
$M\ge10^{12.5}\, h^{-1}M_{\odot}$.

%\begin{figure*}[t!]
%\gridline{\fig{xis_comp_logM_135_diff_z.pdf}{1.\textwidth}{}}
%\caption{
%Similar to Figure~\ref{fig:xi_multipoles}, but we show the results for different redshifts of $z=0.3$ and $1.01$.
%\label{fig:xi_multipoles_diff_z}}
%\end{figure*}

\subsection{Realistic galaxy samples}

\begin{figure*}[ht!]
\gridline{\fig{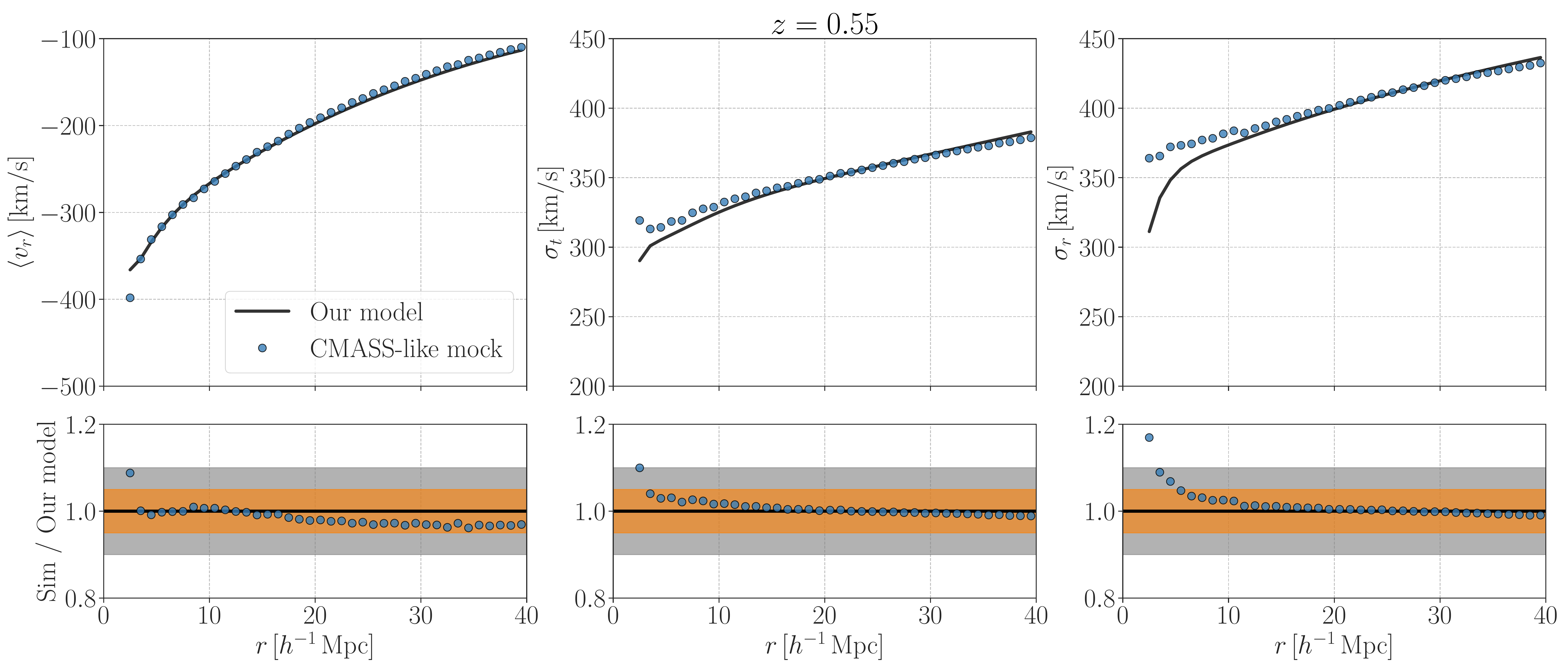}{0.9\textwidth}{}}
\caption{
Similar to Figure~\ref{fig:v_moments_logM_135}, but for a mock sample of CMASS galaxies at $z=0.55$. In each panel, the blue points show the simulation results, while the solid lines represent our model predictions.
Note that the typical halo mass of CMASS galaxies is set to $\sim10^{13}\, h^{-1}M_{\odot}$, but we include the satellite galaxies for $M>10^{13.27}\, h^{-1}M_{\odot}$.
\label{fig:v_moments_CMASS_like}}
\end{figure*}

Here we present the results for a realistic galaxy mock sample with the HOD model in Section~\ref{subsec:mock}.
Figure~\ref{fig:v_moments_CMASS_like} summarizes the comparisons of the velocity-moment profiles in the simulation with our model predictions.
We find that our model can provide a few-percent-level prediction of the mean radial velocity as well as the velocity dispersions for CMASS-like galaxies at $5-40\, h^{-1}\mathrm{Mpc}$.

\begin{figure*}[ht!]
\gridline{\fig{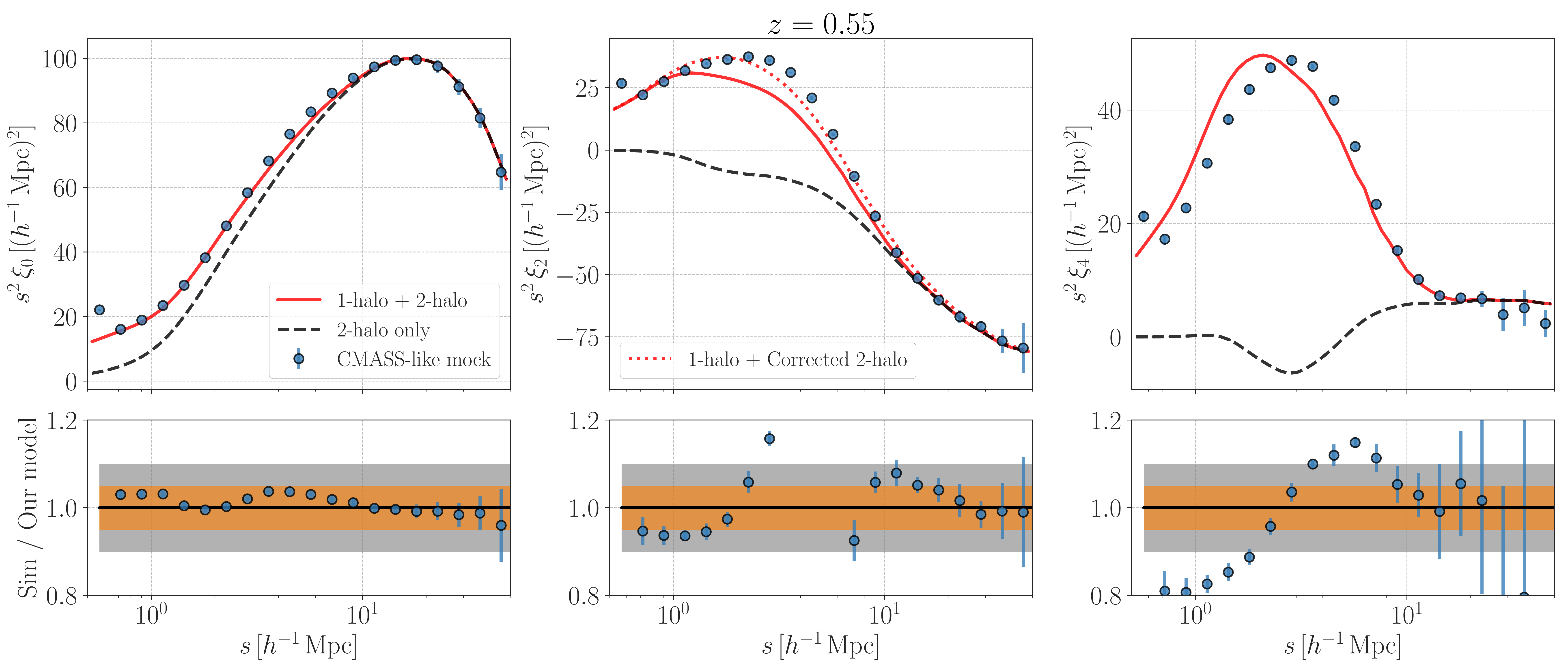}{0.9\textwidth}{}}
\caption{
Similar to Figure~\ref{fig:xi_multipoles}, but for a mock sample of CMASS galaxies at $z=0.55$. In each panel, the blue points with error bars are the simulation results, while the red solid line shows the model predictions based on the HOD and our model of the pairwise velocity PDFs. For a comparison, the dashed lines show the two-halo contribution to the multipole moments.
\ms{In the top middle panel, the red dotted line shows the model with the correction as in Eq.~(\ref{eq:correc_2h_xi2}).}
\label{fig:xi_multipoles_CMASS_like}}
\end{figure*}

For the clustering multipoles, figure~\ref{fig:xi_multipoles_CMASS_like} shows the comparisons between the simulation results and our model predictions. The red solid lines in the figure represent
our model predictions, while the dashed line shows the so-called two-halo terms in a halo-model approach (see Appendix~\ref{apdx:halo_streaming_model} for details).
In our halo model, the redshift-space clustering multipoles can be decomposed into two parts:
\beqa
\xi_{\ell}(s) = \xi_{\ell,\mathrm{1h}}(s) + \xi_{\ell, \mathrm{2h}}(s)
\eeqa
where $\xi_{\ell,\mathrm{1h}}$ represents the two-point correlation in single dark matter halos, 
$\xi_{\ell,\mathrm{2h}}$ is the contribution from the clustering between two neighboring halos.
The one-halo term $\xi_{\ell,\mathrm{1h}}$ can be further divided into two contributions from the central-satellite and satellite-satellite pairs.
The two-halo term $\xi_{\ell,\mathrm{2h}}$ is mostly determined by the clustering and streaming motion of the central-central pairs, but it is also affected by the velocity dispersion of satellites in single halos.
Our model predictions are in good agreement with the simulation results at the scales of $s\simgt10\, h^{-1}\mathrm{Mpc}$, 
i.e. the regime where the two-halo contributions would play a central role.
Note that the inaccurate small-scale two-halo term
can affect the prediction of $\xi_{2}(s)$ at $s\sim3-5\, h^{-1}\mathrm{Mpc}$.
We find that a simple modification in the two halo term of $\xi_{2}$ can provide a better fit to the simulation:
\beqa
\xi_{2, \mathrm{2h}}(s) \rightarrow \exp\left[-\left(\frac{s_{\mathrm{cut}}}{s}\right)^2\right]\xi_{2, \mathrm{2h}}(s), \label{eq:correc_2h_xi2}
\eeqa
where $s_{\mathrm{cut}}$ is a free parameter and we find $s_{\mathrm{cut}}\sim3.5\, h^{-1}\mathrm{Mpc}$ is appropriate for our mock catalog.
The two-halo term $\xi_{2, \mathrm{2h}}$ at $s<10\, h^{-1}\mathrm{Mpc}$
would be affected by the non-Gaussianity in the pairwise velocity PDFs \citep[e.g.][]{2020arXiv200202683C}.
Our model has been calibrated to explain the mean and variance in the pairwise velocity PDF at $5<r\, [h^{-1}\mathrm{Mpc}]<40$.
After the calibration, we found that it fails to provide a fit to the PDF at $r<5\, h^{-1}\mathrm{Mpc}$ (e.g. see the top left panel in Figure~\ref{fig:v_pdf_logM_135}). Hence, we expect that our model can not work for the precise prediction of $\xi_{2, \mathrm{2h}}$ at $s<10\, h^{-1}\mathrm{Mpc}$.
\ms{Nevertheless, including a single nuisance parameter $s_{\mathrm{cut}}$ can improve our model precision for $\xi_{2}$ at $s=1-10 \, h^{-1}\mathrm{Mpc}$.}

\subsection{Information contents of redshift-space clustering multipoles}\label{subsec:information_content}

\begin{figure*}[ht!]
\gridline{\fig{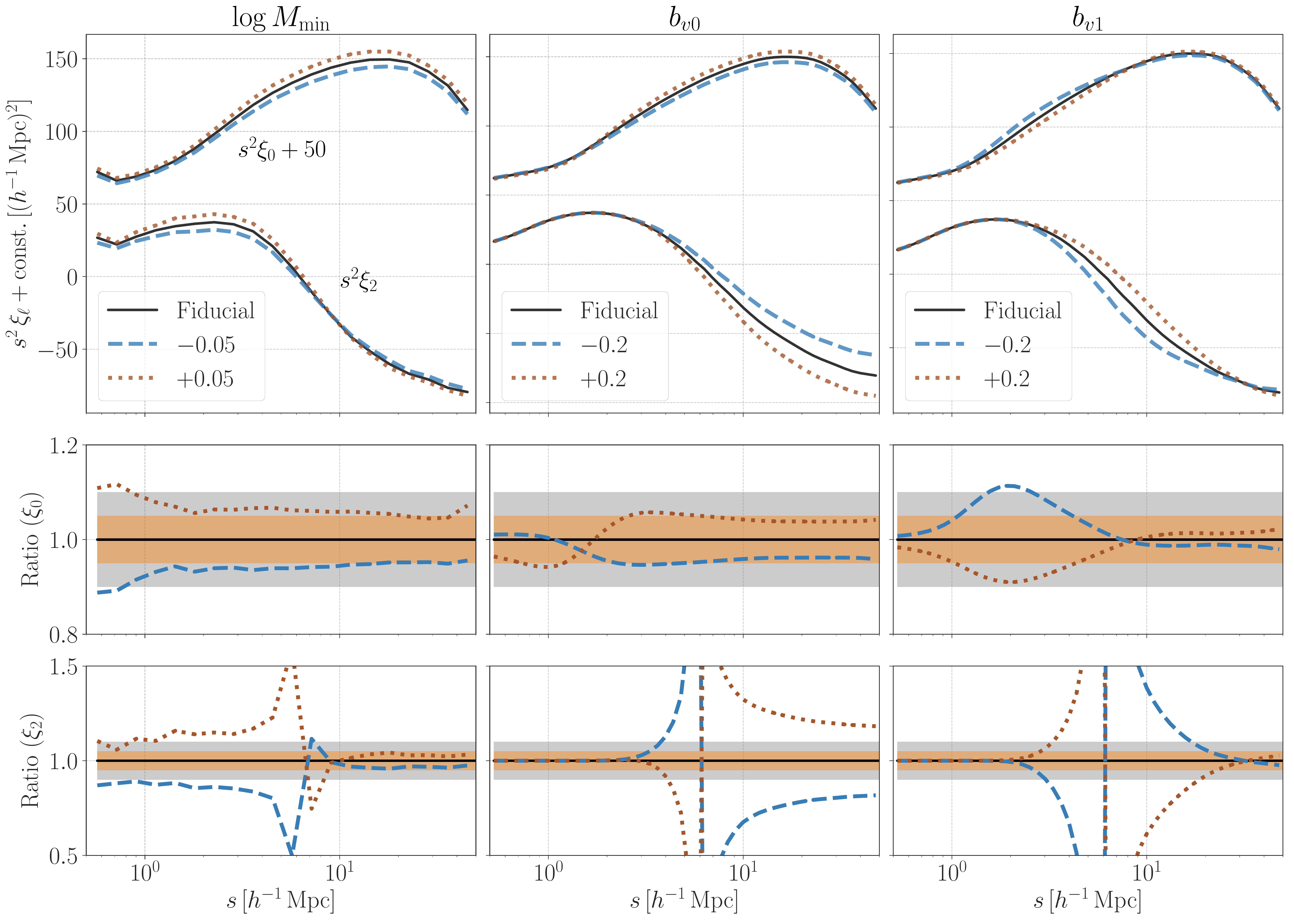}{0.95\textwidth}{}}
\caption{
Dependence of redshift-space clustering multipoles on a HOD parameter and velocity biases.
In the three top panel, we show the monopole $\xi_0$ and quadropole $\xi_2$
for the CMASS-like mock galaxies at $z=0.55$
when varying the HOD parameter $\log M_{\mathrm{min}}$ and 
the velocity-bias parameters $b_{v0}$ and $b_{v1}$ from left to right.
The definitions of $b_{v0}$ and $b_{v1}$ are found in Section~\ref{subsec:information_content}.
In each panel, the black solid lines show the results for our fiducial set of the parameters, while
the blue dashed and red dotted lines represent the responses of $\xi_{0,2}$ when we vary the parameters.
In the three middle panels, we summarize the ratio of $\xi_{0}$ with respect to the fiducial results.
We also show the ratio of $\xi_2$ in the three bottom panels.
For references, the gray and yellow filled regions in the middle and bottom panels show $\pm10$- and 
$\pm5\%$-level differences, respectively.
\label{fig:xi_multipoles_CMASS_like_varied_params}}
\end{figure*}

For an application of our model, 
%\textbf{JASON: 'For an application of our model,' might be better}
we discuss the information content in redshift-space clustering analyses of galaxies.
According to Eq.~(\ref{eq:2pcf_real_to_redshift}), the observed two-point correlation function in redshift space should contain the information about the pairwise velocity statistics of galaxies.
At intermediate scales of $\sim10\, \mathrm{Mpc}$,
a class of modified gravity theory predicts that the mean and dispersion of the pairwise velocity for massive-galaxy-sized halos can differ from the prediction from General Relativity by $10-20\%$ \citep[e.g.][]{2014PhRvL.112v1102H, 2014MNRAS.445.1885Z}.
For the galaxy-halo connection, numerical simulations have shown that the pairwise velocity statistics for realistic galaxies can depend not only on their host halo masses but also the inner mass density profiles of their host halos and ages \citep[e.g.][]{2015MNRAS.451L..45H, 2019MNRAS.486..582P}. This is known as the assembly bias effect.

As a simple example, we study the modified gravity effect and/or the assembly bias effect 
on the halo pairwise velocity
by introducing two free parameters:
\beqa
v^{\mathrm{obs}}_{z} = b_{v0} \langle v_z \rangle
+ b_{v1} \left( v_z - \langle v_z \rangle \right), \label{eq:v_bias}
\eeqa
where $v^{\mathrm{obs}}_{z}$ represents 
the line-of-sight pairwise velocity affected 
by the modified gravity and/or the assembly bias effect,
while $v_{z}$ is a baseline prediction in the $\Lambda$CDM cosmology.
We here assume that statistical properties of
$v_{z}$ can be characterized by halo masses, redshifts, and separation lengths.
Note that $b_{v0}$ changes the mean pairwise velocity for a given galaxy sample, while $b_{v1}$ affects the variance in the pairwise velocity.
The modified gravity and/or the assembly bias can deviate $b_{v0}$ and $b_{v1}$ from unity.
Therefore, it would be interesting to consider the dependence of the clustering multipoles of $\xi_{0,2}$ on the velocity-bias parameters $b_{v0}$ and $b_{v1}$.
We here emphasize that our velocity biases in Eq.~(\ref{eq:v_bias}) have a different meaning from the common definitions in the literature.
Previous studies have mainly focused on the velocity biases
with respect to the core or the kinematics of dark matter inside the single dark matter halo \citep[e.g.][]{2014MNRAS.444..476R, 2015MNRAS.446..578G}.
In contrast, our parameterization of the velocity bias enables us to study the bias in the streaming motion between two neighboring halos\footnote{
Within our framework, non-trivial galaxy-halo connection may induce biases in the streaming motion between two galaxies. An example is the environmental dependence of HODs \citep[e.g.][]{2020MNRAS.493.5506H}. If the HOD depends not only on the halo mass but also the environmental density $\delta$, the pairwise velocity statistics can differ from our predictions. We also expect that a modification of gravity can change the relation of the mean infall velocity and density perturbations \citep[see, e.g.][for the spherical collapse model in a modified gravity theory]{2012MNRAS.421.1431L}, leading to $b_{v0}\neq 1$.}.
Using Eq.~(\ref{eq:v_bias}) and the formulas in Appendix~\ref{apdx:halo_streaming_model}, we compute the expected signal of $\xi_{0,2}$ for the HOD model in Section~\ref{subsec:mock} as a function of $b_{v0}$ and $b_{v1}$.
To check for degeneracy among the HOD parameters, we also vary
a parameter of $\log M_{\mathrm{min}}$ which determines the typical halo mass of galaxy sample of interest.
We generated two additional mock galaxy catalogs using the $\nu^2$GC-L simulation by changing the parameter of $\log M_{\mathrm{min}}$ by $\pm0.05$. In the following, we use these simulation results when studying the effect of $\log M_{\mathrm{min}}$.
When varying the biases of $b_{v0}$ and $b_{v1}$, we use the analytic model of the pairwise velocity PDF as in Appendix~\ref{apdx:halo_streaming_model} and predict the multipoles based on Eq.~(\ref{eq:2pcf_real_to_redshift}).
We also adopt the correction in Eq.~(\ref{eq:correc_2h_xi2})
for our model of the two-halo term in $\xi_{2}$.

Figure~\ref{fig:xi_multipoles_CMASS_like_varied_params}
summarizes the changes in $\xi_{0,2}$ for the CMASS-like galaxy sample at $z=0.55$ caused by differences in 
$\log M_{\mathrm{min}}$, $b_{v0}$ and $b_{v1}$.
In the figure, we set the HOD parameters as in Section~\ref{subsec:mock} and $b_{v0}=b_{v1}=1$ for the fiducial case.
The figure indicates that the effect of $b_{v0}$ and $b_{v1}$ on the redshift-space clustering can not be compensated for by simple changes in the typical host halo mass.
When it comes to other HOD parameters, we find that $\sigma_{\log M}$ and $M_1$ show a strong degeneracy with $\log M_{\mathrm{min}}$, while $M_{\mathrm{cut}}$ and $\alpha_M$ can change the one-halo term while changing two-halo term minimally. 
We also note that the real-space correlation function strongly depends on the HOD parameters, but is independent of the biases of $b_{v0}$ and $b_{v1}$.
The real-space correlation function can be reproduced within the HOD framework for a given cosmological model and the HOD parameters have been tightly constrained with the combined analysis of galaxy-galaxy lensing and projected correlation functions \citep[e.g.][]{2015ApJ...806....2M}.
Therefore, we expect that a joint analysis of $\xi_{0,2}$ with
galaxy-galaxy lensing and projected correlation functions provides an important test of the common HOD framework with no assembly biases at least.
For more details (e.g. expected constraints of $b_{v0}$ and $b_{v1}$ for a given galaxy sample), 
we require a precise estimate of the covariance of $\xi_{0,2}$ and leave it for future studies.

\section{Limitations} \label{sec:limitations}

We summarize the major limitations in
our model of pairwise velocity PDFs of 
dark matter halos. All of the following issues will be addressed in forthcoming studies.

\subsection{Cosmological dependence}

\begin{figure*}[t!]
\gridline{\fig{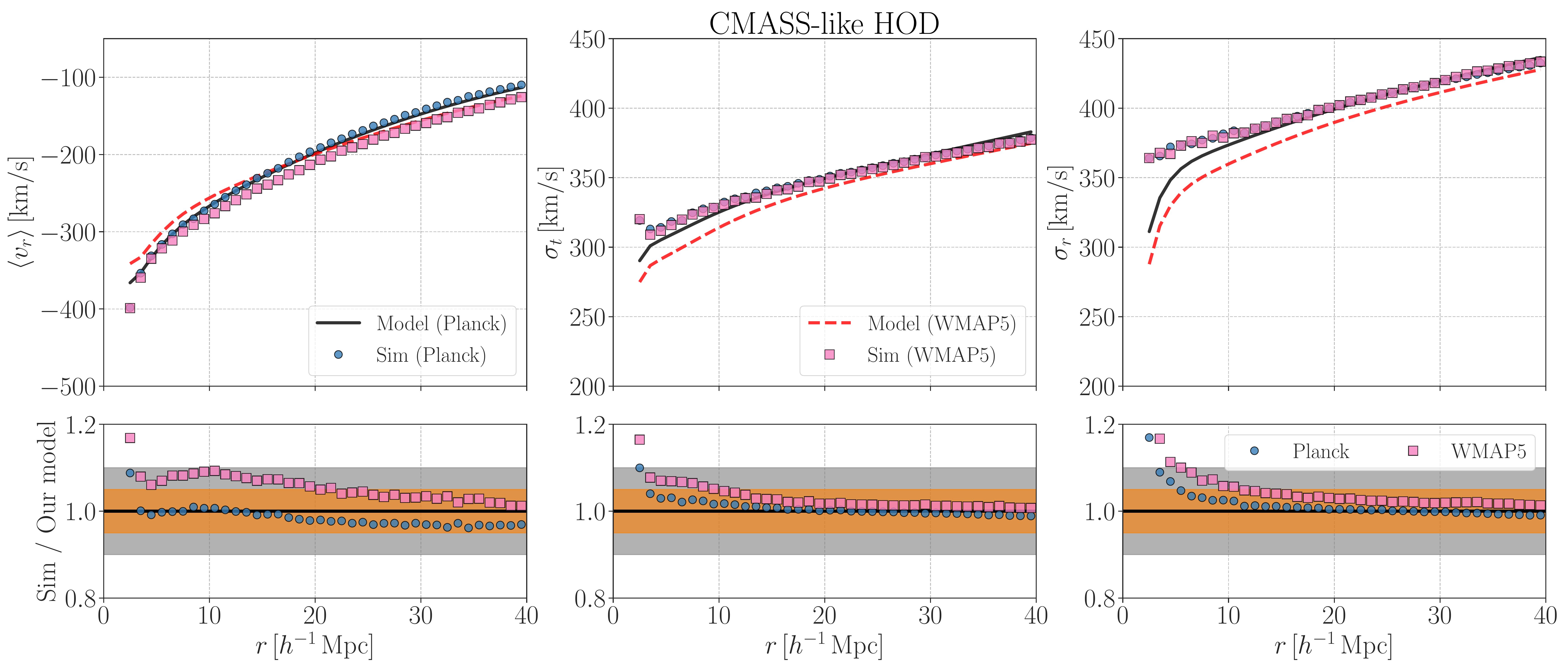}{0.9\textwidth}{}}
\caption{
Similar to Figure~\ref{fig:v_moments_CMASS_like}, but we include the comparison with the results from the MultiDark simulation at $z=0.534$ and our model predictions.
In this figure, we consider the mock galaxy sample with the HOD model in Section~\ref{subsec:mock}.
Note that our fiducial simulation assumes the cosmological model inferred by the Planck satellite ($\Omega_{\rm m0}=0.31$ and $\sigma_{8}=0.83$), while the MultDark run assumes the model with $\Omega_{\rm m0}=0.27$ and $\sigma_{8}=0.82$.
In the upper panels, the blue circle and pink square symbols represent the simulation results in the Planck and WMAP5 cosmology, respectively. The solid line shows our model prediction for the Planck cosmology, while the red dashed line is the model prediction for the WMAP5 cosmology.
\label{fig:v_moments_CMASS_like_cosmo}}
\end{figure*}

Our model of pairwise velocity PDFs is calibrated against $N$-body simulations in the $\Lambda$CDM cosmology
%which is consistent with the measurements of the cosmic microwave background by the Planck satellites (we here refer to this model as the Planck cosmology). 
consistent with Planck.
In terms of studies of large-scale structure, $\Omega_{\mathrm{m0}}$ and $\sigma_8$ are the primary parameters and the simulations in this paper adopt $\Omega_{\mathrm{m0}}=0.31$ and $\sigma_8=0.83$.
Therefore, our functions in Section~\ref{sec:model} and Appendix~\ref{apdx:model_params} may be subject to an overfitting to the specific cosmological model.
To examine the dependence of our model on cosmological models, we use another halo catalog from $N$-body simulations with a different $\Lambda$CDM model.
For this purpose, we use the first MultiDark simulation performed in \citet{2012MNRAS.423.3018P}.
The MultiDark simulation consists of $2048^3$ particles in a volume of $1\, [h^{-1}\mathrm{Gpc}]^3$ and 
assumes the cosmological parameters of
$\Omega_{\mathrm{m0}}= 0.27$,
$\Omega_{\mathrm{b0}}= 0.0469$,
$\Omega_{\Lambda}=1-\Omega_{\mathrm{m0}} = 0.73$,
$h= 0.70$,
$n_s= 0.95$, and $\sigma_8= 0.82$.
These are consistent with the five-year observation of the cosmic microwave background obtained by the WMAP satellite \citep{2009ApJS..180..330K}
and we refer to them as the WMAP5 cosmology.
We use the ROCKSTAR halo catalog at $z=0.534$ from the MultiDark simulation\footnote{The halo catalogs at different redshifts are publicly available at \url{https://slac.stanford.edu/~behroozi/MultiDark_Hlists_Rockstar/}.} and then produce a CMASS-like mock catalog by using the HOD model in Section~\ref{subsec:mock}.
To compute our model prediction for the WMAP5 cosmology,
we fix the functional forms and parameters in Appendix~\ref{apdx:model_params} but include the cosmology-dependence of the log-normal PDF of cosmic mass density, the linear halo bias and the linear growth factor, accordingly.
In other words, we assume the model in Appendix~\ref{apdx:model_params} is universal and valid for different cosmological models.

Figure~\ref{fig:v_moments_CMASS_like_cosmo} summarizes the velocity-moment profiles of the CMASS-like mock catalog for two different cosmologies.
In this figure, the solid and red dashed lines show the predictions by our model for the Planck and WMAP5 cosmologies, respectively.
We find that our model can reproduce simulation results within a 5-10\% level even for the WMAP5 cosmology at $5-40\, h^{-1}\mathrm{Mpc}$.
It is worth noting that the cosmological dependence of the velocity dispersion is small in the simulations, but our model predicts a few percent level difference.
For the mean radial velocity profile, we find a 10\%-level difference between the simulation and our model in the WMAP5 cosmology at $r<20\, h^{-1}\mathrm{Mpc}$, 
while our model provides a better fit to the simulations at larger scales of $r\simgt25\, h^{-1}\mathrm{Mpc}$.
For comparison, our model can predict the mean radial velocity profile at $r=5-40\, h^{-1}\mathrm{Mpc}$ within a 5\%-level precision in the Planck cosmology. 

In summary, our model can not predict the simulation results for the WMAP5 cosmology with the same level as in the Planck cosmology.
The 10\%-level difference in $\Omega_{\mathrm{m0}}$ can cause systematic uncertainties in our model predictions
with a level of 5-10\%.
Note that the velocity dispersions are found to be less sensitive to the change in $\Omega_{\mathrm{m0}}$ in the simulations.
More extensive studies are required to investigate the cosmological dependence of the pairwise velocity statistics.

\subsection{Calibrations with $N$-body particle data}

Our model assumes that the pairwise velocity PDFs can be expressed as a Gaussian at a given environmental density $\delta$. \citet{2007MNRAS.374..477T} already showed that the approximation looks valid by using the $N$-body simulations, while the $\delta$ dependence of the Gaussian parameters (mean and variance) may be different from our 
assumptions in Eqs.~(\ref{eq:Sigma_v_T07}) and (\ref{eq:mean_v_our}). 
We also assume that the conditional PDF of cosmic mass density finding a halo pair is given by the form of Eq.~(\ref{eq:cond_PDF}), but another functional form would provide a better fit to the simulation results at small scales. The calibration with the information of $N$-body particles is important to validate the underlying assumptions in our model.

\begin{figure*}[t!]
\gridline{\fig{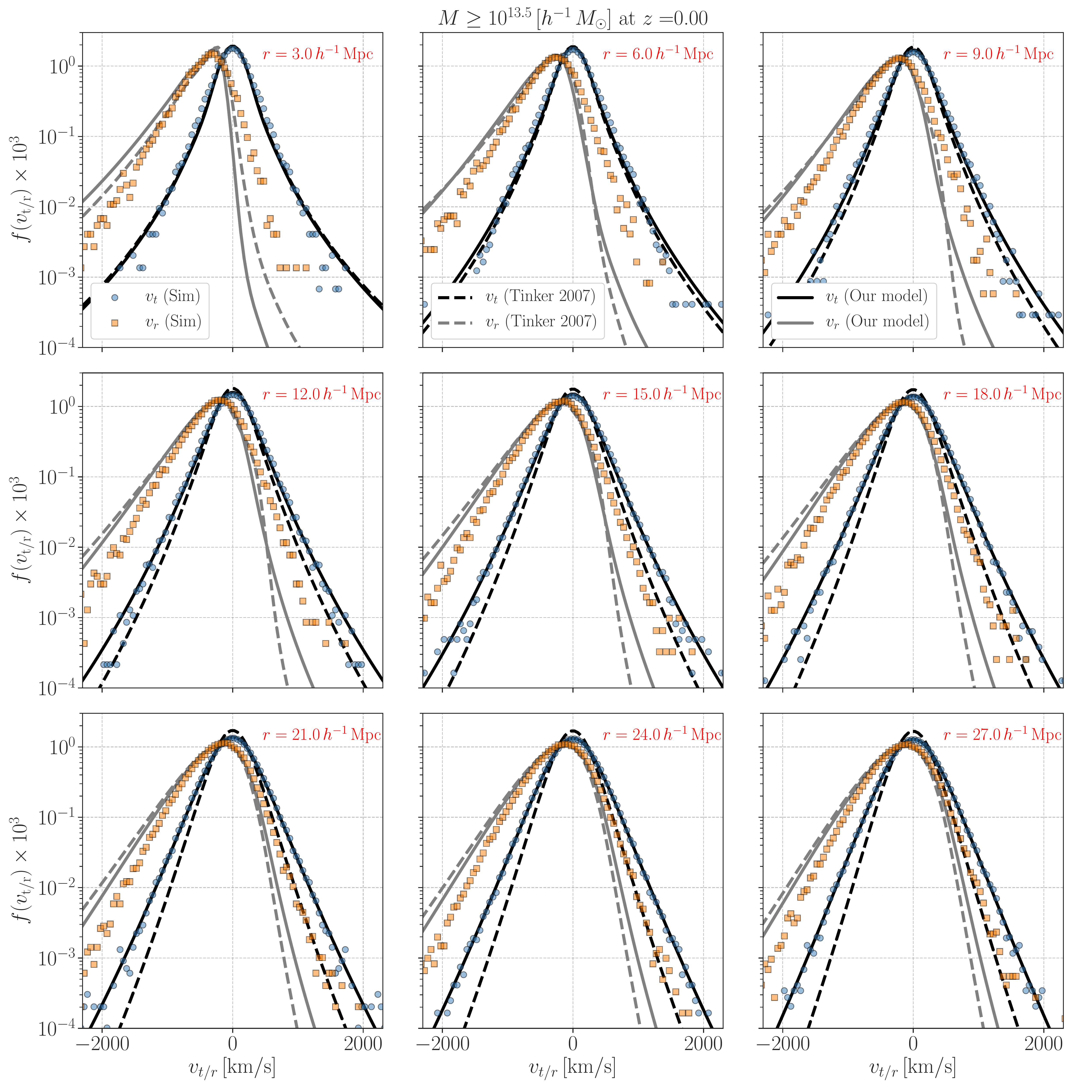}{1.\textwidth}{}}
\caption{
Similar to Figure~\ref{fig:v_pdf_logM_135}, but we here show the 
probability distribution functions (PDFs) 
of the pairwise velocity of dark matter halos with their masses greater than $10^{13.5}\, h^{-1} M_{\odot}$ at 
the redshift of $0$.
\label{fig:v_pdf_logM_135_z0}}
\end{figure*}

After the calibration, we find that our model can not provide a reasonable fit to the pairwise velocity PDF of mass-limited halos at $z=0$.
Figure~\ref{fig:v_pdf_logM_135_z0} summarizes the pairwise velocity PDFs for the mass-limited sample with $M \ge 10^{13.5}\, h^{-1}M_{\odot}$ at $z=0$.
The figure shows sizable differences of radial velocity PDFs between our model and the simulation results.
Note that the standard deviation $\sigma_{r,t}$ can be explained by our model within a 5\%-level precision for this mass-limited sample, but the mean radial velocity profiles in the simulation are larger than our model predictions by $\simeq 30\%$.
Because the mean velocity profile in our model is sensitive to the functional forms in Eqs.~(\ref{eq:cond_PDF}) and (\ref{eq:mean_v_our}), the additional information of $N$-body particles allows us to find more appropriate functional forms.
Also, our model can not explain the velocity-moment profiles at $r\simlt 5\, h^{-1}\mathrm{Mpc}$ for most cases. This also implies that Eqs.~(\ref{eq:Sigma_v_T07}) and (\ref{eq:mean_v_our}) may need some corrections for the velocity statistics at $r\simlt 5\, h^{-1}\mathrm{Mpc}$.
Note that the log-normal approximation for the cosmic mass density PDF can be less accurate at the scale of $r\simlt 5\, h^{-1}\mathrm{Mpc}$ \citep[e.g.][]{2017ApJ...843...73S}.

\subsection{Interpolation errors and more precise modeling}\label{subsec:interpolation_error}

\begin{figure*}[t!]
\gridline{\fig{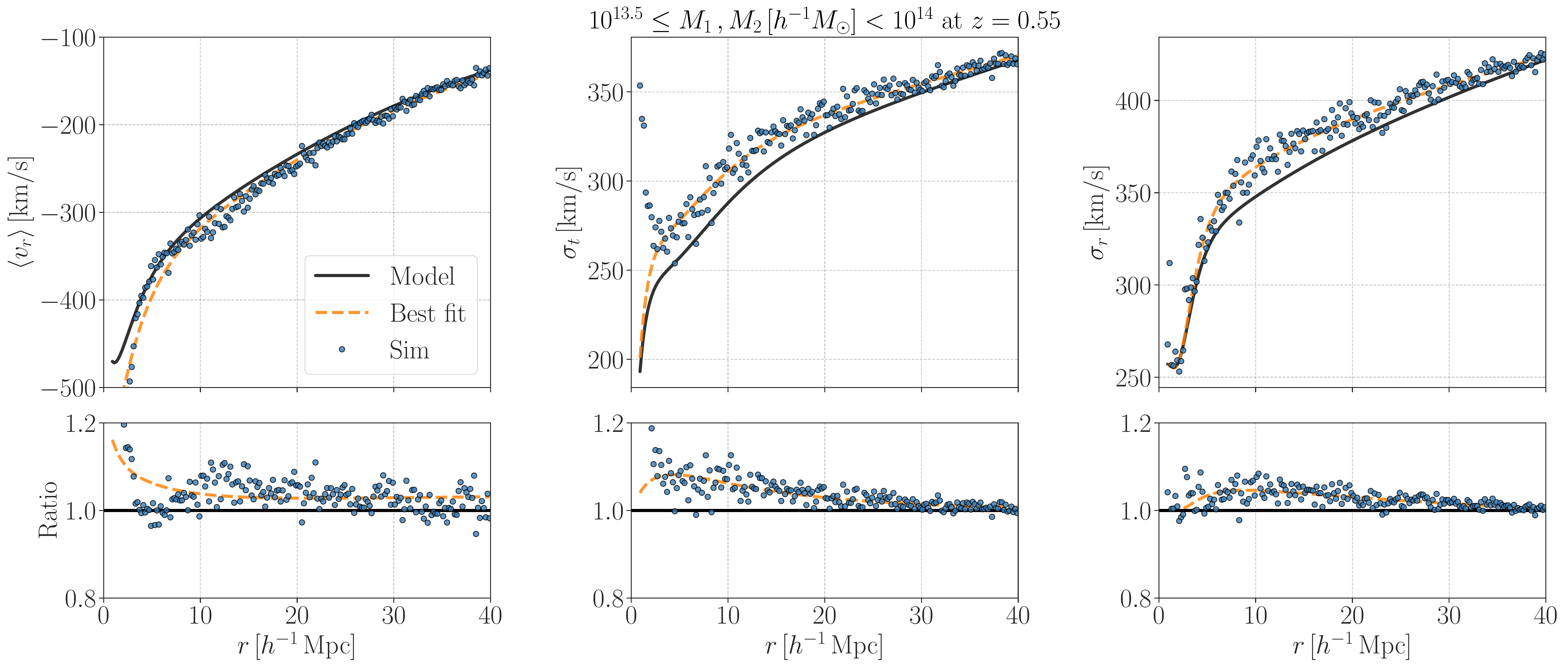}{0.9\textwidth}{}}
\caption{
Interpolation error in our model of the pairwise velocity PDFs. In the upper panels, the blue points show the velocity-moment profiles of $\langle v_r \rangle$, $\sigma_t$,
and $\sigma_r$ for the halos with $M_1=M_2=10^{13.5-14}\, h^{-1}M_{\odot}$ at $z=0.55$, while the solid lines are our model predictions. The orange line shows the best-fit model inferred by the least-square fitting of velocity moment profiles in the calibration process (see Section~\ref{sec:calibration} for details). 
In the bottom, we show the ratio of profiles between simulation results and our model with the blue points,
while the orange line shows the ratio between the best-fit and our model which includes the interpolation over halo masses. \label{fig:interpolation_error}}
\end{figure*}

As in Section~\ref{sec:calibration},
our calibration is based on the least square fitting of
velocity-moment profiles and the interpolation of the best-fit parameters over halo masses and redshifts.
Our interpolation scheme provides the best performance for 
dark matter halos with 
$10^{12.5}<M\, [h^{-1}M_{\odot}]<10^{13.5}$ at $0.3<z<1$, 
but it gives less precise predictions for other ranges of $M$ and $z$.
Figure~\ref{fig:interpolation_error} summarizes 
an example of the interpolation error in our calibration process.
In this figure, we show the velocity-moment profiles for halos with $M=10^{13.5-14}\, h^{-1}M_{\odot}$.
After the fitting process, we find the best-fit expression of each profile as shown in the orange lines of 
%\textbf{JASON: they look orange to me} 
Figure~\ref{fig:interpolation_error}.
Since the final model involves with 
interpolation of model parameters over $M$ and $z$,
sizable residuals can be found in the comparisons of velocity-moment profiles if we use an inaccurate interpolation method.
Figure~\ref{fig:interpolation_error} also highlights that
the best-fit expression reaches a few-percent-level precision
for a given bin of masses and redshift.
This indicates that a more sophisticated interpolation beyond the use of an analytic function will further improve the precision of our model prediction.

A promising approach for the interpolation of our model parameters is the Gaussian Process Regression.
The Gaussian Process Regression allows to interpolate a large-dimensional dataset in a non-parametric way
and it is becoming a standard approach to develop accurate models for various statistics of large-scale structures \citep[e.g.][]{2007PhRvD..76h3503H, 2010ApJ...713.1322L, 2013ApJ...768..123K, 2015ApJ...810...35K, 2019ApJ...872...53M, 2019ApJ...884...29N}.
For the Gaussian Process Regression, one usually needs to reduce the effective numbers of data points in some way such as the Principle Component Analysis.
In our approach, we can reduce the number of model parameters in a physically-motivated way.
Our analyses show that only three functions, $\tilde{\rho}_{0}(r, M_1, M_2, z)$, 
$\tilde{\rho}_{t}(r, M_1, M_2, z)$ 
and $\tilde{\rho}_{r}(r, M_1, M_2, z)$ 
(see Section~\ref{sec:model} for details) 
will be sufficient to fit the pairwise 
velocity PDFs 
for the Planck cosmology.

It would be worth mentioning that this is a huge reduction of the number of dimensions in the model compared to 
other models of pairwise velocity PDFs in the literature.
\citet{2013MNRAS.431.3319Z} introduced a two-dimensional skewed-t distribution with seven functions to explain the pairwise velocity PDFs of galaxies around clusters. The seven functions in \citet{2013MNRAS.431.3319Z} depend on $r, M_1, M_2$ and $z$ in principle.
\citet{2016MNRAS.463.3783B} developed a model of the pairwise velocity PDFs which is valid for both dark matter particles and halos. The model requires the knowledge of the first three moments of the line-of-sight pairwise velocity distribution plus two well-defined dimensionless parameters, and each is a function of $r, M_1, M_2$ and $z$.
\citet{2018MNRAS.479.2256K} found that a mixture of Gaussian PDFs can provide an excellent fit to the pairwise velocity PDFs for the line-of-sight component in the simulations. This model requires five functions to set full properties of the velocity PDFs. These five functions are dependent on
$r_{p}, r_{\pi}, M_1, M_2$ and $z$ for dark matter halos.
Recently, \citet{2020arXiv200202683C} proposed that a one-dimensional skewed-t PDF can provide a sufficient fit to the PDFs of the line-of-sight pairwise velocity for dark matter halos with $M\ge 10^{13}\, h^{-1}M_{\odot}$ at $z=0$.
A skewed-t PDF has four free parameters and each will depend on $r_{p}, r_{\pi}, M_1, M_2$ and $z$ in general.
Most previous studies have not studied the dependence of their PDF model on halo masses, redshifts, and the separation lengths. Future studies should focus on efficient calibrations and emulations of the mass-redshift-scale dependence of pairwise velocity PDFs for dark matter halos.

\section{Discussion and Conclusion} \label{sec:conclusion}

In this paper, we developed a semi-analytic model of the pairwise velocity distributions of dark matter halos.
The model is motivated by the findings and framework in \citet{2007MNRAS.374..477T} and we re-calibrated the model parameters in the relation between the pairwise velocity and 
an environmental density around halo pairs using 
high-resolution $N$-body simulation covering 
a volume of $\sim1\, \mathrm{Gpc}^3$.

Our model has three functions related to the halo formation and the dependence of velocity dispersions on the cosmic mass density. 
By combining the log-normal PDF of cosmic mass density,
our model can realize a significant non-Gaussianity in the pairwise velocity PDF with three parameters alone, while previous non-Gaussian PDF models require more parameters.
We calibrated these three as a function of halo masses ($M_1$ and $M_2$), redshifts $z$ and the separation lengths $r$ using halo catalogs for $10^{12.5}<\, [h^{-1}M_{\odot}]<10^{15}$ and $0<z<1$. We found that our model can reproduce the first three non-zero velocity moments at $5<r\, [h^{-1}\mathrm{Mpc}]<40$
for the halo masses of $10^{12.5}\simlt M \, [h^{-1}M_{\odot}] \simlt 10^{13.5}$ at $0.3<z<1$ with a 5\%-level precision.
For more massive halos or lower redshifts, we expect that our model is still able to explain the mean and dispersions of the pairwise velocity with a precision level of 10-20\%.
Based on the streaming model of two-point correlation functions, we also validated if our model can provide an accurate mapping of the two-point correlations between real and redshift space.
For the mass-limited sample with $M\ge10^{13.5} \,h^{-1} M_{\odot}$ at $z=0.55$, 
we confirmed that our model can explain the redshift-space clustering monopole and quadropole in the range of $5-40$
and $10-30\, h^{-1}\mathrm{Mpc}$ within a 5\%-level precision.
This is valid even for a realistic SDSS-III BOSS CMASS 
galaxy sample based on the framework of a halo occupation distribution (HOD), if we have an accurate model of the real-space correlation function of galaxies.
We then studied the dependence of the clustering multipoles on the velocity biases in the galaxy straming motion by using our model.
We found that a 20\%-level bias in the mean and dispersion of the pairwise velocity of galaxies can induce a characteristic scale dependence of the observables at $\sim10\, \mathrm{Mpc}$.
It would be difficult to reproduce these features by varying the typical halo mass of galaxies alone, but more investigations are needed to make a robust conclusion.
%\textbf{JASON: I did not understand the previous sentence}
 
Although our model of the pairwise velocity PDFs will play an important role in cosmological analyses in redshift surveys of massive galaxies, we require further improvements of the model before applying it to real data sets.
In fact, the statistical uncertainties of the redshift-space clustering monopole and quadropole for the massive galaxies
in BOSS already reach a level of a few percent at $1-10\, h^{-1}\mathrm{Mpc}$ \citep[e.g.][]{2014MNRAS.444..476R}
and our model precision is comparable to them at best.
To improve the model precision, we may require a more sophisticated approach to calibrate model parameters such as Gaussian Process Regression, or some modifications in the functional forms in our model.
Analyses involved with $N$-body particle data would be a key to improve our model, because the relationship between the cosmic mass density and the halo velocity is the essential part in our model.
In addition, our model assumes the specific cosmological model in a $\Lambda$CDM scenario.
We require further investigations to study the cosmological dependence of our model as well as extend our framework to include modified gravity theories.
Upcoming redshift surveys aim at measuring the redshift-space clustering of galaxies with lower masses and higher redshifts than the mass- and redshift ranges explored in this study.
It is thus important to extend our approach so as to be applicable for a wider range of halo masses and redshifts.

The model presented in this paper is an important first step toward statistical inference of the kinematics of galaxies from their clustering information in redshift surveys as well as interpretation of the small-scale measurements of the kinematic Sunyaev-Zel'dovich effect. Precise analyses with current and upcoming redshift surveys enable us to study the motion of several tracers of large-scale structures.
The kinematic information of the tracers 
can provide an independent and important test of the standard cosmological model and allow us to examine possible deviations from General Relativity, if we have an accurate model of the pairwise velocity PDFs of dark matter halos.
Our future work with the model of the pairwise velocity 
include a joint analysis of galaxy-galaxy lensing and the redshift-space clustering to infer the streaming motion of dark matter halos
and investigation of the small-scale information 
in the kinematic Sunyaev-Zel'dovich effect on 
massive galaxies at various redshifts.

\acknowledgments
%The author thanks the anonymous referee for reading the paper carefully and providing thoughtful comments, 
%many of which have resulted in changes to the revised version of the manuscript.
We thank the $\nu^2$GC collaboration for making their simulation data publicly available.
This work is in part supported by MEXT KAKENHI Grant Number (18H04358, 19K14767).
MS is supported by JSPS overseas Research Fellowships during his stay at the Jet Propulsion Laboratory (JPL).
Numerical computations were in part carried out on Cray XC50 at Center 
for Computational Astrophysics, National Astronomical Observatory of Japan. EH, KM, and JR were supported by JPL, which is run by Caltech under a contract with the National Aeronautics and Space Administration (80NM0018D0004).

\appendix
\section{List of model parameters}\label{apdx:model_params}

In this appendix, we provide the fitting functions in our model of the pairwise velocity distribution. The model is summarized in Section~\ref{sec:model} and we introduce 58 parameters to explain the dependence of our model on halo masses, redshifts, separation lengths between halos.

For Eq.~(\ref{eq:density_cut_off_ours}), we find the following forms provide a reasonable fit to the simulation results:
\beqa
{\cal A}_{\rho}(M_1, M_2, z) &=& \frac{a_{1}(z) [D(z) a_{2}(z)\, y]^{a_{3}(z)}}{1+[D(z) a_{2}(z)\, y]^{a_{3}(z)}}, \\
a_{1}(z) &=& 0.0385 \, (1+z)^{-6.47} + 1.04, \\
a_{2}(z) &=& 0.488 \, z^{3.21} + 0.737,\\
a_{3}(z) &=& -0.710 \, (z-0.310)^2 + 5.93,\\
y &\equiv& b_{\mathrm{L}}(M_1, z) + b_{\mathrm{L}}(M_2, z), \\
{\cal B}_{\rho}(M_1, M_2, z) &=&  
(26.7z^2+2.83z+17.4) \, \left(\frac{M_1+M_2}{10^{13}\, h^{-1}M_{\odot}}\right)^{0.631},\\ 
{\cal C}_{\rho}(M_1, M_2, z) &=&
-\left[0.109\,(z-0.189)^2+0.862\right]\,
\left(\frac{M_1+M_2}{10^{13}\, h^{-1}M_{\odot}}\right)^{\left[-0.0223\,(z-0.438)^2+0.204\right]},
\eeqa
where $D(z)$ is the linear growth factor normalized to unity at $z=0$, and $b_{\mathrm{L}}(M,z)$ is the linear halo bias.

For Eq.~(\ref{eq:Sigma_rho_t_r_ours}), we adopt the following forms of
\beqa
{\cal C}^{(0)}_{t}(M_1, M_2, z) &=& 
\left[-47.36\, (z-0.54)^2 +60.6\right]\, (R_{\mathrm{200b},0})^{0.701z^2-1.42z+1.80},\\
{\cal C}^{(1)}_{t}(M_1, M_2, z) &=& 
\left[-8.36\, (z-0.572)^2 +7.72\right]\, (R_{\mathrm{200b},0})^{0.0828z^2-0.509z+0.042}
,\\
{\cal C}^{(2)}_{t}(M_1, M_2, z) &=& 0.45,\\
p_{t}(M_1, M_2, z) &=& 
\left[0.866\, (z-0.715)^2-2.16\right]\, (R_{\mathrm{200b},0})^{0.333\, (1+z)^{-4.82}-0.0943}
,\\
q_{t}(M_1, M_2, z) &=& -0.9,\\
{\cal C}^{(0)}_{r}(M_1, M_2, z) &=& 
\left[786.0\, z^2 -2945 z +2970\right]\, 
\exp\left[-\left(\frac{0.248\, z^2 -1.06z+1.31}{R_{\mathrm{200b}, 0}}\right)^2\right]\, R^{2}_{\mathrm{200b, larger}}
,\\
{\cal C}^{(1)}_{r}(M_1, M_2, z) &=& 
\left[-4.84\, z^2 +0.431 z +26.4\right]
(R_{\mathrm{200b},0})^{-0.0482z^2-0.103z+0.484}
,\\
{\cal C}^{(2)}_{r}(M_1, M_2, z) &=& 
\left[-0.109\, (z-0.66)^2 +0.497\right]
(R_{\mathrm{200b},0})^{-0.161-0.0363z}
,\\
p_{r}(M_1, M_2, z) &=& -4.0,\\
q_{r}(M_1, M_2, z) &=& -1.3,
\eeqa
where $R_{\mathrm{200b}, 0} = R_{\mathrm{200b}}(M_1) + R_{\mathrm{200b}}(M_2)$ and $R_{\mathrm{200b, larger}} = {\rm MAX}\left(R_{\mathrm{200b}}(M_1), R_{\mathrm{200b}}(M_2)\right)$.
The radii $R_{\mathrm{200b}, 0}$ and $R_{\mathrm{200b, larger}}$ are in the unit of comoving $h^{-1}\, \mathrm{Mpc}$.

\section{Halo-based streaming model of redshift-space clustering with a halo occupation distribution}\label{apdx:halo_streaming_model}

In this appendix, we briefly summarize an analytic expression of the redshift-space two point correlation with our model of the pairwise velocity distribution of dark matter halos for a given halo occupation distribution (HOD) \citep[also see][for more details]{2007MNRAS.374..477T}. Note that we omit the redshift $z$ for most parts in the following discussion for simplicity.

\subsection{Setup}\label{subapdx:HOD}

For a galaxy sample of interest, we assume that the galaxies can be decomposed into two types, centrals and satellites.
For the central galaxies, we assume that they reside in the center of their host dark matter halos and individual host halos can have single central galaxies at most.
For the satellite galaxies, we populate satellite galaxies to a halo only when a central galaxy exists.
The HOD represents the mean number of galaxies in host halos with mass $M$ and it is given by
\beqa
\langle N_{\mathrm{gal}}\rangle_M
= \langle N_{\mathrm{cen}}\rangle_M
+ \langle N_{\mathrm{sat}}\rangle_M,
\eeqa
where $\langle N_{\mathrm{cen}}\rangle_M$ and 
$\langle N_{\mathrm{sat}}\rangle_M$ are 
the HODs for centrals and satellites, respectively. 
In the following, 
we assume that the conditional distribution of the number of central galaxies in a given halo follows the Bernoulli distribution (i.e., can take only zero or one) with mean of $\langle N_{\mathrm{cen}}\rangle_M$. On the other hand, the conditional distribution of the number of satellites is set by the Poisson distribution with mean $\lambda_M$. 
In this setup, the HOD for satellites can be expressed as $\langle N_{\mathrm{sat}}\rangle_M = \langle N_{\mathrm{cen}}\rangle_M\, \lambda_M$.
Once the HOD is specified, we can compute the mean number density of the galaxies as
\beqa
\bar{n}_{\mathrm{g}} = \int\mathrm{d}M \frac{\mathrm{d}n}{\mathrm{d}M}\left(\langle N_{\mathrm{cen}}\rangle_M+\langle N_{\mathrm{sat}}\rangle_M\right),
\eeqa
where $\mathrm{d}n/\mathrm{d}M$ is the halo mass function.
In this paper, we adopt the model of halo mass functions in \citet{2008ApJ...688..709T}.
\ms{When comparing simulation results, one can use the mass function directly measured from the simulation. Although this is a better choice, we still adopt the model in \citet{2008ApJ...688..709T} in this paper. For a sanity check, we compared the halo mass function at $z=0.55$ in the $\nu^2$GC simulation with the prediction by \citet{2008ApJ...688..709T}. We found a $10\%$-level difference at $M=10^{13-14}\, h^{-1}M_{\odot}$, which is the most relevant mass range to the CMASS HOD. Nevertheless, this $10\%$-level difference is less dependent on the halo mass. In the clustering, a constant multiplicative bias in the mass function does not affect the model prediction (see Appendix~\ref{subapdx:HOD_2pcf}). Hence, we expect that the halo mass function by \citet{2008ApJ...688..709T} is sufficient in our analyses.}

\subsection{Two-point correlation function}\label{subapdx:HOD_2pcf}

Within the HOD framework, the two-point correlation function of galaxies can be decomposed into two parts known as one-halo and two-halo terms. The one-halo term represents 
the two-point correlation within single halos, 
while two-halo term arises from the clustering among neighboring halos. 
For a given HOD in Appendix~\ref{subapdx:HOD}, the one-halo terms in redshift space can be expressed as \citep{2007MNRAS.374..477T},
\beqa
\xi^{S}_{\mathrm{1h}}(s_p, s_{\pi})&=&
\xi^{S,\mathrm{cs}}_{\mathrm{1h}}(s_p,s_{\pi})
+\xi^{S,\mathrm{ss}}_{\mathrm{1h}}(s_p, s_{\pi}), \\
\xi^{S,\mathrm{cs}}_{\mathrm{1h}}(s_p,s_{\pi}) &=&
\frac{1}{2\pi\bar{n}^2_{\mathrm{g}}}\int\, \mathrm{d}M \frac{\mathrm{d}n}{\mathrm{d}M} 
\langle N_{\mathrm{sat}}\rangle_M\, \int_{-\infty}^{\infty}\,
\frac{H(z)\,\mathrm{d}r_{\pi}}{(1+z)}\, 
\frac{F_{\mathrm{cs}}\left(\sqrt{s^2_{p}+r^2_{\pi}}|M\right)}{s^2_{p}+r^2_{\pi}}\, 
{\cal P}_{\mathrm{cs}}\left(v_z = \frac{H(z)(s_{\pi}-r_{\pi})}{(1+z)}\, \Bigl|\, M\right), \\
%\nonumber \\
%&&\qquad \qquad \qquad \qquad \qquad \qquad
%\qquad \qquad \qquad \qquad \qquad \qquad
%\qquad 
%\times\, 
\xi^{S,\mathrm{ss}}_{\mathrm{1h}}(s_p,s_{\pi}) &=&
\frac{1}{2\pi\bar{n}^2_{\mathrm{g}}}\int\, \mathrm{d}M \frac{\mathrm{d}n}{\mathrm{d}M} \frac{\langle N_{\mathrm{cen}}\rangle \lambda^2_M}{2}\, \int_{-\infty}^{\infty}\,
\frac{H(z)\,\mathrm{d}r_{\pi}}{(1+z)}\, 
\frac{F_{\mathrm{ss}}\left(\sqrt{s^2_{p}+r^2_{\pi}}|M\right)}{s^2_{p}+r^2_{\pi}}\, 
{\cal P}_{\mathrm{ss}}\left(v_z = \frac{H(z)(s_{\pi}-r_{\pi})}{(1+z)}\, \Bigl|\, M\right),
\eeqa
where $F_{\mathrm{cs}}(r|M)$ is the fraction of number of central-satellite pairs at the radius of $r$ 
in a halo with $M$, 
$F_{\mathrm{ss}}(r|M)$ is the fraction of number of satellite-satellite pairs, 
${\cal P}_{\mathrm{cs}}$ and ${\cal P}_{\mathrm{ss}}$
are the PDF of the pairwise velocity along a line of sight
for central-satellite and satellite-satellite pairs, respectively. Note that 
$\int \mathrm{d}r\, F_{\mathrm{cs}}(r|M) = \int \mathrm{d}r\, F_{\mathrm{ss}}(r|M) = 1$.

When assuming the velocity distribution within each halo as
an isotropic, isothermal Gaussian distribution and
the satellite galaxy velocity dispersion in a halo is set to the virial dispersion, one can find
\beqa
{\cal P}_{\mathrm{cs}}(v_z|M) &=& {\cal N}(v_z, 0, \sigma_{\mathrm{vir},M}), \\
{\cal P}_{\mathrm{ss}}(v_z|M) &=& {\cal N}(v_z, 0, \sqrt{2}\sigma_{\mathrm{vir},M}),
\eeqa
where ${\cal N}(x, \mu, \sigma)$ is a Gaussian distribution of a random field $x$ with mean $\mu$ and variance $\sigma^2$,
and $\sigma_{\mathrm{vir},M}$ represents the virival dispersion in a halo with mass $M$.

In addition, it is commonly assumed that the number density profile of satellites follows the mass density profile of its host dark matter halos. When the mass density profile in a halo is described by the (truncated) NFW profile \citep{1996ApJ...462..563N}, 
the fraction of number of galaxy pairs is given by
\beqa
F_{\mathrm{cs}}(r|M) &=& \frac{1}{f(c)}\frac{r}{(r+r_s)^2},\\
F_{\mathrm{ss}}(r|M) &=& \frac{r}{2\, f^2(c)\, r^2_s}\, \int_{0}^{c}\, \mathrm{d}{x}_1\, Q(x_1, r/r_s, c),
\eeqa
where $c$ and $r_{s}$ are the halo concentration and scaled radius for the NFW profile, $f(c)=\ln(1+c)-c/(1+c)$, and
\beqa
Q(x_1, x, c) = 
\left\{
\begin{array}{ll}
0 & \,\,\,\, (|x-x_1| > c) \\
(1+x_1)^{-2}\left[(1+|x-x_1|)^{-1}-(1+x+x_1)^{-1}\right] & \,\,\,\, (x+x_1 < c \, , |x-x_1| \le c) \\
(1+x_1)^{-2}\left[(1+|x-x_1|)^{-1}-(1+c)^{-1}\right] & \,\,\,\, (x+x_1 \ge c \, , |x-x_1| \le c)
\end{array} \right. .
\eeqa

The two-halo term is then modeled by 
\beqa
1+\xi^{S}_{\mathrm{2h}}(s_p, s_{\pi})
 = \int_{-\infty}^{\infty} \frac{H(z)\,\mathrm{d}r_{\pi}}{(1+z)} \, 
 {\cal P}_{\mathrm{2h,g}}\left(v_z=\frac{H(z)(s_{\pi}-r_{\pi})}{(1+z)} \, \Bigl| \, s_p, r_{\pi}\right)\, \left[1+\xi_{\mathrm{2h}}(\sqrt{s^2_p+r^2_{\pi}})\right],
\eeqa
where $\xi_{\mathrm{2h}}(r)$ is the two-halo term of real-space correlation function, and ${\cal P}_{\mathrm{2h,g}}$ is the pairwise velocity PDF of galaxies for two separated halos.
We here suppose that $\xi_{\mathrm{2h}}(r)$ is accurately predicted by some approach such as perturbation-theory-based models \citep[e.g.][for a recent review]{2018PhR...733....1D}, semi-analytic models \citep[e.g.][]{2001ApJ...561L.143H, 2005ApJ...631...41T, 2013MNRAS.430..725V}, and simulation-based models \citep[e.g.][]{2015ApJ...810...35K, 2019ApJ...884...29N, 2019ApJ...874...95Z}.
For the pairwise velocity PDF, we first compute the pairwise velocity PDF of dark matter halos for the line-of-sight component by using Eq.~(\ref{eq:T07_model_joint_PDF}):
\beqa
{\cal P}(v_z \, | \, r_p, r_{\pi}, M_1, M_2) &=& \int \mathrm{d} v_t \, {\cal P}(v_r, v_t \, | \, r, M_1, M_2) \, \delta_{\mathrm{D}}\left(v_t - \frac{v_r\, \cos \theta - v_z}{\sin \theta}\right) \nonumber \\
&=& \int \mathrm{d} \delta \, {\cal N}\left(v_z, \mu_r[\delta]\cos\theta, 
\sqrt{\Sigma^2_{r}[\delta]\, \cos^2\theta + \Sigma^2_{t}[\delta]\, \sin^2\theta}\right)\, {\cal F}(\delta\, |\, r, M_1, M_2), \label{eq:vlos_PDF_halos}
\eeqa
where $r=\sqrt{r^2_{p}+r^2_{\pi}}$,
$\cos \theta = r_{\pi} / r$, 
${\cal F}$ is the condtional PDF of cosmic mass density having a halo pair with masses of $M_1$ and $M_2$ within $r$,
$\mu_r$ is the mean radial velocity at a given environmental density $\delta$, and $\Sigma_{t,r}$ represents the velocity dispersion at a given $\delta$. The details of these functions are found in Section~\ref{sec:model} and Appendix~\ref{apdx:model_params}.
We then incorporate Eq.~(\ref{eq:vlos_PDF_halos}) with the HOD framework by assuming the Gaussian velocity distribution of satellites with the virial dispersion of $\sigma_{\mathrm{vir}}$. The final expression of 
${\cal P}_{\mathrm{2h,g}}$ is given by
\beqa
{\cal P}_{\mathrm{2h,g}}(v_z \, | r_p, r_{\pi}) &=&
\left(n^{\prime}_{\mathrm{g}}\right)^{-2}\,
\int_{M_{\mathrm{min,0}}}^{M_{\mathrm{lim},1}}\, \mathrm{d}M_1\, \frac{\mathrm{d}n}{\mathrm{d}M_1}\, \langle N_{\mathrm{gal}}\rangle_{M_1}\,
\int_{M_{\mathrm{min,0}}}^{M_{\mathrm{lim},2}}\, \mathrm{d}M_2\, \frac{\mathrm{d}n}{\mathrm{d}M_2}\, \langle N_{\mathrm{gal}}\rangle_{M_2}\nonumber \\
&&
\qquad \qquad \qquad \qquad \qquad \qquad
\qquad \qquad \qquad
\times\,
{\cal P}_{\mathrm{g+h}}(v_z\, |\, r_p, r_{\pi}, M_{1}, M_{2}),\label{eq:2h_vlos_PDF_gals}\\
{\cal P}_{\mathrm{g+h}}(v_z\, |\, r_p, r_{\pi}, M_{1}, M_{2})
&=&
\int \mathrm{d} \delta \, \sum_{i=1}^{4} w_{i}\, 
{\cal N}(v_z, \mu_{r}\cos\theta, \sigma_{i})
\, {\cal F}(\delta\, |\, r, M_1, M_2), \\
\left(n^{\prime}_{\mathrm{g}}\right)^{2}
&=&
\int_{M_{\mathrm{min,0}}}^{M_{\mathrm{lim},1}}\, \mathrm{d}M_1\, \frac{\mathrm{d}n}{\mathrm{d}M_1}\, \langle N_{\mathrm{gal}}\rangle_{M_1}\,
\int_{M_{\mathrm{min,0}}}^{M_{\mathrm{lim},2}}\, \mathrm{d}M_2\, \frac{\mathrm{d}n}{\mathrm{d}M_2}\, \langle N_{\mathrm{gal}}\rangle_{M_2},\label{eq:2h_vlos_PDF_gals_norm}
\eeqa
where $M_{\mathrm{min,0}}$ is the minimum halo mass that can host a galaxy (usually set by a sufficient small value), 
and
\beqa
w_{1} &=& 
\frac{\langle N_{\mathrm{cen}} \rangle_{M_1} \langle N_{\mathrm{cen}} \rangle_{M_2}}{\langle N_{\mathrm{gal}}\rangle_{M_1}\langle N_{\mathrm{gal}}\rangle_{M_2}},
\,\,
w_{2} =
\frac{\langle N_{\mathrm{cen}} \rangle_{M_1} \langle N_{\mathrm{sat}} \rangle_{M_2}}{\langle N_{\mathrm{gal}}\rangle_{M_1}\langle N_{\mathrm{gal}}\rangle_{M_2}},
\,\,
w_{3} =
\frac{\langle N_{\mathrm{sat}} \rangle_{M_1} \langle N_{\mathrm{cen}} \rangle_{M_2}}{\langle N_{\mathrm{gal}}\rangle_{M_1}\langle N_{\mathrm{gal}}\rangle_{M_2}},
\,\,
w_{4} =
\frac{\langle N_{\mathrm{sat}} \rangle_{M_1} \langle N_{\mathrm{sat}} \rangle_{M_2}}{\langle N_{\mathrm{gal}}\rangle_{M_1}\langle N_{\mathrm{gal}}\rangle_{M_2}},
\\
\sigma^2_{1} &=& \Sigma^2_{r}(\delta, r, M_1, M_2)\, \cos^2\theta + \Sigma^2_{t}(\delta, r, M_1, M_2)\, \sin^2\theta, \\
\sigma^2_{2} &=& \sigma^2_{1} + \sigma^2_{\mathrm{vir},M_2}, 
\qquad
\sigma^2_{3} = \sigma^2_{1} + \sigma^2_{\mathrm{vir},M_1},
\qquad
\sigma^2_{4} = \sigma^2_{1} + \sigma^2_{\mathrm{vir},M_1}+\sigma^2_{\mathrm{vir},M_2}.
\eeqa
In Eqs.~(\ref{eq:2h_vlos_PDF_gals}) and (\ref{eq:2h_vlos_PDF_gals_norm}), we set the upper limits of the integral to $R_{\mathrm{200b}}(M_{\mathrm{lim},1}) = r-R_{\mathrm{200b}}(M_{\mathrm{min,0}})$ and 
$R_{\mathrm{200b}}(M_{\mathrm{lim},2}) = r-R_{\mathrm{200b}}(M_{1})$ by taking into account the effect of halo exclusion.

\section{Performance evaluation of our model for pairwise velocity distribution of dark matter halos}\label{apdx:performance_model}

\begin{figure*}[ht!]
\gridline{\fig{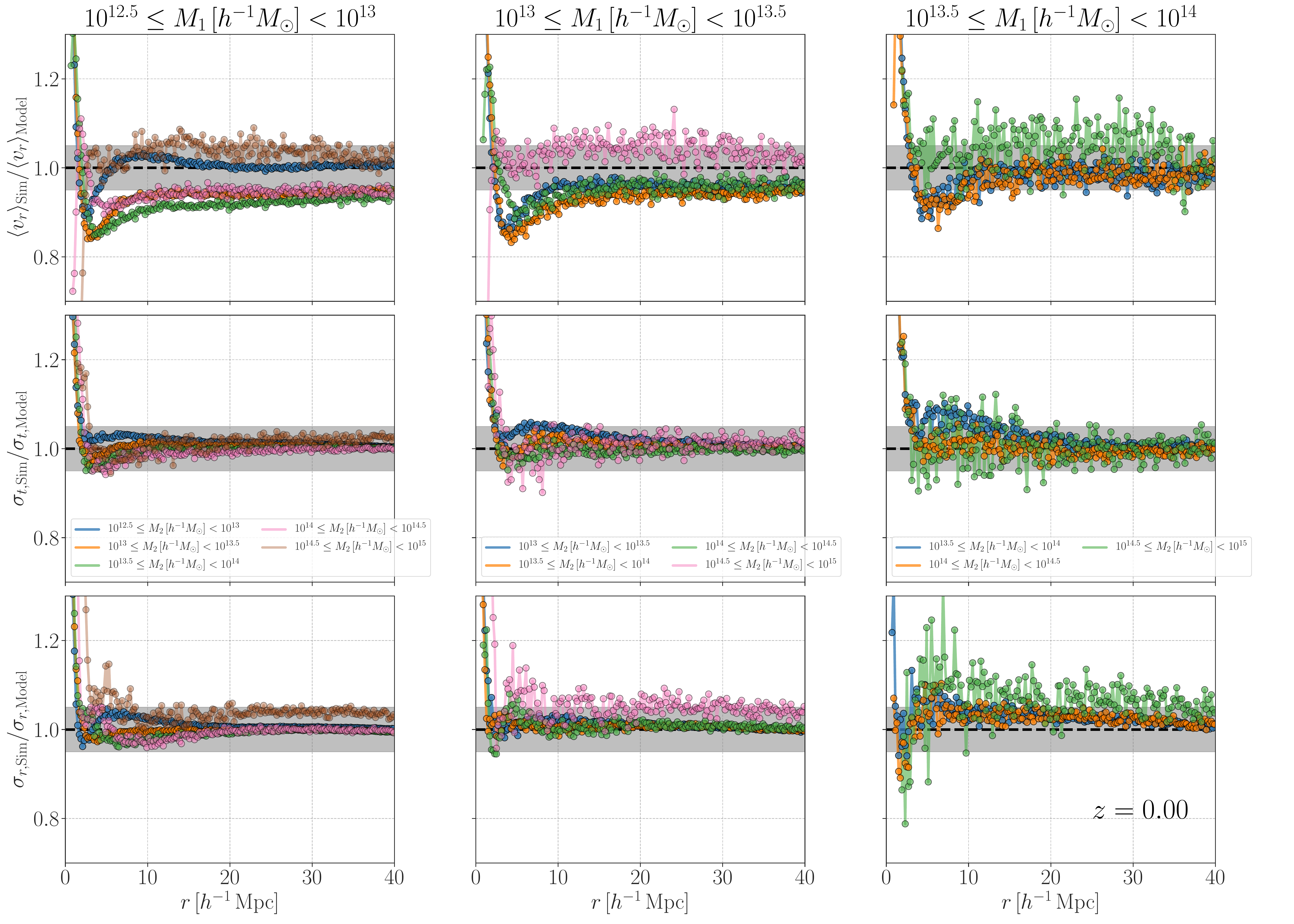}{1.\textwidth}{}}
\caption{
The mean and dispersion of pairwise-velocity of dark matter halos with various masses at the redshift of $z=0$.
Each panel shows the ratio of the velocity-moment profiles for different halo masses. The three left panels show the ratio of $\langle v_{r} \rangle$, $\sigma_t$, and $\sigma_r$ for the halo masses of $M_1 = 10^{12.5-13}\, h^{-1}M_{\odot}$ and $M_2 \ge M_1$ from top to bottom.
The middle and right panels represent the results for $M_1 = 10^{13-13.5}\, h^{-1}M_{\odot}$ and $M_1 = 10^{13.5-14}\, h^{-1}M_{\odot}$, respectively. For a reference, the gray filled region in each panel shows $\pm5$\%-level differences.
\label{fig:v_moments_tests_z0}}
\end{figure*}

\begin{figure*}[ht!]
\gridline{\fig{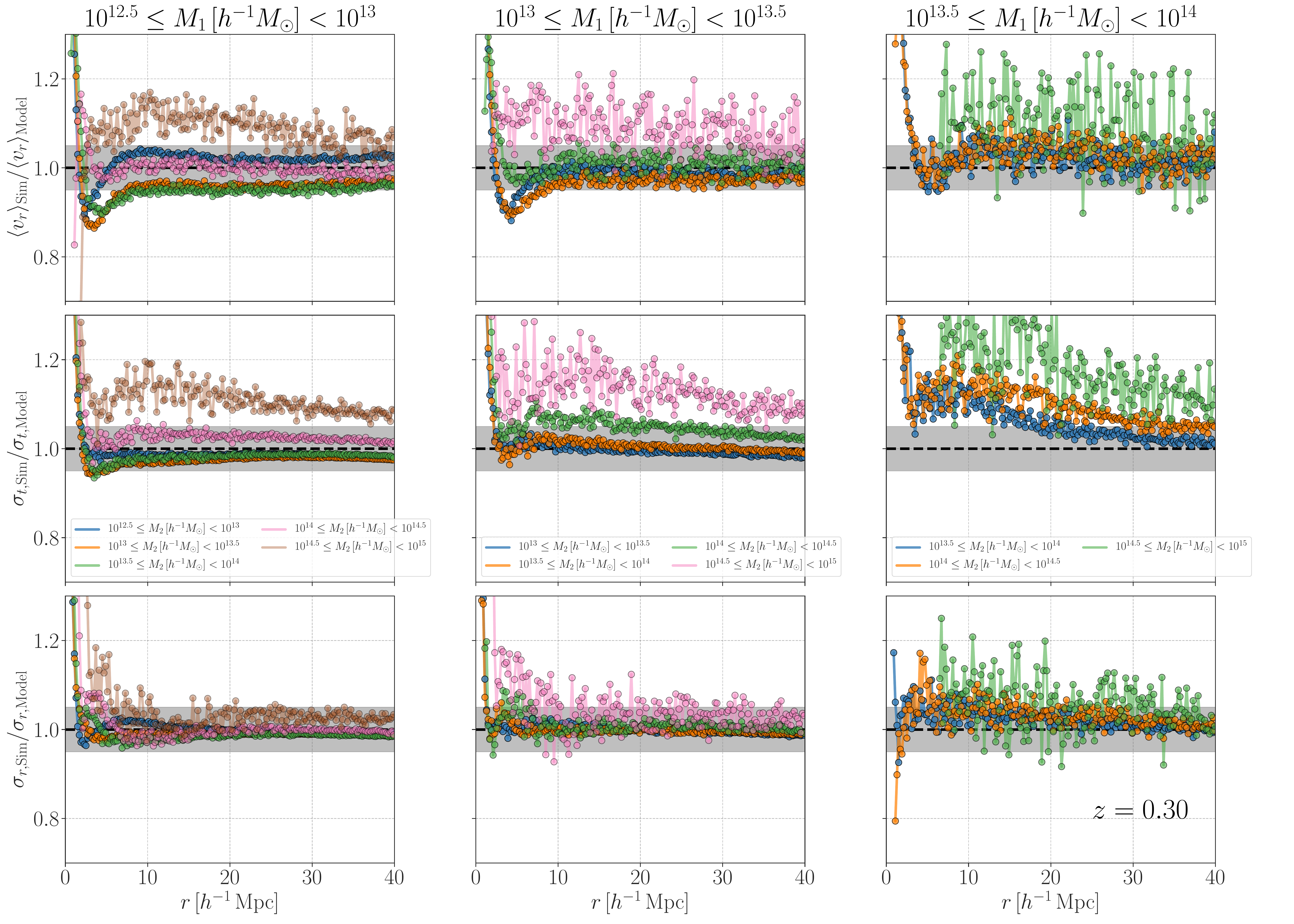}{0.7\textwidth}{}}
\caption{
Similar to Figure~\ref{fig:v_moments_tests_z0},
but this figure presents the results at the redshift of $z=0.30$.
\label{fig:v_moments_tests_z03}}
\end{figure*}

\begin{figure*}[ht!]
\gridline{\fig{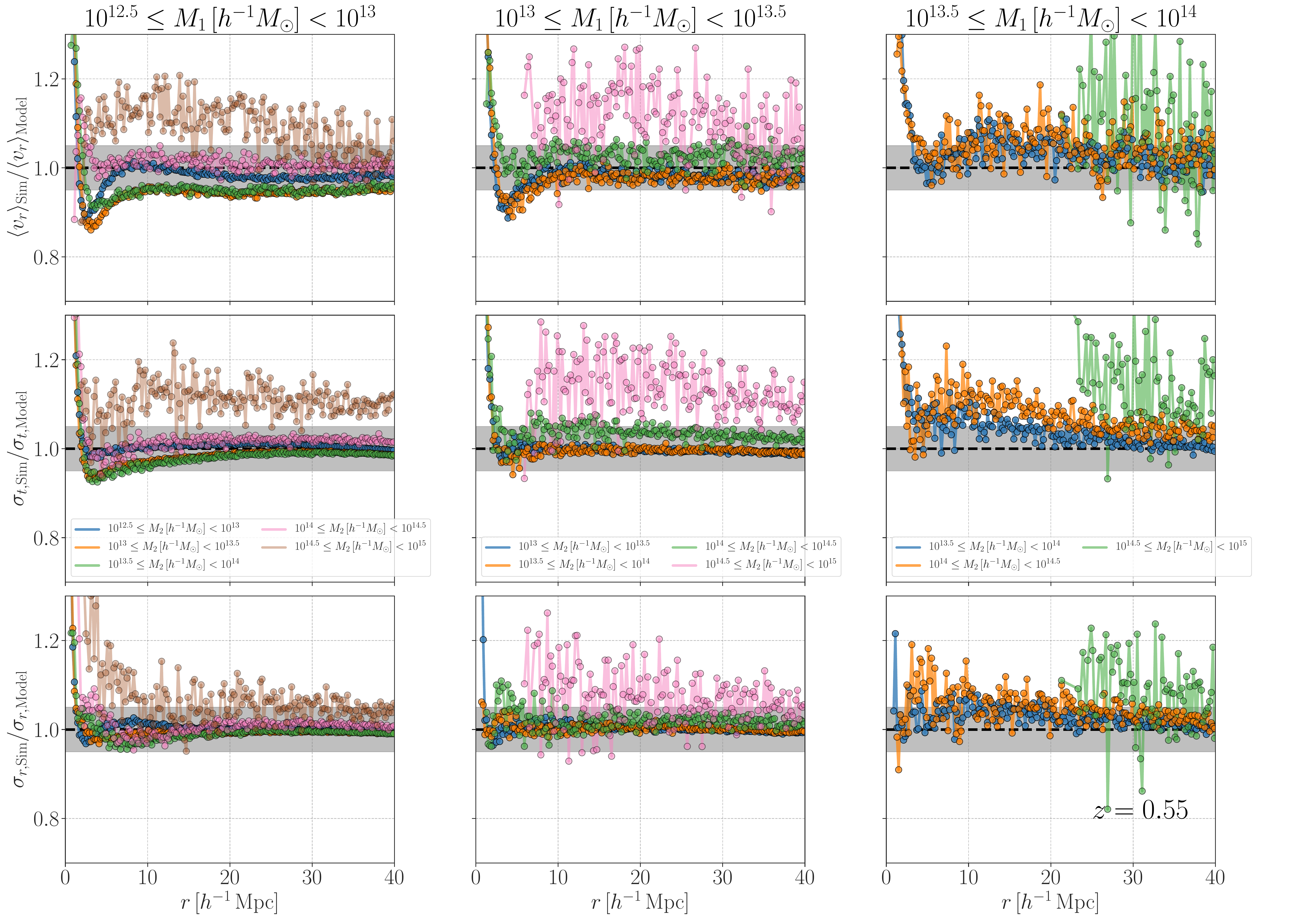}{0.7\textwidth}{}}
\caption{
Similar to Figure~\ref{fig:v_moments_tests_z0},
but this figure presents the results at the redshift of $z=0.55$.
\label{fig:v_moments_tests_z05}}
\end{figure*}

\begin{figure*}[ht!]
\gridline{\fig{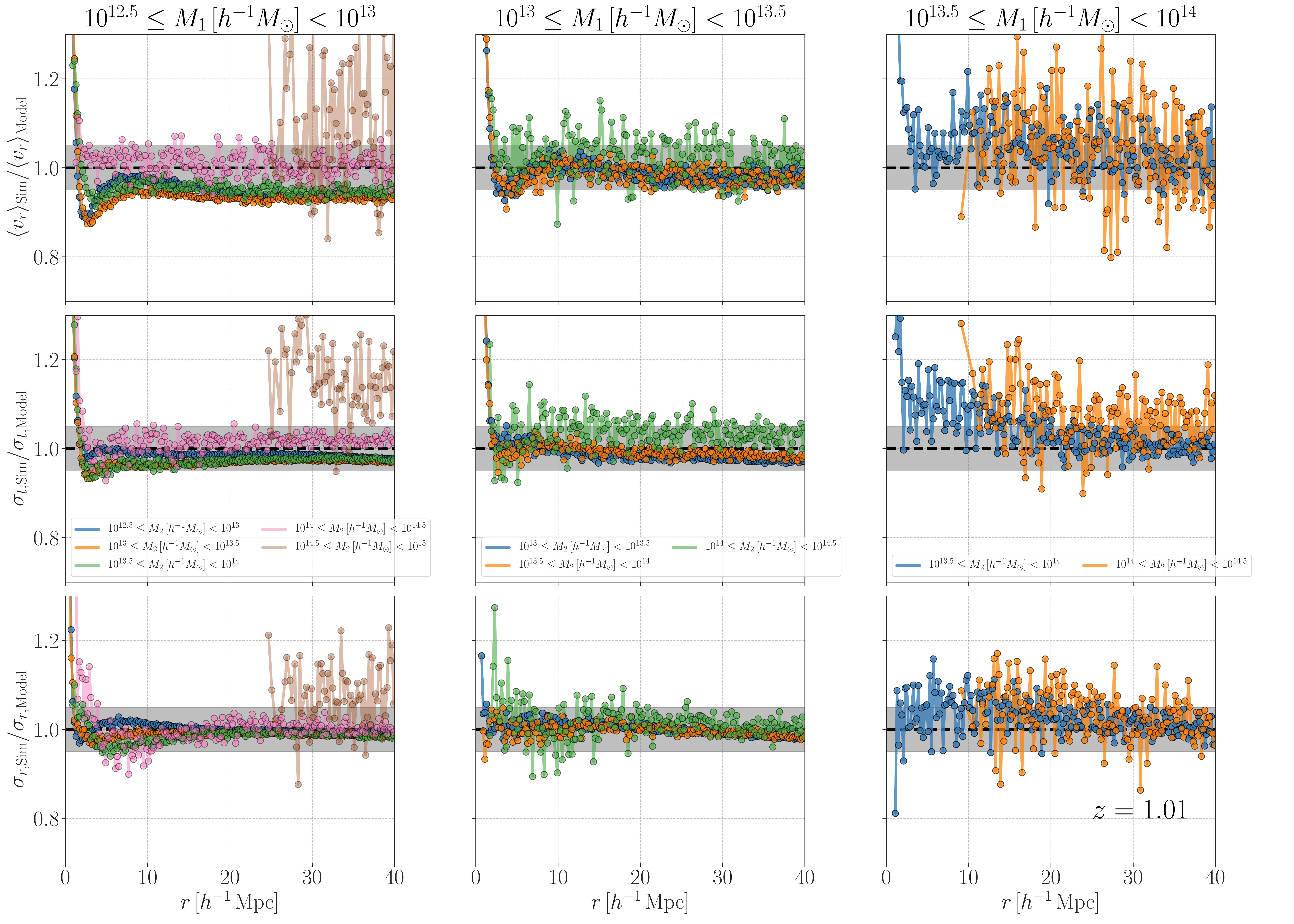}{0.7\textwidth}{}}
\caption{
Similar to Figure~\ref{fig:v_moments_tests_z0},
but this figure presents the results at the redshift of $z=1.01$.
\label{fig:v_moments_tests_z1}}
\end{figure*}

In this appendix, we evaluate our model precision for the profiles of the mean and dispersion in the pairwise velocity of dark matter halos in a wide range of halo masses and redshifts.
Figures~\ref{fig:v_moments_tests_z0}-\ref{fig:v_moments_tests_z1} summarize the ratio of the velocity moments 
between the simulation results and our model predictions
for different halo masses ($M_2 \ge M_1$) and redshifts.
In each figure, the three left panels show the results 
for $M_1 = 10^{12.5-13}\, h^{-1}M_{\odot}$.
From top to bottom, each panel shows the ratio of 
$\langle v_r \rangle$, $\sigma_t$, and $\sigma_r$, respectively.
The three middle panels present the results 
for $M_1 = 10^{13-13.5}\, h^{-1}M_{\odot}$,
while the three right panels are for $M_1 = 10^{13.5-14}\, h^{-1}M_{\odot}$.
Figures~\ref{fig:v_moments_tests_z0}, 
\ref{fig:v_moments_tests_z03}, 
\ref{fig:v_moments_tests_z05},
and \ref{fig:v_moments_tests_z1}
provide the results at $z=0$, 0.30, 0.55, and 1.01, respectively.
There are 10-20\%-level differences 
for halo masses greater than $\sim10^{13.5}\, h^{-1}M_{\odot}$, but our model can reproduce the simulation results for $10^{12.5} < M \,[h^{-1}M_{\odot}] < 10^{13.5}$ at $0.3\simlt z < 1$ with a 5\%-level precision.
We thus expect that our model would be suitable for analyses of massive-galaxy-sized dark matter halos at $z<1$ targeted in various redshift surveys.

%\bibliography{sample63}
\bibliographystyle{aasjournal}
\bibliography{refs}

\begin{thebibliography}{}
\expandafter\ifx\csname natexlab\endcsname\relax\def\natexlab#1{#1}\fi
\providecommand{\url}[1]{\href{#1}{#1}}
\providecommand{\dodoi}[1]{doi:~\href{http://doi.org/#1}{\nolinkurl{#1}}}
\providecommand{\doeprint}[1]{\href{http://ascl.net/#1}{\nolinkurl{http://ascl.net/#1}}}
\providecommand{\doarXiv}[1]{\href{https://arxiv.org/abs/#1}{\nolinkurl{https://arxiv.org/abs/#1}}}

\bibitem[{{Behroozi} {et~al.}(2013){Behroozi}, {Wechsler}, \&
  {Wu}}]{2013ApJ...762..109B}
{Behroozi}, P.~S., {Wechsler}, R.~H., \& {Wu}, H.-Y. 2013, \apj, 762, 109,
  \dodoi{10.1088/0004-637X/762/2/109}

\bibitem[{{Bianchi} {et~al.}(2016){Bianchi}, {Percival}, \&
  {Bel}}]{2016MNRAS.463.3783B}
{Bianchi}, D., {Percival}, W.~J., \& {Bel}, J. 2016, \mnras, 463, 3783,
  \dodoi{10.1093/mnras/stw2243}

\bibitem[{{Bond} \& {Myers}(1996)}]{1996ApJS..103....1B}
{Bond}, J.~R., \& {Myers}, S.~T. 1996, \apjs, 103, 1, \dodoi{10.1086/192267}

\bibitem[{{Coles} \& {Jones}(1991)}]{1991MNRAS.248....1C}
{Coles}, P., \& {Jones}, B. 1991, \mnras, 248, 1, \dodoi{10.1093/mnras/248.1.1}

\bibitem[{{Cooray}(2006)}]{2006MNRAS.365..842C}
{Cooray}, A. 2006, \mnras, 365, 842, \dodoi{10.1111/j.1365-2966.2005.09747.x}

\bibitem[{{Cooray} \& {Sheth}(2002)}]{2002PhR...372....1C}
{Cooray}, A., \& {Sheth}, R. 2002, \physrep, 372, 1,
  \dodoi{10.1016/S0370-1573(02)00276-4}

\bibitem[{{Crocce} {et~al.}(2006){Crocce}, {Pueblas}, \&
  {Scoccimarro}}]{2006MNRAS.373..369C}
{Crocce}, M., {Pueblas}, S., \& {Scoccimarro}, R. 2006, \mnras, 373, 369,
  \dodoi{10.1111/j.1365-2966.2006.11040.x}

\bibitem[{{Croton} {et~al.}(2007){Croton}, {Gao}, \&
  {White}}]{2007MNRAS.374.1303C}
{Croton}, D.~J., {Gao}, L., \& {White}, S. D.~M. 2007, \mnras, 374, 1303,
  \dodoi{10.1111/j.1365-2966.2006.11230.x}

\bibitem[{{Cuesta-Lazaro} {et~al.}(2020){Cuesta-Lazaro}, {Li}, {Eggemeier},
  {Zarrouk}, {Baugh}, {Nishimichi}, \& {Takada}}]{2020arXiv200202683C}
{Cuesta-Lazaro}, C., {Li}, B., {Eggemeier}, A., {et~al.} 2020, arXiv e-prints,
  arXiv:2002.02683.
\newblock \doarXiv{2002.02683}

\bibitem[{{De Bernardis} {et~al.}(2017){De Bernardis}, {Aiola}, {Vavagiakis},
  {Battaglia}, {Niemack}, {Beall}, {Becker}, {Bond}, {Calabrese}, {Cho},
  {Coughlin}, {Datta}, {Devlin}, {Dunkley}, {Dunner}, {Ferraro}, {Fox},
  {Gallardo}, {Halpern}, {Hand}, {Hasselfield}, {Henderson}, {Hill}, {Hilton},
  {Hilton}, {Hincks}, {Hlozek}, {Hubmayr}, {Huffenberger}, {Hughes}, {Irwin},
  {Koopman}, {Kosowsky}, {Li}, {Louis}, {Lungu}, {Madhavacheril}, {Maurin},
  {McMahon}, {Moodley}, {Naess}, {Nati}, {Newburgh}, {Nibarger}, {Page},
  {Partridge}, {Schaan}, {Schmitt}, {Sehgal}, {Sievers}, {Simon}, {Spergel},
  {Staggs}, {Stevens}, {Thornton}, {van Engelen}, {Van Lanen}, \&
  {Wollack}}]{2017JCAP...03..008D}
{De Bernardis}, F., {Aiola}, S., {Vavagiakis}, E.~M., {et~al.} 2017, \jcap,
  2017, 008, \dodoi{10.1088/1475-7516/2017/03/008}

\bibitem[{{Desjacques} {et~al.}(2018){Desjacques}, {Jeong}, \&
  {Schmidt}}]{2018PhR...733....1D}
{Desjacques}, V., {Jeong}, D., \& {Schmidt}, F. 2018, \physrep, 733, 1,
  \dodoi{10.1016/j.physrep.2017.12.002}

\bibitem[{{Diemer} \& {Kravtsov}(2015)}]{2015ApJ...799..108D}
{Diemer}, B., \& {Kravtsov}, A.~V. 2015, \apj, 799, 108,
  \dodoi{10.1088/0004-637X/799/1/108}

\bibitem[{{Eisenstein} \& {Hu}(1998)}]{1998ApJ...496..605E}
{Eisenstein}, D.~J., \& {Hu}, W. 1998, \apj, 496, 605, \dodoi{10.1086/305424}

\bibitem[{{Guo} {et~al.}(2015){Guo}, {Zheng}, {Zehavi}, {Dawson}, {Skibba},
  {Tinker}, {Weinberg}, {White}, \& {Schneider}}]{2015MNRAS.446..578G}
{Guo}, H., {Zheng}, Z., {Zehavi}, I., {et~al.} 2015, \mnras, 446, 578,
  \dodoi{10.1093/mnras/stu2120}

\bibitem[{{Habib} {et~al.}(2007){Habib}, {Heitmann}, {Higdon}, {Nakhleh}, \&
  {Williams}}]{2007PhRvD..76h3503H}
{Habib}, S., {Heitmann}, K., {Higdon}, D., {Nakhleh}, C., \& {Williams}, B.
  2007, \prd, 76, 083503, \dodoi{10.1103/PhysRevD.76.083503}

\bibitem[{{Hadzhiyska} {et~al.}(2020){Hadzhiyska}, {Bose}, {Eisenstein},
  {Hernquist}, \& {Spergel}}]{2020MNRAS.493.5506H}
{Hadzhiyska}, B., {Bose}, S., {Eisenstein}, D., {Hernquist}, L., \& {Spergel},
  D.~N. 2020, \mnras, 493, 5506, \dodoi{10.1093/mnras/staa623}

\bibitem[{{Hamana} {et~al.}(2001){Hamana}, {Yoshida}, {Suto}, \&
  {Evrard}}]{2001ApJ...561L.143H}
{Hamana}, T., {Yoshida}, N., {Suto}, Y., \& {Evrard}, A.~E. 2001, \apjl, 561,
  L143, \dodoi{10.1086/324677}

\bibitem[{{Hand} {et~al.}(2012){Hand}, {Addison}, {Aubourg}, {Battaglia},
  {Battistelli}, {Bizyaev}, {Bond}, {Brewington}, {Brinkmann}, {Brown}, {Das},
  {Dawson}, {Devlin}, {Dunkley}, {Dunner}, {Eisenstein}, {Fowler}, {Gralla},
  {Hajian}, {Halpern}, {Hilton}, {Hincks}, {Hlozek}, {Hughes}, {Infante},
  {Irwin}, {Kosowsky}, {Lin}, {Malanushenko}, {Malanushenko}, {Marriage},
  {Marsden}, {Menanteau}, {Moodley}, {Niemack}, {Nolta}, {Oravetz}, {Page},
  {Palanque-Delabrouille}, {Pan}, {Reese}, {Schlegel}, {Schneider}, {Sehgal},
  {Shelden}, {Sievers}, {Sif{\'o}n}, {Simmons}, {Snedden}, {Spergel}, {Staggs},
  {Swetz}, {Switzer}, {Trac}, {Weaver}, {Wollack}, {Yeche}, \&
  {Zunckel}}]{2012PhRvL.109d1101H}
{Hand}, N., {Addison}, G.~E., {Aubourg}, E., {et~al.} 2012, \prl, 109, 041101,
  \dodoi{10.1103/PhysRevLett.109.041101}

\bibitem[{{Hearin}(2015)}]{2015MNRAS.451L..45H}
{Hearin}, A.~P. 2015, \mnras, 451, L45, \dodoi{10.1093/mnrasl/slv064}

\bibitem[{{Hellwing} {et~al.}(2014){Hellwing}, {Barreira}, {Frenk}, {Li}, \&
  {Cole}}]{2014PhRvL.112v1102H}
{Hellwing}, W.~A., {Barreira}, A., {Frenk}, C.~S., {Li}, B., \& {Cole}, S.
  2014, \prl, 112, 221102, \dodoi{10.1103/PhysRevLett.112.221102}

\bibitem[{{Ishiyama} {et~al.}(2015){Ishiyama}, {Enoki}, {Kobayashi}, {Makiya},
  {Nagashima}, \& {Oogi}}]{2015PASJ...67...61I}
{Ishiyama}, T., {Enoki}, M., {Kobayashi}, M. A.~R., {et~al.} 2015, \pasj, 67,
  61, \dodoi{10.1093/pasj/psv021}

\bibitem[{{Ishiyama} {et~al.}(2009){Ishiyama}, {Fukushige}, \&
  {Makino}}]{2009PASJ...61.1319I}
{Ishiyama}, T., {Fukushige}, T., \& {Makino}, J. 2009, \pasj, 61, 1319,
  \dodoi{10.1093/pasj/61.6.1319}

\bibitem[{{Ishiyama} {et~al.}(2012){Ishiyama}, {Nitadori}, \&
  {Makino}}]{2012arXiv1211.4406I}
{Ishiyama}, T., {Nitadori}, K., \& {Makino}, J. 2012, arXiv e-prints,
  arXiv:1211.4406.
\newblock \doarXiv{1211.4406}

\bibitem[{{Jain} \& {Zhang}(2008)}]{2008PhRvD..78f3503J}
{Jain}, B., \& {Zhang}, P. 2008, \prd, 78, 063503,
  \dodoi{10.1103/PhysRevD.78.063503}

\bibitem[{{Kayo} {et~al.}(2001){Kayo}, {Taruya}, \&
  {Suto}}]{2001ApJ...561...22K}
{Kayo}, I., {Taruya}, A., \& {Suto}, Y. 2001, \apj, 561, 22,
  \dodoi{10.1086/323227}

\bibitem[{{Kofman} {et~al.}(1994){Kofman}, {Bertschinger}, {Gelb}, {Nusser}, \&
  {Dekel}}]{1994ApJ...420...44K}
{Kofman}, L., {Bertschinger}, E., {Gelb}, J.~M., {Nusser}, A., \& {Dekel}, A.
  1994, \apj, 420, 44, \dodoi{10.1086/173541}

\bibitem[{{Komatsu} {et~al.}(2009){Komatsu}, {Dunkley}, {Nolta}, {Bennett},
  {Gold}, {Hinshaw}, {Jarosik}, {Larson}, {Limon}, {Page}, {Spergel},
  {Halpern}, {Hill}, {Kogut}, {Meyer}, {Tucker}, {Weiland}, {Wollack}, \&
  {Wright}}]{2009ApJS..180..330K}
{Komatsu}, E., {Dunkley}, J., {Nolta}, M.~R., {et~al.} 2009, \apjs, 180, 330,
  \dodoi{10.1088/0067-0049/180/2/330}

\bibitem[{{Kuruvilla} \& {Porciani}(2018)}]{2018MNRAS.479.2256K}
{Kuruvilla}, J., \& {Porciani}, C. 2018, \mnras, 479, 2256,
  \dodoi{10.1093/mnras/sty1654}

\bibitem[{{Kwan} {et~al.}(2013){Kwan}, {Bhattacharya}, {Heitmann}, \&
  {Habib}}]{2013ApJ...768..123K}
{Kwan}, J., {Bhattacharya}, S., {Heitmann}, K., \& {Habib}, S. 2013, \apj, 768,
  123, \dodoi{10.1088/0004-637X/768/2/123}

\bibitem[{{Kwan} {et~al.}(2015){Kwan}, {Heitmann}, {Habib}, {Padmanabhan},
  {Lawrence}, {Finkel}, {Frontiere}, \& {Pope}}]{2015ApJ...810...35K}
{Kwan}, J., {Heitmann}, K., {Habib}, S., {et~al.} 2015, \apj, 810, 35,
  \dodoi{10.1088/0004-637X/810/1/35}

\bibitem[{{Lam} {et~al.}(2012){Lam}, {Nishimichi}, {Schmidt}, \&
  {Takada}}]{2012PhRvL.109e1301L}
{Lam}, T.~Y., {Nishimichi}, T., {Schmidt}, F., \& {Takada}, M. 2012, \prl, 109,
  051301, \dodoi{10.1103/PhysRevLett.109.051301}

\bibitem[{{Lam} \& {Sheth}(2008)}]{2008MNRAS.389.1249L}
{Lam}, T.~Y., \& {Sheth}, R.~K. 2008, \mnras, 389, 1249,
  \dodoi{10.1111/j.1365-2966.2008.13621.x}

\bibitem[{{Landy} \& {Szalay}(1993)}]{1993ApJ...412...64L}
{Landy}, S.~D., \& {Szalay}, A.~S. 1993, \apj, 412, 64, \dodoi{10.1086/172900}

\bibitem[{{Lawrence} {et~al.}(2010){Lawrence}, {Heitmann}, {White}, {Higdon},
  {Wagner}, {Habib}, \& {Williams}}]{2010ApJ...713.1322L}
{Lawrence}, E., {Heitmann}, K., {White}, M., {et~al.} 2010, \apj, 713, 1322,
  \dodoi{10.1088/0004-637X/713/2/1322}

\bibitem[{{Leauthaud} {et~al.}(2012){Leauthaud}, {Tinker}, {Bundy}, {Behroozi},
  {Massey}, {Rhodes}, {George}, {Kneib}, {Benson}, {Wechsler}, {Busha},
  {Capak}, {Cort{\^e}s}, {Ilbert}, {Koekemoer}, {Le F{\`e}vre}, {Lilly},
  {McCracken}, {Salvato}, {Schrabback}, {Scoville}, {Smith}, \&
  {Taylor}}]{2012ApJ...744..159L}
{Leauthaud}, A., {Tinker}, J., {Bundy}, K., {et~al.} 2012, \apj, 744, 159,
  \dodoi{10.1088/0004-637X/744/2/159}

\bibitem[{Lewis {et~al.}(2000)Lewis, Challinor, \& Lasenby}]{Lewis:1999bs}
Lewis, A., Challinor, A., \& Lasenby, A. 2000, \apj, 538, 473,
  \dodoi{10.1086/309179}

\bibitem[{{Li} \& {Efstathiou}(2012)}]{2012MNRAS.421.1431L}
{Li}, B., \& {Efstathiou}, G. 2012, \mnras, 421, 1431,
  \dodoi{10.1111/j.1365-2966.2011.20404.x}

\bibitem[{{Masters} {et~al.}(2011){Masters}, {Maraston}, {Nichol}, {Thomas},
  {Beifiori}, {Bundy}, {Edmondson}, {Higgs}, {Leauthaud}, {Mandelbaum},
  {Pforr}, {Ross}, {Ross}, {Schneider}, {Skibba}, {Tinker}, {Tojeiro}, {Wake},
  {Brinkmann}, \& {Weaver}}]{2011MNRAS.418.1055M}
{Masters}, K.~L., {Maraston}, C., {Nichol}, R.~C., {et~al.} 2011, \mnras, 418,
  1055, \dodoi{10.1111/j.1365-2966.2011.19557.x}

\bibitem[{{McClintock} {et~al.}(2019){McClintock}, {Rozo}, {Becker}, {DeRose},
  {Mao}, {McLaughlin}, {Tinker}, {Wechsler}, \& {Zhai}}]{2019ApJ...872...53M}
{McClintock}, T., {Rozo}, E., {Becker}, M.~R., {et~al.} 2019, \apj, 872, 53,
  \dodoi{10.3847/1538-4357/aaf568}

\bibitem[{{More} {et~al.}(2015){More}, {Miyatake}, {Mandelbaum}, {Takada},
  {Spergel}, {Brownstein}, \& {Schneider}}]{2015ApJ...806....2M}
{More}, S., {Miyatake}, H., {Mandelbaum}, R., {et~al.} 2015, \apj, 806, 2,
  \dodoi{10.1088/0004-637X/806/1/2}

\bibitem[{{Navarro} {et~al.}(1996){Navarro}, {Frenk}, \&
  {White}}]{1996ApJ...462..563N}
{Navarro}, J.~F., {Frenk}, C.~S., \& {White}, S. D.~M. 1996, \apj, 462, 563,
  \dodoi{10.1086/177173}

\bibitem[{{Nishimichi} {et~al.}(2019){Nishimichi}, {Takada}, {Takahashi},
  {Osato}, {Shirasaki}, {Oogi}, {Miyatake}, {Oguri}, {Murata}, {Kobayashi}, \&
  {Yoshida}}]{2019ApJ...884...29N}
{Nishimichi}, T., {Takada}, M., {Takahashi}, R., {et~al.} 2019, \apj, 884, 29,
  \dodoi{10.3847/1538-4357/ab3719}

\bibitem[{{Padilla} {et~al.}(2019){Padilla}, {Contreras}, {Zehavi}, {Baugh}, \&
  {Norberg}}]{2019MNRAS.486..582P}
{Padilla}, N., {Contreras}, S., {Zehavi}, I., {Baugh}, C.~M., \& {Norberg}, P.
  2019, \mnras, 486, 582, \dodoi{10.1093/mnras/stz824}

\bibitem[{{Peebles}(1980)}]{1980lssu.book.....P}
{Peebles}, P.~J.~E. 1980, {The large-scale structure of the universe}

\bibitem[{{Planck Collaboration} {et~al.}(2016){Planck Collaboration}, {Ade},
  {Aghanim}, {Arnaud}, {Ashdown}, {Aumont}, {Baccigalupi}, {Banday},
  {Barreiro}, {Bartlett}, {Bartolo}, {Battaner}, {Battye}, {Benabed},
  {Beno{\^\i}t}, {Benoit-L{\'e}vy}, {Bernard}, {Bersanelli}, {Bielewicz},
  {Bock}, {Bonaldi}, {Bonavera}, {Bond}, {Borrill}, {Bouchet}, {Boulanger},
  {Bucher}, {Burigana}, {Butler}, {Calabrese}, {Cardoso}, {Catalano},
  {Challinor}, {Chamballu}, {Chary}, {Chiang}, {Chluba}, {Christensen},
  {Church}, {Clements}, {Colombi}, {Colombo}, {Combet}, {Coulais}, {Crill},
  {Curto}, {Cuttaia}, {Danese}, {Davies}, {Davis}, {de Bernardis}, {de Rosa},
  {de Zotti}, {Delabrouille}, {D{\'e}sert}, {Di Valentino}, {Dickinson},
  {Diego}, {Dolag}, {Dole}, {Donzelli}, {Dor{\'e}}, {Douspis}, {Ducout},
  {Dunkley}, {Dupac}, {Efstathiou}, {Elsner}, {En{\ss}lin}, {Eriksen},
  {Farhang}, {Fergusson}, {Finelli}, {Forni}, {Frailis}, {Fraisse},
  {Franceschi}, {Frejsel}, {Galeotta}, {Galli}, {Ganga}, {Gauthier}, {Gerbino},
  {Ghosh}, {Giard}, {Giraud-H{\'e}raud}, {Giusarma}, {Gjerl{\o}w},
  {Gonz{\'a}lez-Nuevo}, {G{\'o}rski}, {Gratton}, {Gregorio}, {Gruppuso},
  {Gudmundsson}, {Hamann}, {Hansen}, {Hanson}, {Harrison}, {Helou},
  {Henrot-Versill{\'e}}, {Hern{\'a}ndez-Monteagudo}, {Herranz}, {Hildebrand t},
  {Hivon}, {Hobson}, {Holmes}, {Hornstrup}, {Hovest}, {Huang}, {Huffenberger},
  {Hurier}, {Jaffe}, {Jaffe}, {Jones}, {Juvela}, {Keih{\"a}nen}, {Keskitalo},
  {Kisner}, {Kneissl}, {Knoche}, {Knox}, {Kunz}, {Kurki-Suonio}, {Lagache},
  {L{\"a}hteenm{\"a}ki}, {Lamarre}, {Lasenby}, {Lattanzi}, {Lawrence}, {Leahy},
  {Leonardi}, {Lesgourgues}, {Levrier}, {Lewis}, {Liguori}, {Lilje},
  {Linden-V{\o}rnle}, {L{\'o}pez-Caniego}, {Lubin}, {Mac{\'\i}as-P{\'e}rez},
  {Maggio}, {Maino}, {Mandolesi}, {Mangilli}, {Marchini}, {Maris}, {Martin},
  {Martinelli}, {Mart{\'\i}nez-Gonz{\'a}lez}, {Masi}, {Matarrese}, {McGehee},
  {Meinhold}, {Melchiorri}, {Melin}, {Mendes}, {Mennella}, {Migliaccio},
  {Millea}, {Mitra}, {Miville-Desch{\^e}nes}, {Moneti}, {Montier}, {Morgante},
  {Mortlock}, {Moss}, {Munshi}, {Murphy}, {Naselsky}, {Nati}, {Natoli},
  {Netterfield}, {N{\o}rgaard-Nielsen}, {Noviello}, {Novikov}, {Novikov},
  {Oxborrow}, {Paci}, {Pagano}, {Pajot}, {Paladini}, {Paoletti}, {Partridge},
  {Pasian}, {Patanchon}, {Pearson}, {Perdereau}, {Perotto}, {Perrotta},
  {Pettorino}, {Piacentini}, {Piat}, {Pierpaoli}, {Pietrobon}, {Plaszczynski},
  {Pointecouteau}, {Polenta}, {Popa}, {Pratt}, {Pr{\'e}zeau}, {Prunet},
  {Puget}, {Rachen}, {Reach}, {Rebolo}, {Reinecke}, {Remazeilles}, {Renault},
  {Renzi}, {Ristorcelli}, {Rocha}, {Rosset}, {Rossetti}, {Roudier},
  {Rouill{\'e} d'Orfeuil}, {Rowan-Robinson}, {Rubi{\~n}o-Mart{\'\i}n},
  {Rusholme}, {Said}, {Salvatelli}, {Salvati}, {Sandri}, {Santos},
  {Savelainen}, {Savini}, {Scott}, {Seiffert}, {Serra}, {Shellard}, {Spencer},
  {Spinelli}, {Stolyarov}, {Stompor}, {Sudiwala}, {Sunyaev}, {Sutton},
  {Suur-Uski}, {Sygnet}, {Tauber}, {Terenzi}, {Toffolatti}, {Tomasi},
  {Tristram}, {Trombetti}, {Tucci}, {Tuovinen}, {T{\"u}rler}, {Umana},
  {Valenziano}, {Valiviita}, {Van Tent}, {Vielva}, {Villa}, {Wade}, {Wandelt},
  {Wehus}, {White}, {White}, {Wilkinson}, {Yvon}, {Zacchei}, \&
  {Zonca}}]{2016A&A...594A..13P}
{Planck Collaboration}, {Ade}, P.~A.~R., {Aghanim}, N., {et~al.} 2016, \aap,
  594, A13, \dodoi{10.1051/0004-6361/201525830}

\bibitem[{{Prada} {et~al.}(2012){Prada}, {Klypin}, {Cuesta}, {Betancort-Rijo},
  \& {Primack}}]{2012MNRAS.423.3018P}
{Prada}, F., {Klypin}, A.~A., {Cuesta}, A.~J., {Betancort-Rijo}, J.~E., \&
  {Primack}, J. 2012, \mnras, 423, 3018,
  \dodoi{10.1111/j.1365-2966.2012.21007.x}

\bibitem[{{Reid} {et~al.}(2014){Reid}, {Seo}, {Leauthaud}, {Tinker}, \&
  {White}}]{2014MNRAS.444..476R}
{Reid}, B.~A., {Seo}, H.-J., {Leauthaud}, A., {Tinker}, J.~L., \& {White}, M.
  2014, \mnras, 444, 476, \dodoi{10.1093/mnras/stu1391}

\bibitem[{{Reid} \& {Spergel}(2009)}]{2009ApJ...698..143R}
{Reid}, B.~A., \& {Spergel}, D.~N. 2009, \apj, 698, 143,
  \dodoi{10.1088/0004-637X/698/1/143}

\bibitem[{{Scoccimarro}(2004)}]{2004PhRvD..70h3007S}
{Scoccimarro}, R. 2004, \prd, 70, 083007, \dodoi{10.1103/PhysRevD.70.083007}

\bibitem[{{Shen} {et~al.}(2006){Shen}, {Abel}, {Mo}, \&
  {Sheth}}]{2006ApJ...645..783S}
{Shen}, J., {Abel}, T., {Mo}, H.~J., \& {Sheth}, R.~K. 2006, \apj, 645, 783,
  \dodoi{10.1086/504513}

\bibitem[{{Shin} {et~al.}(2017){Shin}, {Kim}, {Pichon}, {Jeong}, \&
  {Park}}]{2017ApJ...843...73S}
{Shin}, J., {Kim}, J., {Pichon}, C., {Jeong}, D., \& {Park}, C. 2017, \apj,
  843, 73, \dodoi{10.3847/1538-4357/aa74b9}

\bibitem[{{Smith} {et~al.}(2018){Smith}, {Madhavacheril}, {M{\"u}nchmeyer},
  {Ferraro}, {Giri}, \& {Johnson}}]{2018arXiv181013423S}
{Smith}, K.~M., {Madhavacheril}, M.~S., {M{\"u}nchmeyer}, M., {et~al.} 2018,
  arXiv e-prints, arXiv:1810.13423.
\newblock \doarXiv{1810.13423}

\bibitem[{{Sugiyama} {et~al.}(2017){Sugiyama}, {Okumura}, \&
  {Spergel}}]{2017JCAP...01..057S}
{Sugiyama}, N.~S., {Okumura}, T., \& {Spergel}, D.~N. 2017, \jcap, 2017, 057,
  \dodoi{10.1088/1475-7516/2017/01/057}

\bibitem[{{Sugiyama} {et~al.}(2018){Sugiyama}, {Okumura}, \&
  {Spergel}}]{2018MNRAS.475.3764S}
---. 2018, \mnras, 475, 3764, \dodoi{10.1093/mnras/stx3362}

\bibitem[{{Takahashi} {et~al.}(2012){Takahashi}, {Sato}, {Nishimichi},
  {Taruya}, \& {Oguri}}]{2012ApJ...761..152T}
{Takahashi}, R., {Sato}, M., {Nishimichi}, T., {Taruya}, A., \& {Oguri}, M.
  2012, \apj, 761, 152, \dodoi{10.1088/0004-637X/761/2/152}

\bibitem[{{Tinker} {et~al.}(2008){Tinker}, {Kravtsov}, {Klypin}, {Abazajian},
  {Warren}, {Yepes}, {Gottl{\"o}ber}, \& {Holz}}]{2008ApJ...688..709T}
{Tinker}, J., {Kravtsov}, A.~V., {Klypin}, A., {et~al.} 2008, \apj, 688, 709,
  \dodoi{10.1086/591439}

\bibitem[{{Tinker}(2007)}]{2007MNRAS.374..477T}
{Tinker}, J.~L. 2007, \mnras, 374, 477,
  \dodoi{10.1111/j.1365-2966.2006.11157.x}

\bibitem[{{Tinker} {et~al.}(2010){Tinker}, {Robertson}, {Kravtsov}, {Klypin},
  {Warren}, {Yepes}, \& {Gottl{\"o}ber}}]{2010ApJ...724..878T}
{Tinker}, J.~L., {Robertson}, B.~E., {Kravtsov}, A.~V., {et~al.} 2010, \apj,
  724, 878, \dodoi{10.1088/0004-637X/724/2/878}

\bibitem[{{Tinker} {et~al.}(2005){Tinker}, {Weinberg}, {Zheng}, \&
  {Zehavi}}]{2005ApJ...631...41T}
{Tinker}, J.~L., {Weinberg}, D.~H., {Zheng}, Z., \& {Zehavi}, I. 2005, \apj,
  631, 41, \dodoi{10.1086/432084}

\bibitem[{{Tinker} {et~al.}(2017){Tinker}, {Brownstein}, {Guo}, {Leauthaud},
  {Maraston}, {Masters}, {Montero-Dorta}, {Thomas}, {Tojeiro}, {Weiner},
  {Zehavi}, \& {Olmstead}}]{2017ApJ...839..121T}
{Tinker}, J.~L., {Brownstein}, J.~R., {Guo}, H., {et~al.} 2017, \apj, 839, 121,
  \dodoi{10.3847/1538-4357/aa6845}

\bibitem[{{van den Bosch} {et~al.}(2013){van den Bosch}, {More}, {Cacciato},
  {Mo}, \& {Yang}}]{2013MNRAS.430..725V}
{van den Bosch}, F.~C., {More}, S., {Cacciato}, M., {Mo}, H., \& {Yang}, X.
  2013, \mnras, 430, 725, \dodoi{10.1093/mnras/sts006}

\bibitem[{{Weinberg} {et~al.}(2013){Weinberg}, {Mortonson}, {Eisenstein},
  {Hirata}, {Riess}, \& {Rozo}}]{2013PhR...530...87W}
{Weinberg}, D.~H., {Mortonson}, M.~J., {Eisenstein}, D.~J., {et~al.} 2013,
  \physrep, 530, 87, \dodoi{10.1016/j.physrep.2013.05.001}

\bibitem[{{White} {et~al.}(2011){White}, {Blanton}, {Bolton}, {Schlegel},
  {Tinker}, {Berlind}, {da Costa}, {Kazin}, {Lin}, {Maia}, {McBride},
  {Padmanabhan}, {Parejko}, {Percival}, {Prada}, {Ramos}, {Sheldon}, {de
  Simoni}, {Skibba}, {Thomas}, {Wake}, {Zehavi}, {Zheng}, {Nichol},
  {Schneider}, {Strauss}, {Weaver}, \& {Weinberg}}]{2011ApJ...728..126W}
{White}, M., {Blanton}, M., {Bolton}, A., {et~al.} 2011, \apj, 728, 126,
  \dodoi{10.1088/0004-637X/728/2/126}

\bibitem[{{Xu} \& {Zheng}(2018)}]{2018MNRAS.479.1579X}
{Xu}, X., \& {Zheng}, Z. 2018, \mnras, 479, 1579, \dodoi{10.1093/mnras/sty1547}

\bibitem[{{Zehavi} {et~al.}(2005){Zehavi}, {Zheng}, {Weinberg}, {Frieman},
  {Berlind}, {Blanton}, {Scoccimarro}, {Sheth}, {Strauss}, {Kayo}, {Suto},
  {Fukugita}, {Nakamura}, {Bahcall}, {Brinkmann}, {Gunn}, {Hennessy},
  {Ivezi{\'c}}, {Knapp}, {Loveday}, {Meiksin}, {Schlegel}, {Schneider},
  {Szapudi}, {Tegmark}, {Vogeley}, {York}, \& {SDSS
  Collaboration}}]{2005ApJ...630....1Z}
{Zehavi}, I., {Zheng}, Z., {Weinberg}, D.~H., {et~al.} 2005, \apj, 630, 1,
  \dodoi{10.1086/431891}

\bibitem[{{Zehavi} {et~al.}(2011){Zehavi}, {Zheng}, {Weinberg}, {Blanton},
  {Bahcall}, {Berlind}, {Brinkmann}, {Frieman}, {Gunn}, {Lupton}, {Nichol},
  {Percival}, {Schneider}, {Skibba}, {Strauss}, {Tegmark}, \&
  {York}}]{2011ApJ...736...59Z}
---. 2011, \apj, 736, 59, \dodoi{10.1088/0004-637X/736/1/59}

\bibitem[{{Zentner} {et~al.}(2014){Zentner}, {Hearin}, \& {van den
  Bosch}}]{2014MNRAS.443.3044Z}
{Zentner}, A.~R., {Hearin}, A.~P., \& {van den Bosch}, F.~C. 2014, \mnras, 443,
  3044, \dodoi{10.1093/mnras/stu1383}

\bibitem[{{Zhai} {et~al.}(2019){Zhai}, {Tinker}, {Becker}, {DeRose}, {Mao},
  {McClintock}, {McLaughlin}, {Rozo}, \& {Wechsler}}]{2019ApJ...874...95Z}
{Zhai}, Z., {Tinker}, J.~L., {Becker}, M.~R., {et~al.} 2019, \apj, 874, 95,
  \dodoi{10.3847/1538-4357/ab0d7b}

\bibitem[{{Zheng} {et~al.}(2007){Zheng}, {Coil}, \&
  {Zehavi}}]{2007ApJ...667..760Z}
{Zheng}, Z., {Coil}, A.~L., \& {Zehavi}, I. 2007, \apj, 667, 760,
  \dodoi{10.1086/521074}

\bibitem[{{Zheng} \& {Guo}(2016)}]{2016MNRAS.458.4015Z}
{Zheng}, Z., \& {Guo}, H. 2016, \mnras, 458, 4015, \dodoi{10.1093/mnras/stw523}

\bibitem[{{Zheng} {et~al.}(2009){Zheng}, {Zehavi}, {Eisenstein}, {Weinberg}, \&
  {Jing}}]{2009ApJ...707..554Z}
{Zheng}, Z., {Zehavi}, I., {Eisenstein}, D.~J., {Weinberg}, D.~H., \& {Jing},
  Y.~P. 2009, \apj, 707, 554, \dodoi{10.1088/0004-637X/707/1/554}

\bibitem[{{Zu} \& {Weinberg}(2013)}]{2013MNRAS.431.3319Z}
{Zu}, Y., \& {Weinberg}, D.~H. 2013, \mnras, 431, 3319,
  \dodoi{10.1093/mnras/stt411}

\bibitem[{{Zu} {et~al.}(2014){Zu}, {Weinberg}, {Jennings}, {Li}, \&
  {Wyman}}]{2014MNRAS.445.1885Z}
{Zu}, Y., {Weinberg}, D.~H., {Jennings}, E., {Li}, B., \& {Wyman}, M. 2014,
  \mnras, 445, 1885, \dodoi{10.1093/mnras/stu1739}

\end{thebibliography}



\end{document}